\documentclass[12pt]{article}
\pdfoutput=1
\usepackage{color,amsmath,amssymb,exscale,psfrag,epsfig}
\usepackage{cite,color,url}
\usepackage{pdfpages}
\usepackage{graphicx}
\usepackage{slashed}
\usepackage[utf8]{inputenc}
\usepackage[T1]{fontenc}
\usepackage{xcolor}
\usepackage{accents}
\usepackage{centernot}
\usepackage{enumerate}
\usepackage[toc]{appendix}
\usepackage{subfig}
\usepackage{float}

\usepackage[colorlinks=true
,urlcolor=blue
,anchorcolor=blue
,citecolor=blue
,filecolor=blue
,linkcolor=blue
,menucolor=blue
,linktocpage=true
,pdfproducer=medialab
]{hyperref}
\input epsf
\usepackage{lineno}
\def\bea{\begin{eqnarray}}
\def\eea{\end{eqnarray}}

\definecolor{nicered}{rgb}{0.7,0.1,0.1}
\definecolor{nicegreen}{rgb}{0.1,0.5,0.1}
\hypersetup{colorlinks,citecolor= nicegreen,linkcolor= nicered}

\catcode`\@=11
\def\lsim{\mathrel{\mathpalette\@versim<}}
\def\gsim{\mathrel{\mathpalette\@versim>}}
\def\@versim#1#2{\vcenter{\offinterlineskip
\ialign{$\m@th#1\hfil##\hfil$\crcr#2\crcr\sim\crcr } }}
\catcode`\@=12
\catcode`@=12
\begin{document}
\thispagestyle{empty}
\begin{flushright}
ICAS 055/20
\end{flushright}
\vspace{0.1in}
\begin{center}
{\Large \bf Topping-up multilepton plus b-jets anomalies\\ at the LHC with a $Z'$ boson} \\
\vspace{0.2in}
{\bf Ezequiel Alvarez$^{(a)\dagger}$,
Aurelio Juste$^{(b,c)\star}$,
Manuel Szewc$^{(a)\diamond}$,\\
Tamara Vazquez Schroeder$^{(d)\ast}$
}
\vspace{0.2in} \\
	{\sl $^{(a)}$ International Center for Advanced Studies (ICAS), UNSAM \& CONICET\\
25 de Mayo y Francia, (1650) Buenos Aires, Argentina }
\\[1ex]
{\sl $^{(b)}$ 
Institut de F\'isica d'Altes Energies (IFAE), Edifici Cn, Facultat de Ciencies,\\
Universitat Aut\`onoma de Barcelona, E-08193 Bellaterra, Barcelona, Spain}
\\[1ex]
{\sl $^{(c)}$ 
Instituci\'o Catalana de Recerca i Estudis Avan\c{c}ats (ICREA), E-08010 Barcelona, Spain}
\\[1ex]
{\sl $^{(d)}$ 
CERN, CH-1211 Geneva, Switzerland}
\end{center}
\vspace{0.1in}

\begin{abstract}
	During the last years ATLAS and CMS have reported a number of slight to mild discrepancies in signatures of multileptons plus $b$-jets in analyses such as $t\bar t H$, $t\bar t W^\pm$, $t\bar t Z$ and $t\bar t t\bar t$.  Among them, a recent ATLAS result on $t\bar t H$ production has also reported an excess in the charge asymmetry in the same-sign dilepton channel with two or more $b$-tagged jets. Motivated by these tantalizing discrepancies, we study a phenomenological New Physics model consisting of a $Z'$ boson that couples to up-type quarks via right-handed currents: $t_R\gamma^\mu \bar t_R$, $t_R\gamma^\mu \bar c_R$, and $t_R \gamma^\mu \bar u_R$.  The latter vertex allows to translate the charge asymmetry at the LHC initial state protons to a final state with top quarks which, decaying to a positive lepton and a $b$-jet, provides a crucial contribution to some of the observed discrepancies. Through an analysis at a detector level, we select the region in parameter space of our model that best reproduces the data in the aforementioned $t\bar t H$ study, and in a recent ATLAS $t\bar t t \bar t$ search. We find that our model provides a better fit to the experimental data than the Standard Model for a New Physics scale of approximately $\sim$500 GeV, and with a hierarchical coupling of the $Z'$ boson that favours the top quark and the presence of FCNC currents.  In order to estimate the LHC sensitivity to this signal, we design a broadband search featuring many kinematic regions with different signal-to-background ratio, and perform a global analysis.  We also define signal-enhanced regions and study observables that could further distinguish signal from background. We find that the region in parameter space of our model that best fits the analysed data could be probed with a significance exceeding 3 standard deviations with just the full Run-2 dataset.
\end{abstract}

\vspace*{2mm}
\noindent {\footnotesize E-mail:
{\tt 
$\dagger$ \href{mailto:sequi@unsam.edu.ar}{sequi@unsam.edu.ar},
$\star$ \href{mailto:juste@ifae.es}{juste@ifae.es},\\
$\diamond$ \href{mailto:mszewc@unsam.edu.ar}{mszewc@unsam.edu.ar}
$\ast$ \href{mailto:tamara.vazquez.schroeder@cern.ch}{tamara.vazquez.schroeder@cern.ch},
}}

%%%%%%%%%%%%%%%%%%%%%%%%%%%%%%%%%%%%%%%%%%%%%%%%%%%

\newpage
\section{Introduction}
\label{section:1}

Since the beginning of operation of the Large Hadron Collider (LHC), the ATLAS and CMS collaborations have carried out a broad program of precision measurements of Standard Model (SM) parameters and processes, as well as sensitive searches for new phenomena beyond the SM (BSM), all of which have furthered our understanding of the fundamental interactions of Nature. However, we know the SM to be an incomplete theory and we still hope to find New Physics at the LHC. Being a proton-proton collider, the LHC is a discovery machine with an energy range that would ideally produce BSM resonances at the TeV scale. These resonances have been searched for in many different channels, motivated by a variety of BSM models. As these resonance searches have excluded a larger portion of the parameter space for the simplest models, the attention has turned to precision measurements where deviations from SM predictions could provide hints to BSM signatures. Such BSM could still be produced in a resonant way but, without specific dedicated searches, this resonant feature could go unnoticed.

Among the most recent SM benchmarks being explored, the LHC program has made impressive progress in the measurement of the cross-sections of $t\overline{t}W^{\pm}$\cite{Aaboud:2019njj, Sirunyan:2017uzs}, $t\overline{t}H$~\cite{ATLAS-CONF-2019-045, Sirunyan:2020icl} and four-top-quarks ($t\overline{t}t\overline{t}$)~\cite{Aad:2020klt,Sirunyan:2019wxt} production. To measure these cross-sections, a careful choice of the final state must be made. The same-sign dilepton and multilepton final states with a high $b$-jet multiplicity present a good balance between low SM backgrounds and high signal yield. In these final states, a persistent discrepancy has been present in the $t\overline{t}W^{\pm}$ normalization, as the observed yields are consistently larger than those expected from state-of-art theoretical predictions. This discrepancy has been found both in dedicated $t\overline{t}W^{\pm}$ cross-section measurements~\cite{Aaboud:2019njj, Sirunyan:2017uzs} and in several searches that consider the $t\overline{t}W^{\pm}$ process as a background~\cite{ATLAS-CONF-2019-045, Sirunyan:2020icl,Aad:2020klt, Sirunyan:2019wxt}, as summarised in Table~\ref{table:ttW}. We see that the ATLAS and CMS $t\overline{t}H$ measurements~\cite{ATLAS-CONF-2019-045, Sirunyan:2020icl} analyse a larger dataset than the $t\overline{t}W^{\pm}$ cross-section measurements~\cite{Aaboud:2019njj, Sirunyan:2017uzs}. If we consider the same reference cross-section for all searches, using for example the reported cross-section of Ref.~\cite{deFlorian:2016spz}, we see that the $t\overline{t}H$ measurements show a larger tension than the $t\overline{t}W^{\pm}$ measurements. Because of this two facts, we focus on the $t\overline{t}H$ searches. Despite having similar strategies to suppress non-prompt lepton background, there are several differences in the $t\overline{t}H$ analysis approach between ATLAS and CMS, such as different fake estimation techniques (simultaneous profile likelihood template fit in ATLAS vs misidentification probability method in CMS) and different analysis strategies (jet multiplicity, total lepton charge, and Boosted Decision Trees (BDT) categorisation in ATLAS vs Deep Neural Networks in CMS). A summary of the event selection used in the ATLAS and CMS $t\overline{t}H$ analyses can be found in Table~\ref{table:ttH_sel}. We focus on the two same-sign dilepton and trilepton channels where hadronically decaying $\tau$s are vetoed.

We choose the ATLAS $t\overline{t}H$ measurement for the re-interpretation in this work since it provides complete information on the multilepton discrepancies observed as a function of the total lepton charge and the b-jet multiplicity of the event. A recent CMS search for new physics within the effective field theory (EFT) framework~\cite{Sirunyan:2020tqm} using 41.5 fb$^{-1}$ of integrated luminosity exploits a similar categorisation. However, since there is no information on the two same-sign dilepton selection with exactly 1 b-jet and given the lower data statistics, we do not include the CMS EFT results in the current work.

In addition, a four-top-quarks search by the ATLAS Collaboration in the same final states has measured a four-top-quarks cross-section a factor of two higher than the SM prediction~\cite{Aad:2020klt}. A summary of the signal region event selection used in the ATLAS and CMS four-top-quarks analyses can be found in Table~\ref{table:tttt_sel}. 

It should be pointed out that these measurements are individually consistent with the SM and thus the observed discrepancies could be merely the result of statistical fluctuations, and/or unaccounted experimental or theoretical uncertainties. However, when combined, they paint an interesting picture that is worth exploring, as it may open the door to new exciting discoveries.

\begin{table}[h!]
\centering
 \begin{tabular}{|p{5cm}|p{2cm}|p{3cm}|p{2.5cm}|p{2.5cm}|} 
 \hline
 Search & $\mathcal{L}$ $[\text{fb}^{-1}]$ & $\sigma_{\text{ref}}$ [$\text{pb}$] & $\mu$& $\mu_{\text{YR4}}$\\ [0.5ex] 
 \hline\hline
 $t\overline{t}W^{\pm}$ ATLAS~\cite{Aaboud:2019njj} & 36.1 & $0.60\pm0.07$ & $1.44\pm0.32$ & $1.44\pm0.32$\\ 
 $t\overline{t}W^{\pm}$ CMS~\cite{Sirunyan:2017uzs} & 35.9 & $0.628\pm0.082$ & $1.23^{+0.30}_{-0.28}$ & $1.29^{+0.31}_{-0.29}$\\%^{+0.19}_{-0.18}\left(\text{stat}\right)^{+0.20}_{-0.18}\left(\text{syst}\right)^{+0.13}_{-0.12}\left(\text{theo}\right)$ \\
 $t\overline{t}H$ ATLAS~\cite{ATLAS-CONF-2019-045}& 80 & $0.727\pm0.092$ & $1.39^{+0.17}_{-0.16}$ & $1.68^{+0.21}_{-0.19}$\\
 $t\overline{t}H$ CMS~\cite{Sirunyan:2020icl}& 137 & $0.650$ & $1.43\pm0.21$ & $1.55\pm0.23$\\
 four-top-quarks ATLAS~\cite{Aad:2020klt}& 139 & $0.601$ & $1.6\pm0.3$ & $1.6\pm0.3$\\
 four-top-quarks CMS~\cite{Sirunyan:2019wxt} & 137 & $0.610$ & $1.3\pm0.2$ & $1.3\pm0.2$  \\ [1ex] 
 \hline
 \end{tabular}
 \caption{This table lists the $t\overline{t}W^{\pm}$ reference cross-sections and the corresponding signal strengths for different searches. The last column is the signal strength corresponding to the reference cross-section listed in Ref.~\cite{deFlorian:2016spz}, $600.8$ fb. The reference cross-section was not listed in Ref.~\cite{Sirunyan:2019wxt} and so it was taken from Ref.~\cite{Banelli:2020iau}.}
 \label{table:ttW}
\end{table}

\begin{table}[h!]
\centering
 \begin{tabular}{|p{4.5cm}|p{1.25cm}|p{4cm}|p{1.25cm}|p{4cm}|} 
 \hline
 $t\overline{t}H$ & \multicolumn{2}{c}{ATLAS} & \multicolumn{2}{|c|}{CMS} \\ [0.5ex] 
 \hline
 	                      & 2LSS & 3L & 2LSS & 3L \\
 \hline\hline
Total lepton charge & $\pm 2$ & $\pm 1$ & $\pm 2$ & $\pm 1$ \\
\hline
Lepton $p_{T}$ [GeV] & 20/20 & 15/15/10 & 25/15 & 25/15/10 \\
\hline
Number of jets & \multicolumn{2}{c|}{$\geq 2$} & $\geq 3$ & $\geq 2$ \\
\hline
Number of b-jets & \multicolumn{2}{c}{$\geq 1$ (70\% eff.)} & \multicolumn{2}{|c|}{$\geq 1$ (70\% eff.) OR} \\
 & \multicolumn{2}{c}{} & \multicolumn{2}{|c|}{$\geq 2$ (84\% eff.)} \\
 \hline
$|m_{\ell \ell}|$ (2LSS) or & \multicolumn{4}{c|}{> 12} \\
$|m_{OS SF}|$ (3L) [GeV]         & \multicolumn{4}{c|}{ } \\
\hline
$|m_{e^{\pm}e^{\pm}} - m_Z|$ (2LSS) or & - & \multicolumn{3}{c|}{> 10} \\
$|m_{OS SF} - m_Z|$  (3L) [GeV]    & & \multicolumn{3}{c|}{ }\\
\hline
Other      & -  & $|m_{\ell \ell \ell} - m_Z| > 10$ GeV & \multicolumn{2}{|c|}{Missing transverse} \\
      &   & & \multicolumn{2}{|c|}{momentum cuts} \\
\hline
 \end{tabular}
 \caption{Comparison of event selections between the ATLAS~\cite{ATLAS-CONF-2019-045} and CMS~\cite{Sirunyan:2020icl} $t\overline{t}H$ analyses. }
 \label{table:ttH_sel}
\end{table}

\begin{table}[h!]
\centering
 \begin{tabular}{|p{4.5cm}|p{2.25cm}|p{2.25cm}|p{3.0cm}|p{3.0cm}|} 
 \hline
 four-top-quarks & \multicolumn{2}{c}{ATLAS} & \multicolumn{2}{|c|}{CMS} \\ [0.5ex] 
 \hline
 	                      & 2LSS & $\geq$ 3L & 2LSS & $\geq$ 3L \\
 \hline\hline
Total lepton charge & $\pm 2$ & - & $\pm 2$ & - \\
\hline
Lepton $p_{T}$ [GeV] & \multicolumn{2}{c|}{28 (all $\ell$)} & 25/20 & 25/20/20(/20) \\
\hline
Number of jets & \multicolumn{2}{c|}{$\geq 6$j} & $\geq 6$j $\geq 2$bj OR & $\geq 5$j $\geq 2$bj OR \\
and b-jets & \multicolumn{2}{c|}{$\geq 2$bj (77\% eff.)} & $5$j $\geq 3$bj & $4$j $\geq 3$bj \\
           & \multicolumn{2}{c|}{ } &  (55-70\% eff.) &  (55-70\% eff.) \\
 \hline
$H_{T}$ [GeV] & \multicolumn{2}{c|}{> 500} & \multicolumn{2}{c|}{> 300} \\ 
\hline
$|m_{e^{\pm}e^{\pm}}|$ (2LSS) or & > 15 & - & \multicolumn{2}{c|}{> 12} \\
$|m_{OS SF}|$ (3L) [GeV]         &  &  & \multicolumn{2}{c|}{ } \\
\hline
$|m_{e^{\pm}e^{\pm}} - m_Z|$ (2LSS) or & \multicolumn{2}{c|}{> 10} & - & > 15 \\
$|m_{OS SF} - m_Z|$  (3L) [GeV]    & \multicolumn{2}{c|}{ } & & \\
\hline
Other      &  \multicolumn{2}{c}{-} & \multicolumn{2}{|c|}{Missing transverse} \\
           &  \multicolumn{2}{c}{} & \multicolumn{2}{|c|}{momentum cuts} \\
\hline
 \end{tabular}
 \caption{Comparison of event selections between the ATLAS~\cite{Aad:2020klt} and CMS~\cite{Sirunyan:2019wxt} four-top-quarks analyses. $H_{T}$ is the scalar $p_{T}$ sum of jets, leptons and b-jets. }
 \label{table:tttt_sel}
\end{table}

In this work, we propose a purely phenomenological BSM model, where a $Z'$ spin-1 boson with mass below 1 TeV is responsible for the reported discrepancies in $t\overline{t}H$ and four-top-quarks searches. We choose this model with a particular coupling structure as it is able to provide both charge asymmetry and high $b$-jet multiplicity, key ingredients to accommodate the different experimental signatures. 

This paper is organized as follows. We present our phenomenological model in Sect.~\ref{section:2}. We then study its experimental imprints in Sect.~\ref{section:3}, including estimating the region of parameter space that best fits the $t\overline{t}H$ and four-top-quarks ATLAS data, and possible constraints from other observables on this parameter space. In Sect.~\ref{section:4} we propose a global search strategy to discriminate signal from background. Finally, we present our conclusions in Sect.~\ref{section:5}.

\section{The model}
\label{section:2}

We aim to construct a model that can account for several slight excesses in multileptons plus $b$-jets final states at the LHC.  More precisely, we require the model to be capable of producing more positive than negative leptons to better fit the imbalance suggested by the results in Ref.~\cite{ATLAS-CONF-2019-045}. We motivate the model from a phenomenological point of view.% and then examine its theoretical strengths. 

Since the imbalance in the final states reported in Ref.~\cite{ATLAS-CONF-2019-045} has more positive than negative leptonic charge, the model needs to capture the excess in positive charge present in the colliding protons.  Thus the BSM should couple to {\it up} quarks, while being safe to low-energy physics observables.  If the new particle coupling to {\it up} quarks would be charged, then it would also couple to bottom quarks (a $W'$ boson) or to leptons (a Leptoquark).  If the new particle were a $W'$, then it would be difficult to produce an excess in positive multileptons with $b$-quarks in the final state.  Were the new particle a Leptoquark then, being charged under $SU(3)_C$, the bounds on its mass from pair production would make it more difficult to reproduce the observed excesses.  We are then left with neutral particles coupling to {\it up} quarks ({\it u, c } and {\it t}) and with non-negligible Flavour Changing Neutral Currents (FCNC) to account for the deviations.  Among the three usual spins 0, 1 or 2, we find suitable to study spin-1 since, if color neutral, its gluon fusion production is protected through the Landau-Yang theorem \cite{Landau:1948kw, Yang:1950rg, Pleitez:2015cpa}, which also guarantees that the restrictive bounds in di-photon do not apply~\cite{Aad:2014ioa, Aaboud:2017yyg, ATLAS:2018xad, Khachatryan:2015qba, Sirunyan:2018aui}. Examples of studies of a FCNC spin 0 scalar boson phenomenology at the LHC, can be seen in Refs.~\cite{vonBuddenbrock:2019ajh, Hou:2020ciy}.

A color neutral spin-1 particle with non-diagonal couplings is known as a FCNC $Z'$.  Stringent constraints from the LEP II and Drell-Yan experiments require tiny or null coupling to leptons, thus for the purposes of this work we restrict to a leptophobic $Z'$.  Moreover, in light of limits set by low energy physics experiments and precision tests \cite{10.1093/ptep/ptaa104}, we restrict $Z'$ to only couple to right quark flavour changing $ut$ and $ct$ and flavour conserving $tt$ currents,
\begin{equation}
       {\cal L}_{int} \supseteq Z^\prime_\mu \left(g_{ut}^R \, \bar t_R \gamma^\mu u_R + g_{ct}^R \, \bar t_R \gamma^\mu c_R +g_{tt}^R \, \bar t_R \gamma^\mu t_R \right) + \mbox{h.c.}.
       \label{lagrangian1}
\end{equation}
In addition, one should consider a kinetic Lagrangian and an eventually negligible $Z$--$Z'$ mixing because of the restrictive bounds imposed by LEP \cite{Erler:2009jh}.  For the sake of simplifying notation, in the following of this article --except in Appendix~\ref{appendix_d0mixing}--, we drop the $R$ supraindex from the couplings.  

In Section \ref{section:3_other} we study bounds to the parameter space of this BSM Lagrangian coming from low energy physics and collider phenomenology.  Effective theories similar to the one described by the interaction in Eq.~\ref{lagrangian1} have been studied in different contexts, as for instance in Refs.~\cite{Cox:2015afa,Fox:2018ldq,Alvarez:2016nrz,Alvarez:2019uxp,Ko:2012ud}. In particular, a very similar set-up has been implemented in Ref.~\cite{Cho_2020}, albeit in a different mass range and assuming a hidden sector that increases the width-to-mass ratio, and in Ref.~\cite{Alvarez:2012ca} where a similar model was implemented to study the forward-backward $t\bar{t}$ asymmetry at the Tevatron.

\subsection{Phenomenology}\label{section:2_pheno}

\begin{figure}[h]
	\begin{center}
	\subfloat[]{\includegraphics[width=0.21\textwidth, keepaspectratio]{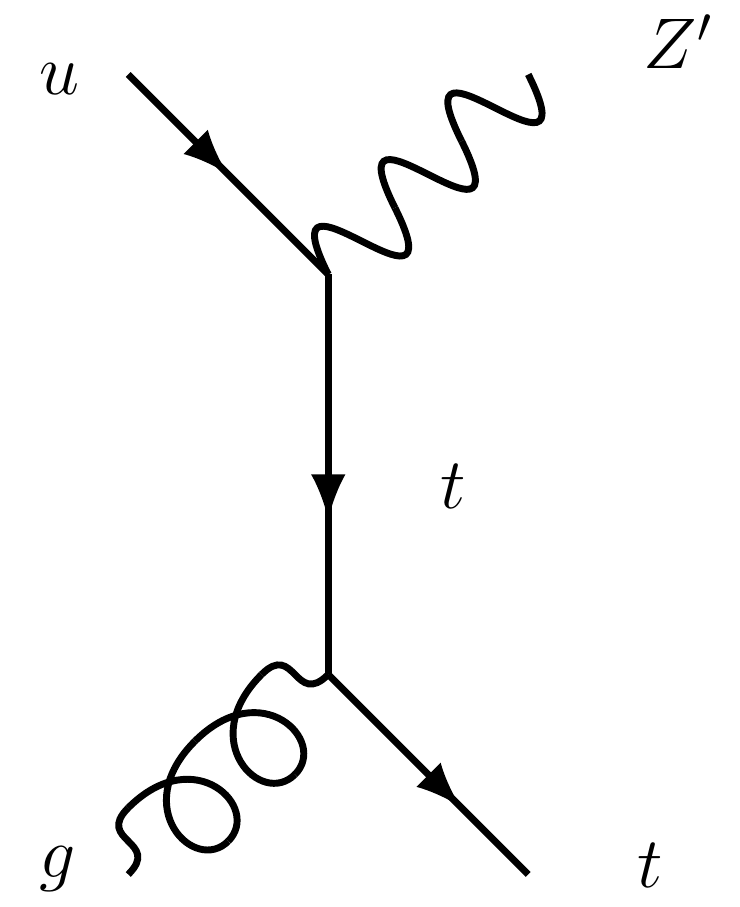}}\hspace{20mm}
	\subfloat[]{\includegraphics[width=0.2\textwidth, keepaspectratio]{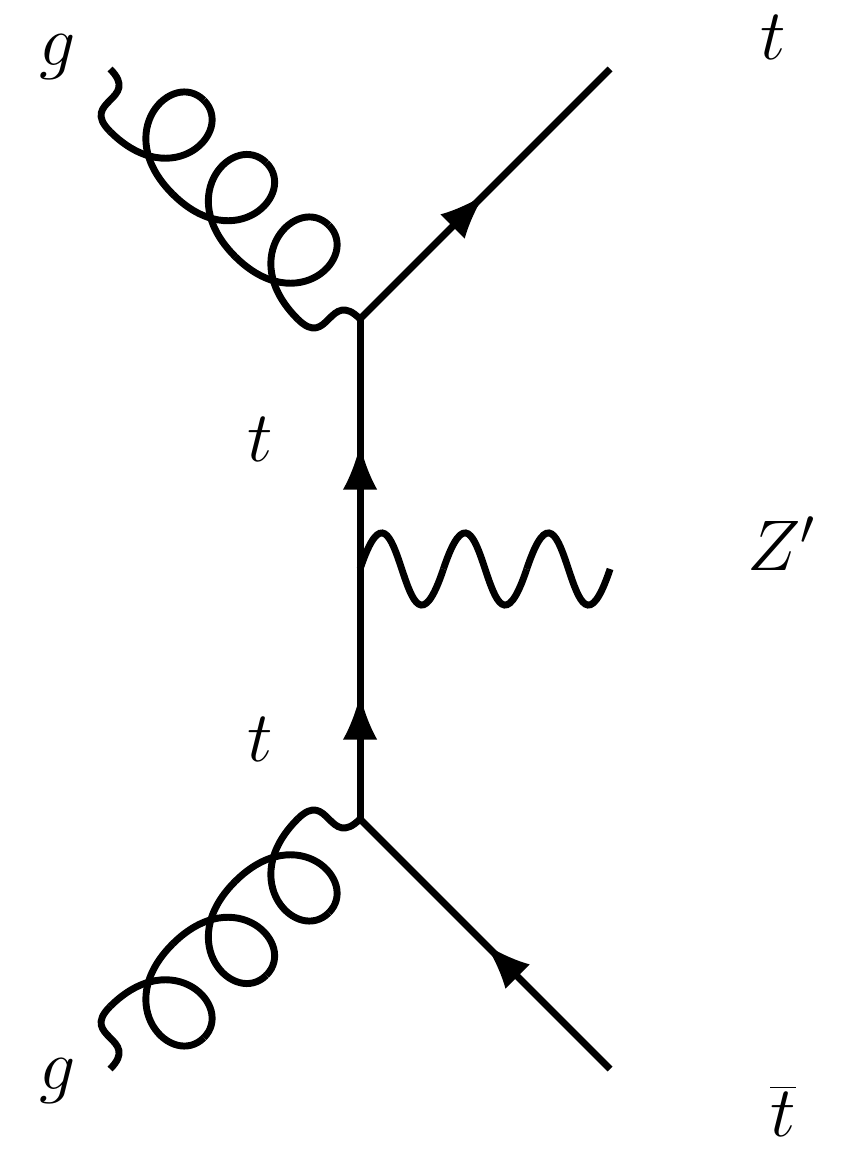}}
	\end{center}
	\caption{Representative diagrams for some of the most relevant processes: (a) $tZ'$ production and (b) $t\overline{t}Z'$ production.}
\label{feyndiag}
\end{figure}

\begin{figure}[h]
	\begin{center}
	\subfloat[\label{cross}]{\includegraphics[width=0.45\textwidth]{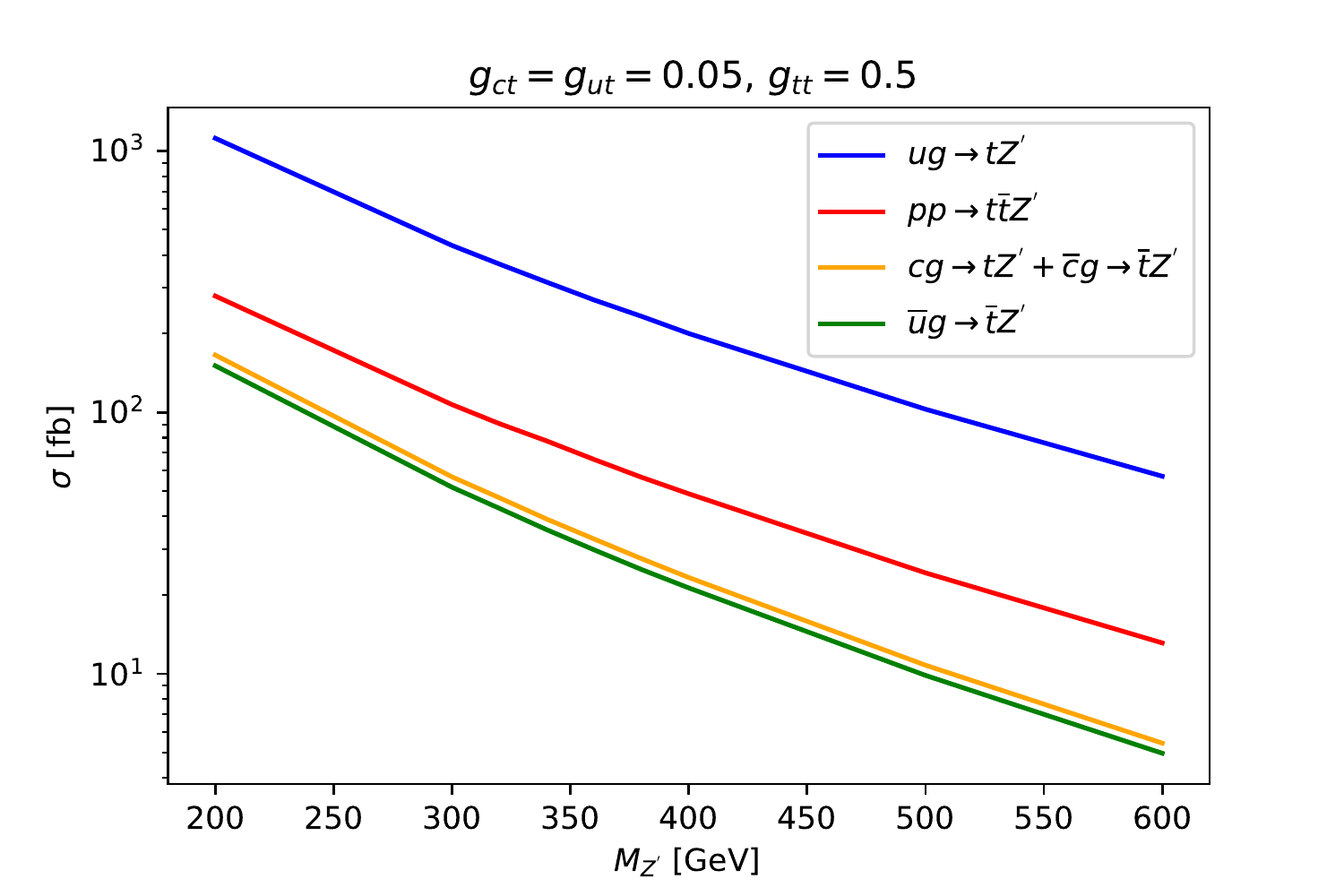}}\hspace{3mm}
	\subfloat[\label{brs}]{\includegraphics[width=0.45\textwidth]{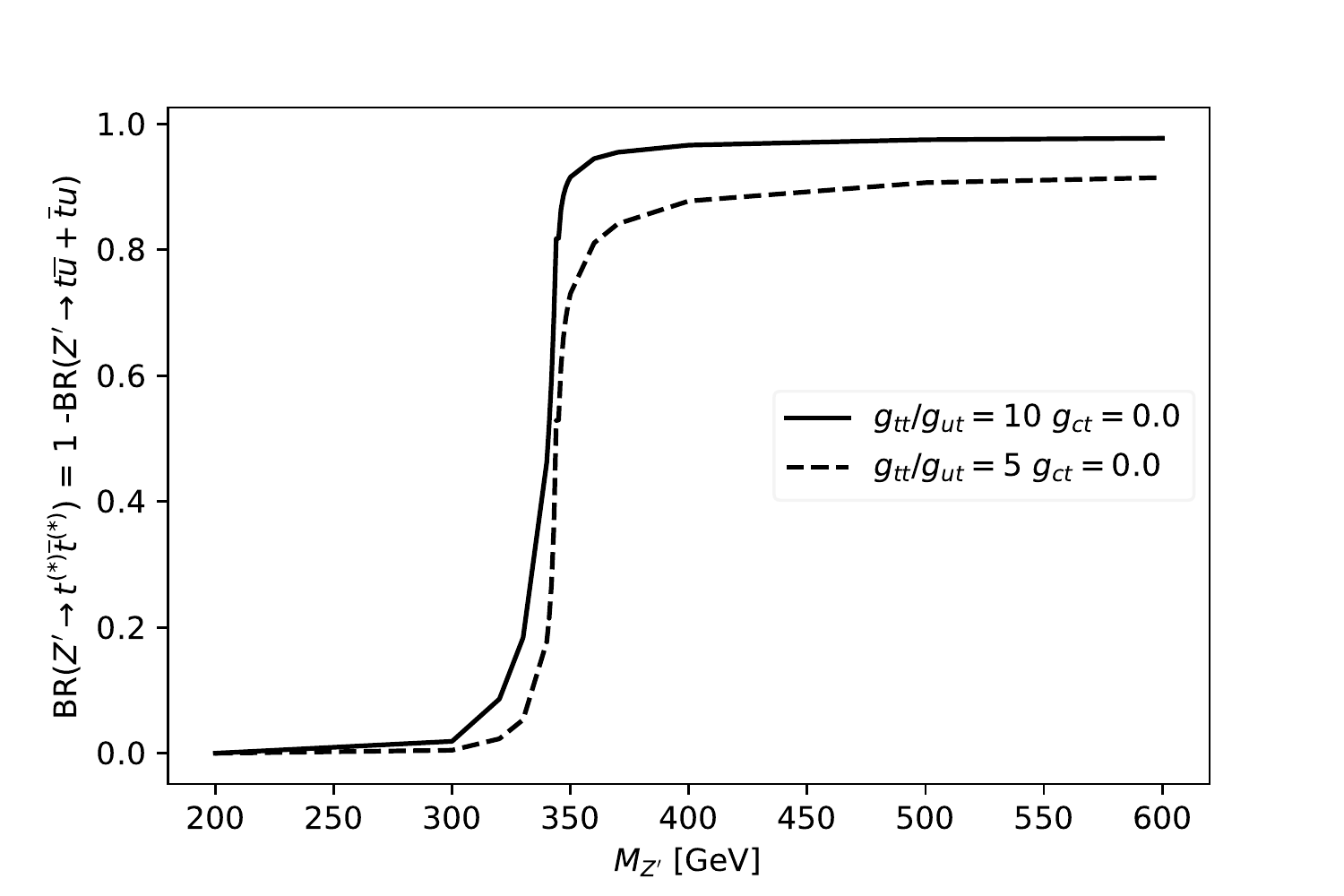}}
	\end{center}
	\caption{(a) Production cross-sections at $\sqrt{s}=13 \text{ TeV}$ as a function of $M_{Z'}$ for the most relevant signal processes, assuming a benchmark set of couplings. (b) Branching ratios for the dominant $Z'$ decay modes as a function of $M_{Z'}$ for two different benchmark sets of couplings.}
\label{cross_and_brs}
\end{figure}

To account for the observed data (see Sect.~\ref{section:3_multi} for more details), we need to produce same electric charge dilepton (denoted $2LSS$, with SS standing for same-sign) and multilepton (at least three leptons, denoted $3L$) final states with charge asymmetry and high $b$-jet multiplicity. To accomplish this, we consider two relevant $Z'$-induced processes, $tZ'+\overline{t}Z'$ (denoted $tZ'$ in the following) and $t\overline{t}Z'$, with a hierarchy between the relevant couplings to enforce a high probability of three- and four-top-quarks final states. We show two examples of the Feynman diagrams in Fig.~\ref{feyndiag}, and the relevant cross-sections and branching ratios in Fig.~\ref{cross_and_brs} as a function of $M_{Z'}$ for a given set of couplings. The cross-sections for a different set of couplings can be obtained by simple re-scaling. The branching ratios for a different set of couplings can be obtained by using $\Gamma(Z'\rightarrow t\overline{u})/(g_{ut})^{2}=\Gamma(Z'\rightarrow t\overline{c})/(g_{ct})^{2}$ and re-scaling.

Other processes that could be tested in these channels (see Sect.~\ref{section:3_other}) are either numerically irrelevant such as same-sign top-quark pair production ($tt,\ \overline{t}\,\overline{t}$) \cite{Cho_2020}, or chirality suppressed such as radiative $Z'$ production ($tZ'j,\ \overline{t}Z'j$)~\cite{Craig_2017,Cho_2020}. Note that these results can be fairly specific to our model, see e.g. Ref.~\cite{Hou:2017ozb} for a different $Z'$ model where $cg\rightarrow tZ'$ is the main production channel. In particular, non-resonant effects are suppressed both by the small $Z'$ width we consider by neglecting hidden sector decays, and by our particular choice of coupling structure.

From Fig.~\ref{feyndiag}, we see that contributions to $tZ'$ are proportional to $(g_{ut})^{2}$ and to $(g_{ct})^{2}$ while $t\overline{t}Z'$ is proportional to $(g_{tt})^{2}$. From Fig.~\ref{cross}, we see that even for $g_{tt}\gg g_{ut},g_{ct}$, the cross-section for $tZ'$ production is larger than for $t\overline{t}Z'$, 
mainly due to the kinematic requirements that must be met to produce each of the final-state particles on-shell. 
When comparing the four possible $tZ'$ processes, the largest cross-section is $ug\rightarrow tZ'$, as expected from the model motivations. This is due to $u$-quark abundance in the proton, which ensures that the $g_{ut}$-induced processes yield a considerable charge asymmetry that is not present in the other production processes. We are interested in this charge asymmetry and how it is reflected in current experimental searches.

If we take the possible relevant decays obtained from the Lagrangian in Eq.~\ref{lagrangian1} into account, and assuming they are all kinematically accessible, we see that $tZ'$  can yield the following final states: $ttj,\ \overline{t}\,\overline{t}j,\ tt\overline{t}$ and $\overline{t}t\overline{t}$. We are discarding the $t\overline{t}j$ final state because of the 2LSS and 3L selection criteria. On the other hand, $t\overline{t}Z'$ can yield $t\overline{t}tj,\ \overline{t}t\overline{t}j$ and $t\overline{t}t\overline{t}$ final states.

If we want events enriched with leptonic charge asymmetry and $b$-jets, we need $tZ'$ to decay mostly to three top quarks while being as charge asymmetric as possible. That is, we need $BR(Z'\rightarrow t\overline{t}) > BR(Z'\rightarrow tj+\overline{t}j)$. As we consider relatively low $M_{Z'}$ masses, we need $g_{tt}\gg g_{ut}$ to avoid phase-space suppression. If $M_{Z'} < 2 m_{t}$, we consider the three-body decay $Z'\rightarrow tW^{-}\overline{b},\ \overline{t}W^{+}b$. From Fig.~\ref{cross_and_brs} we see that for the benchmarks points we display, when $M_{Z'}$ is large enough the $t\overline{t}$ decay mode dominates. When combined with $tZ'$ and $t\overline{t}Z'$ production, these decays produce three- and four-top-quark final states.

After this overview of the basic phenomenology of the proposed model, we turn to studying its effect on the relevant observables, and how these observables determine the region in its parameter space most compatible with the experimental results.

\section{Experimental imprints and model tuning}\label{section:3}

In this section we study how the $Z'$ model detailed in Sect.~\ref{section:2} is probed by different existing experimental results. A $Z'$ as in Eq.~\ref{lagrangian1} affects both high- and low-energy observables. In light of recent experimental results from $t\overline{t}H$ and $t\overline{t}t\overline{t}$ searches in ATLAS and CMS, we are specially interested in multilepton-plus-$b$-jets final states. In particular, multilepton final states with non-zero total leptonic charge are highly sensitive to $Z'$. Taking this into account, we detail in Sect.~\ref{section:3_multi} how on-shell $Z'$ production in association either with a single top quark ($tZ'$ or $\overline{t}Z'$) or with a top quark pair ($t\overline{t}Z'$) could explain the need for $t\overline{t}W^{\pm}$ re-scaling to account for tensions in data in recent $t\overline{t}H$ results \cite{ATLAS-CONF-2019-045}, while yielding interesting signatures in four-top-quarks. In Sect.~\ref{section:3_other} we test whether the parameter space indicated by multilepton and high $b$-jet multiplicity is safe to other observables that could be affected by our model. The more relevant observables we consider are $D^0-\bar D^0$-meson mixing, top-quark rare decays through flavour-changing neutral currents (FCNC), top-quark pair production, same-sign top-quark pair production, resonant $tj$ production in $t\overline{t}$+jets events, and $Z'$ radiative production $tZ'j$. Other processes involving resonant $Z'$ production, such as single $Z'$, $Z'j$ and single top quark production $tj+\overline{t}j$, are absent due to having approximated as null the $g_{uu}$ and $g_{cc}$ couplings.

\subsection{Fits to experimental data}
\label{section:3_multi}

We study how our phenomenological $Z'$ model detailed in Sect.~\ref{section:2} accommodates recent results reported for $t\overline{t}H$ production while avoiding constraints and yielding potentially interesting results in $t\overline{t}t\overline{t}$ searches. Regarding $t\overline{t}H$, we focus on the ATLAS preliminary results reported in Ref.~\cite{ATLAS-CONF-2019-045}, although results from CMS~\cite{Sirunyan:2020icl} are consistent with our conclusions. Regarding four-top-quarks we consider the results reported by ATLAS in Ref.~\cite{Aad:2020klt}, having corroborated that the results obtained are compatible with the combination of the ATLAS and CMS~\cite{Sirunyan:2019wxt} results. Both searches target same-sign dilepton and multilepton processes but their channel definitions are not the same and the reconstructed objects, both leptons and jets, have different kinematic cuts and tagging efficiencies, which we take into account in our study.

For the case of $t\overline{t}H$, the results reported in Ref.~\cite{ATLAS-CONF-2019-045} are particularly interesting because of the difficulties reported when dealing with the irreducible $t\overline{t}W^{\pm}$ background. Figure 2 of Ref.~\cite{ATLAS-CONF-2019-045} highlights the need for a missing charge asymmetric contribution to match the data, which yields a normalization factor for $t\overline{t}W^{\pm}$ larger than one. This is consistent with other measurements, for example Refs.~\cite{Sirunyan:2017uzs, Aaboud:2019njj}, and has motivated a push for higher theoretical accuracy in the $t\overline{t}W^{\pm}$ calculations~\cite{Frixione:2015zaa, Frederix:2017wme, Broggio:2019ewu, Kulesza:2020nfh, Frederix:2020jzp,Bevilacqua:2020pzy, Denner:2020hgg, vonBuddenbrock:2020ter}, which nevertheless have not fully explained the discrepancy between the expected  and the observed $t\overline{t}W^{\pm}$ event yields.

If we treat the $t\overline{t}W^{\pm}$ background as well modelled, and thus constrained to have a normalization factor consistent with unity within the uncertainty in its theoretical cross-section, we are faced with charge asymmetric anomalous events with high $b$-jet multiplicity. Our $Z'$ model is designed to accommodate these two features. A recent example of BSM effects in $t\overline{t}W$ can be found in Ref.~\cite{Banelli:2020iau}, where, in contrast to a resonant BSM physics model, they study an Effective Field Theory in the top-quark sector.

As detailed in Sect.~\ref{section:2_pheno}, we study $Z'$ production in association either with a single top quark or with a top-quark pair. These processes, $tZ'$ and $t\overline{t}Z'$, along with a considerable $Z'\rightarrow t\overline{t}$ branching ratio achieved with a suitable coupling hierarchy, provide the necessary same-sign dilepton and multilepton signatures, with both charge asymmetry and high $b$-jet multiplicity.

As we consider three-top-quarks production and four-top-quarks production, we need to make sure that our results are compatible with four-top-quarks limits. Even if four-top-quarks production will be mostly sensitive to $(g_{tt})^{4}$, three-top-quarks production is proportional to $(g_{ut})^{2}(g_{tt})^{2}$ and will introduce a charge asymmetry.

To see how well our model can agree with the data, we obtain the event yields expected from $tZ'$ and $t\overline{t}Z'$ in the different reported bins in Fig.~2 of Ref.~\cite{ATLAS-CONF-2019-045}, and the expected events in four-top-quarks searches with the selection criteria of Ref.~\cite{Aad:2020klt} as a function of $(g_{ut},g_{ct},g_{tt})$ for different values of $M_{Z'}$. To do this, we have implemented the $Z'$ model in Eq.~\ref{lagrangian1} using {\tt Feynrules}~\cite{Alloul:2013bka} and then simulated a fixed set of points through {\tt Madgraph\;5}~\cite{Alwall:2014hca} for production and decay of $tZ'$ and $t\overline{t}Z'$.  The signal events have been generated using a leading-order matrix element and the NN23LO1 PDF set~\cite{Ball:2012cx}, and have been processed 
through {\tt Pythia\;8}~\cite{Sjostrand:2014zea} for the modelling of parton showering and hadronization, as well as through a simulation 
of the detector response as implemented in {\tt Delphes}~\cite{deFavereau:2013fsa}. We use the Monash tune~\cite{Skands:2014pea} of {\tt Pythia\;8} with a few changes aimed to reproduce the $t\overline{t}W^{\pm}$ $N_{j}$ distribution as faithfully as possible. These changes are detailed in Table~\ref{table:pythia_params} in Appendix~\ref{appendix_tunes}. We also modify the default {\tt Delphes} card, producing two new cards that allow us to reproduce the reported expected events in Refs.~\cite{ATLAS-CONF-2019-045} and~\cite{Aad:2020klt}. These changes are detailed in Appendix~\ref{appendix_tunes}. The only difference between the two {\tt Delphes} cards is the b-tagging efficiency. The four-top-quarks search~\cite{Aad:2020klt} uses a higher b-tagging efficiency working point ($77\%$ average b-tagging efficiency) than the one used by the $t\overline{t}H$ search~\cite{ATLAS-CONF-2019-045} ($70\%$ average b-tagging efficiency).

After simulating the events, we implement the event selection cuts and obtain the event yields. Each search has a different event selection for each channel as detailed in Tables~\ref{table:ttH_sel} and~\ref{table:tttt_sel}. Additionally, we require at least 4 jets in the 2LSS $t\overline{t}H$ selection and at least 3 b-jets in the 2LSS and 3L four-top-quarks selection. The former modification is needed to compare our results to Fig.~2 in Ref.~\cite{ATLAS-CONF-2019-045} and the latter is needed to obtain a signal-enhanced selection similar to the one defined by the use of the BDT in Ref.~\cite{Aad:2020klt}.

We also incorporate specific trigger selection efficiencies for each leptonic channel to the {\tt ROOT}~\cite{Brun:1997pa} code used to analyse the {\tt Delphes} output. The simulated signal samples have been normalized using k-factors obtained from simulating similar events to NLO with the same set-up, and which are consistent with those in the literature~\cite{Guzzi:2019ucs}.

After simulating a fixed set of events, we observe that the expected number of events from a given process in an analysis channel, $N^{\text{channel}}_{\text{process}}$, can be parametrized as follows\footnote{This parametrization neglects non-resonant $Z'$ effects.}:
\begin{eqnarray}
	\begin{aligned}
		N^{\text{ch}}_{tZ'}=&A^{\text{ch}}_{1}\cdot BR(Z'\rightarrow t\overline{u}+\overline{t}u)\cdot(g_{ut})^{2}+A^{\text{ch}}_{2}\cdot BR(Z'\rightarrow t\overline{c}+\overline{t}c)\cdot(g_{ut})^{2}\\
&+A^{\text{ch}}_{3}\cdot BR(Z'\rightarrow t\overline{t})\cdot(g_{ut})^{2}+A^{\text{ch}}_{4}\cdot BR(Z'\rightarrow t\overline{u}+\overline{t}u)\cdot(g_{ct})^{2}\\
&+A^{\text{ch}}_{5}\cdot BR(Z'\rightarrow t\overline{c}+\overline{t}c)\cdot(g_{ct})^{2}+A^{\text{ch}}_{6}\cdot BR(Z'\rightarrow t\overline{t})\cdot(g_{ct})^{2},\\
		N^{\text{ch}}_{t\overline{t}Z'}=&B^{\text{ch}}_{1}\cdot BR(Z'\rightarrow t\overline{u}+\overline{t}u)\cdot(g_{tt})^{2}+B^{\text{ch}}_{2}\cdot BR(Z'\rightarrow t\overline{c}+\overline{t}c)\cdot(g_{tt})^{2}\\
&+B^{\text{ch}}_{3}\cdot BR(Z'\rightarrow t\overline{t}) \cdot(g_{tt})^{2},
	\end{aligned}
\label{eq:interpol}
\end{eqnarray}
where all the coefficients $A^{\text{ch}}_{i},B^{\text{ch}}_{i}$ are functions of $M_{Z'}$ and absorb the acceptance of the channel and the cross-section for the specific process normalized to the corresponding coupling set to unity. The channels we consider are the different bins of Fig.~2 of Ref.~\cite{ATLAS-CONF-2019-045} and the four-top-quarks event yield with $Q>0$ and with $Q<0$ with the selection criteria of Ref.~\cite{Aad:2020klt}. We obtain the $A^{\text{ch}}_{i}$ and $B^{\text{ch}}_{i}$ with Weighted Least Squares, where the uncertainty of each measurement is due to the Monte Carlo finite sampling, and with them we generate arbitrary points in the parameter space.

The likelihood fit to the eight bins of Fig.~2 of Ref.~\cite{ATLAS-CONF-2019-045} is performed using the \texttt{HistFitter} package \cite{Baak:2014wma}, which relies on \texttt{RooFit} \cite{RooFit:manual} and the minimization algorithms from \texttt{MINUIT} \cite{James:1994vla}. Additionally, systematic uncertainties affecting the overall normalization of the SM processes are included in the fit as nuisance parameters (NP) with Gaussian constraints: 20\% uncertainty is assigned to $t\overline{t}H$, $t\overline{t}W$, $t\overline{t}Z$, and signal, and 50\% uncertainty to diboson, while the various fake lepton components have uncertainties assigned corresponding to the normalisation factor precision reported in the ATLAS result. 

\begin{figure}[th!]
	\begin{center}
	\subfloat[]{\includegraphics[width=0.45\textwidth]{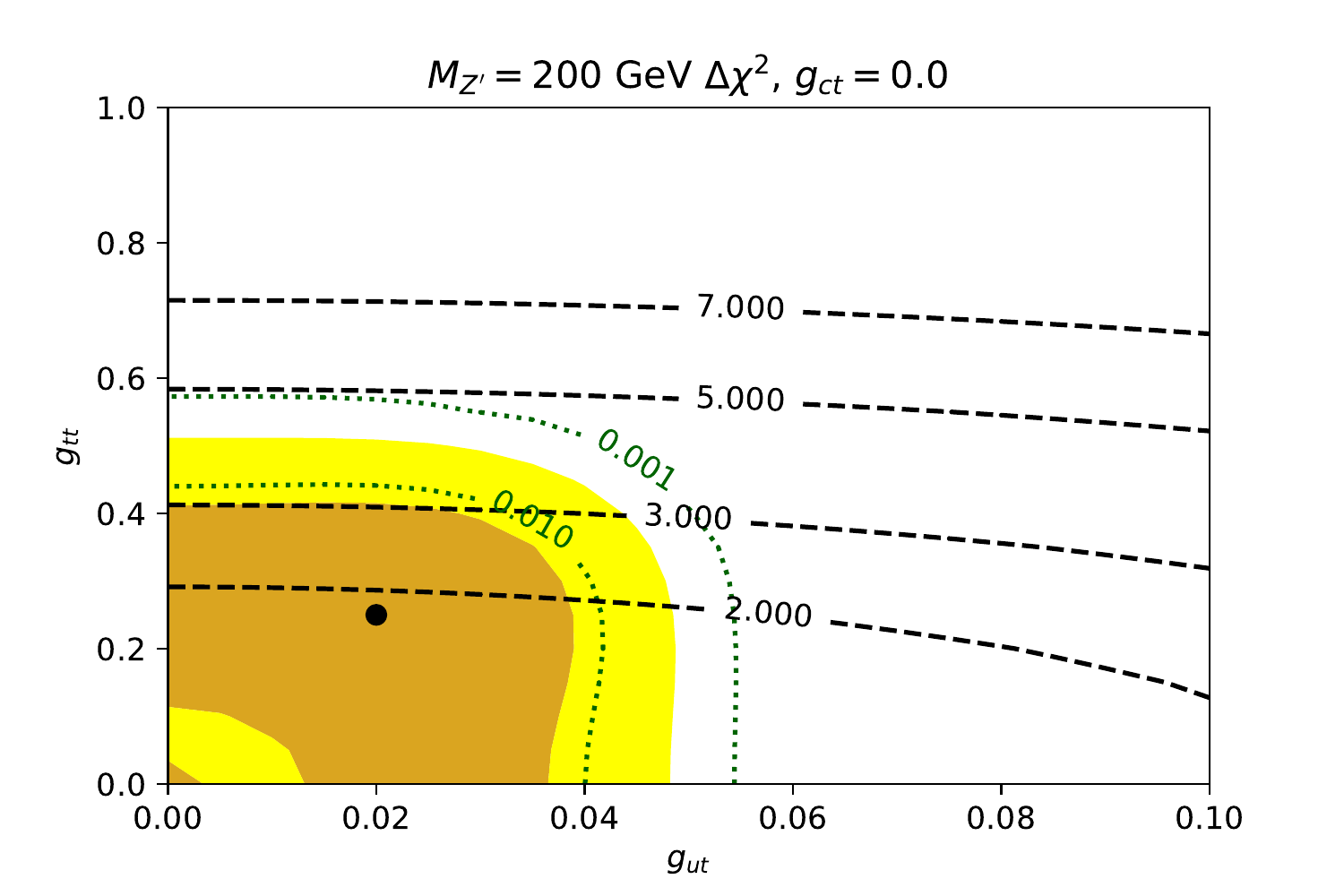}}\hspace{3mm}
	\subfloat[]{\includegraphics[width=0.45\textwidth]{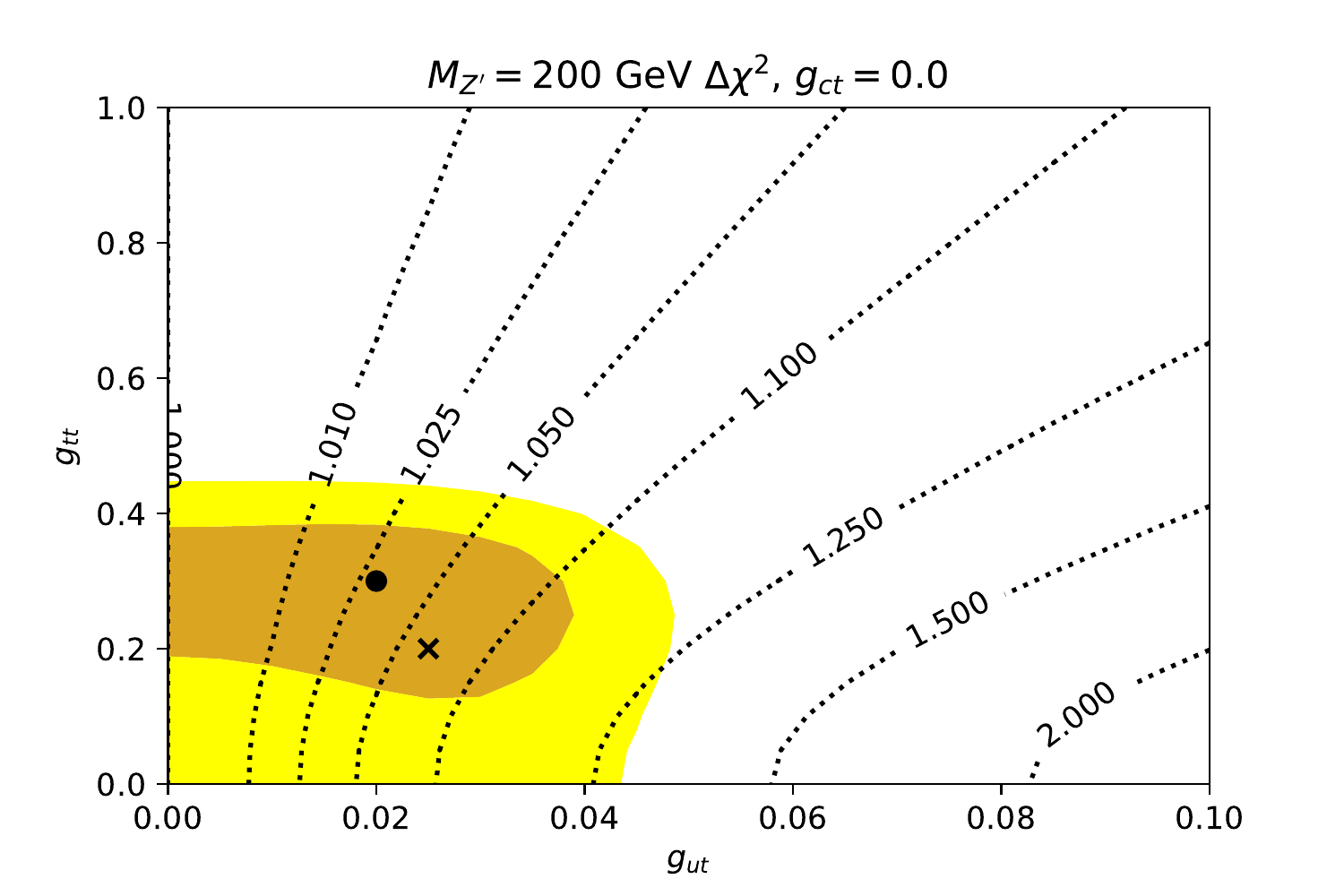}}\\
	\subfloat[]{\includegraphics[width=0.45\textwidth]{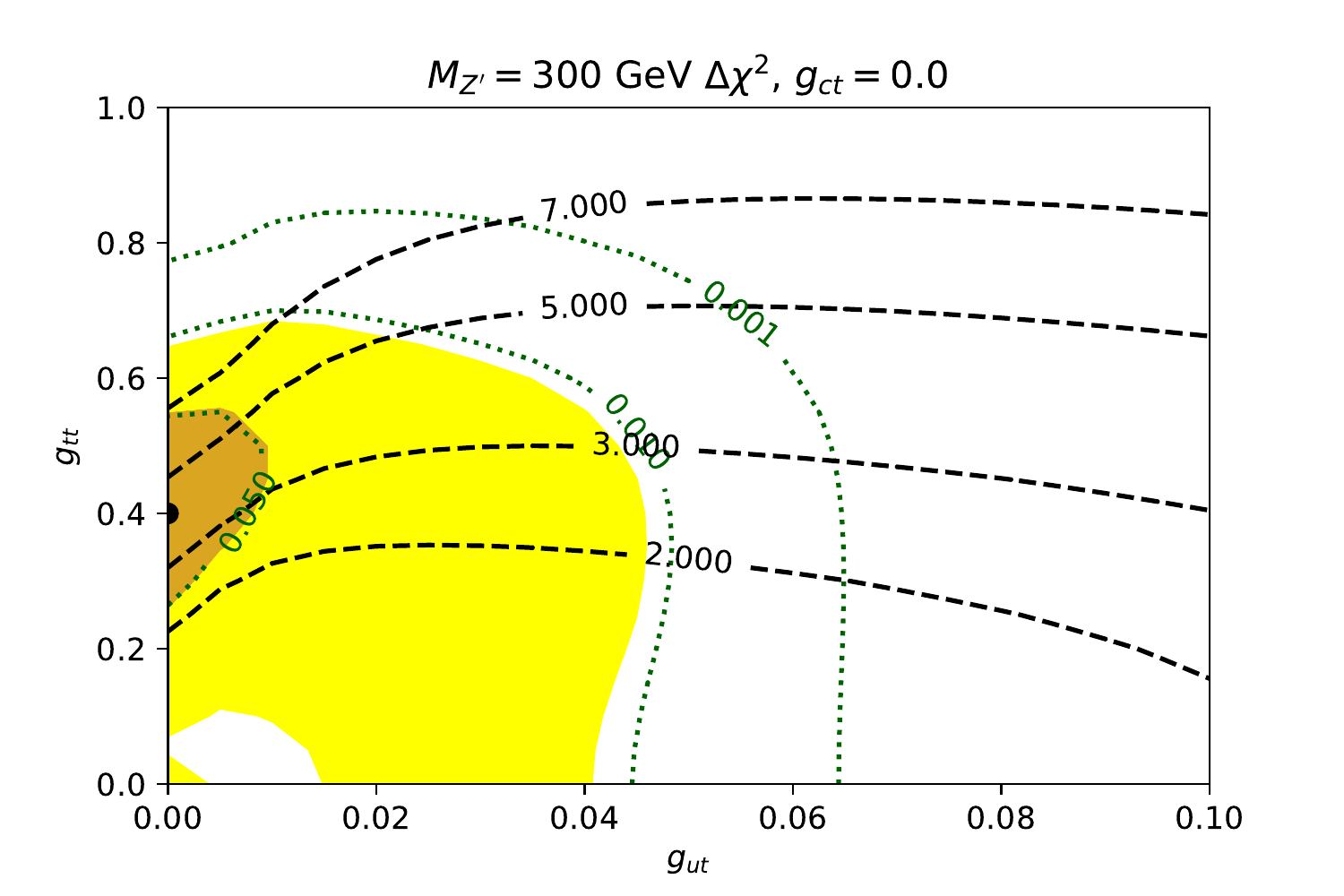}}\hspace{3mm}
	\subfloat[]{\includegraphics[width=0.45\textwidth]{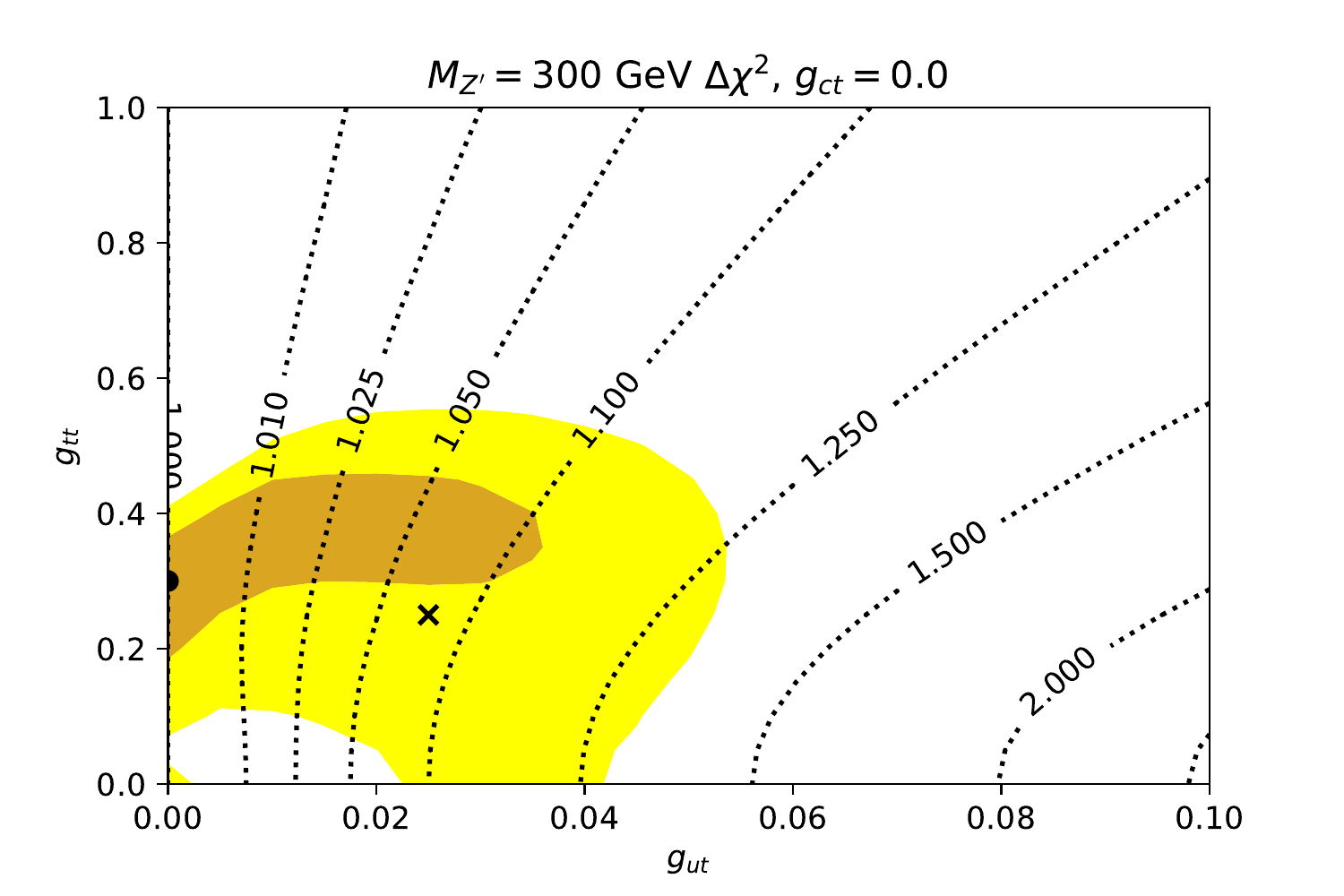}}
	\end{center}
	\caption{Fits to data for $M_{Z'}<2m_t$ in the $(g_{ut},g_{tt})$ plane. Left column: ellipses correspond to 1 s.d. and 2 s.d. from the $t\overline{t}H$ NLL minima with the green dotted curve being the goodness-of-fit and the black dashed curve corresponding to $N^{\text{four-tops}}_{SM+BSM}/N^{\text{four-tops}}_{SM}$. Right column: ellipses correspond to 1 s.d. and 2 s.d. from the $t\overline{t}H+$ four-top-quarks NLL minima with the black dotted curve corresponding to the charge imbalance ratio of yields $r(4t)$ defined in Eq.~\ref{r4t}. The $t\overline{t}H+$ ATLAS and CMS average four-top-quarks NLL minima is shown with a black cross.
		}
\label{zp-fits-1}
\end{figure}

\begin{figure}[th!]
	\begin{center}
	\subfloat[]{\includegraphics[width=0.45\textwidth]{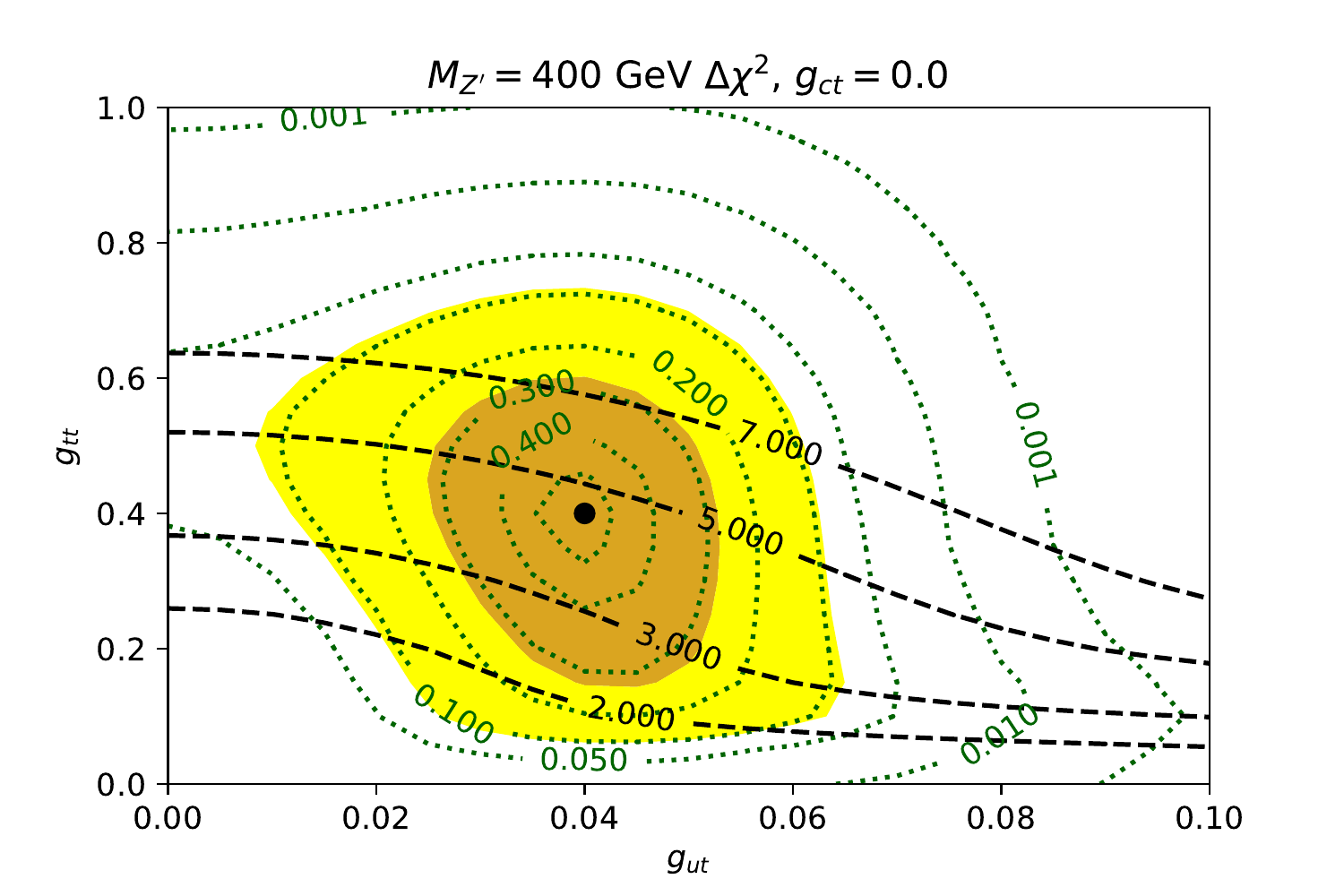}}\hspace{3mm}
	\subfloat[]{\includegraphics[width=0.45\textwidth]{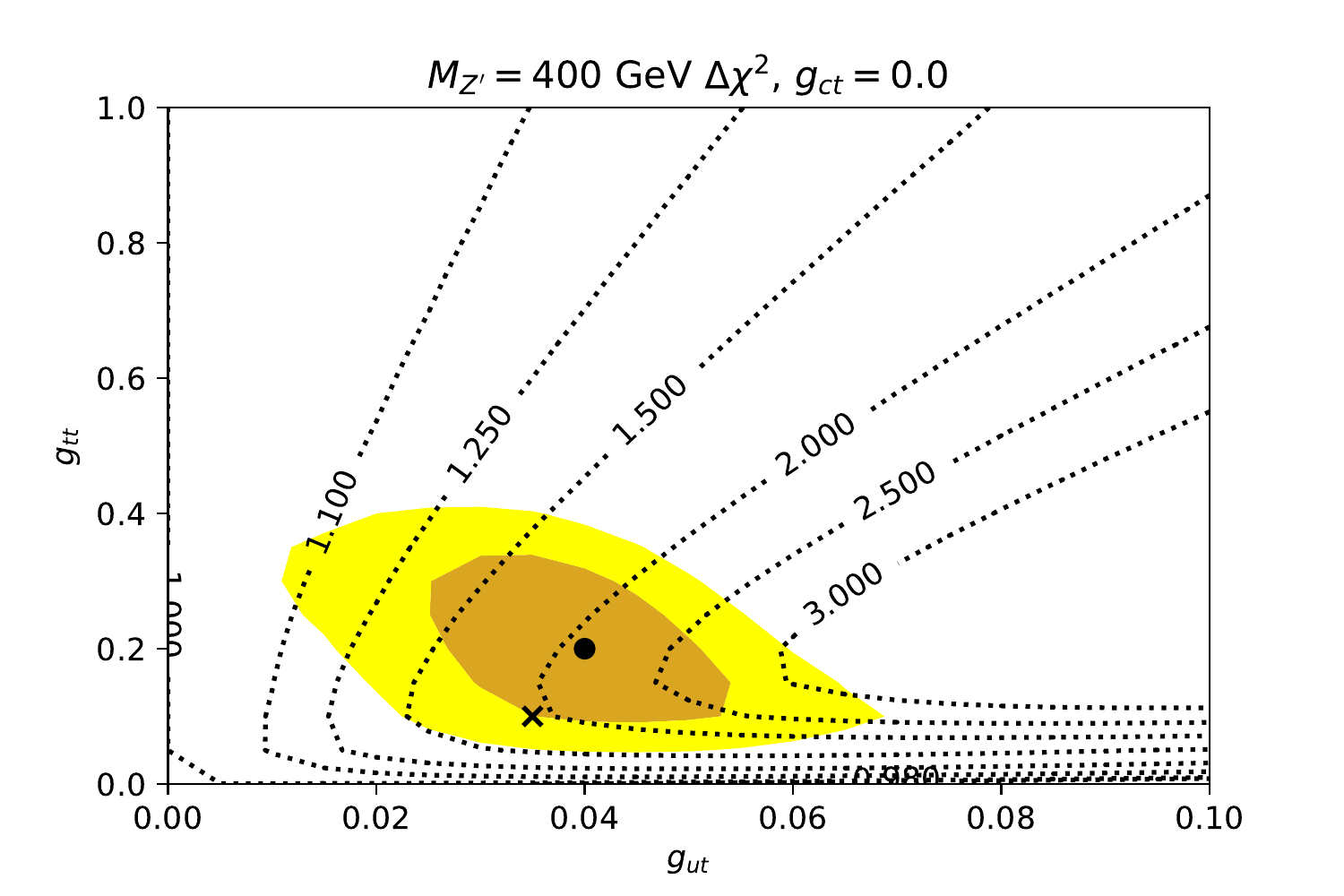}}\\
	\subfloat[]{\includegraphics[width=0.45\textwidth]{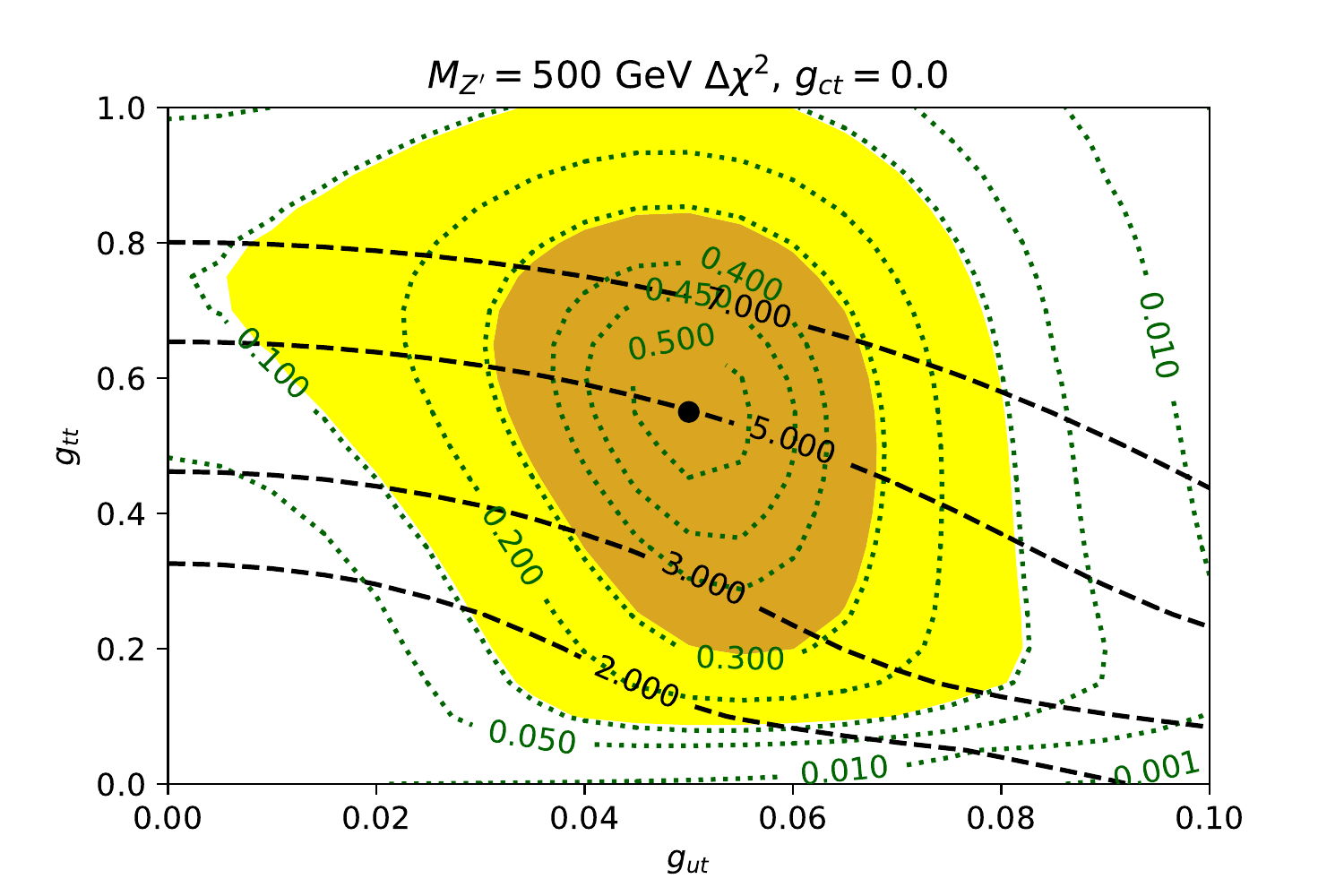}}\hspace{3mm}
	\subfloat[]{\includegraphics[width=0.45\textwidth]{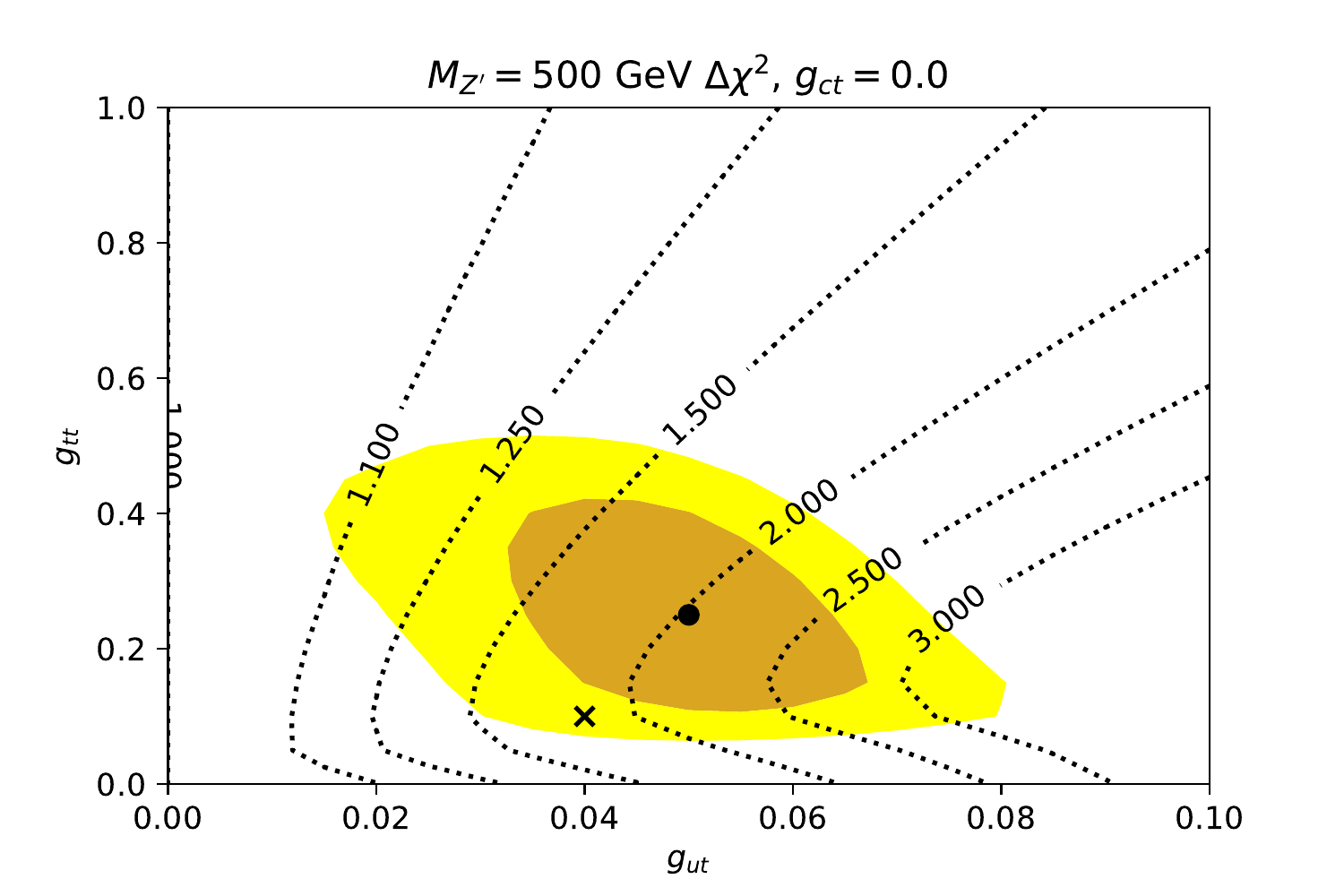}}\\
	\subfloat[]{\includegraphics[width=0.45\textwidth]{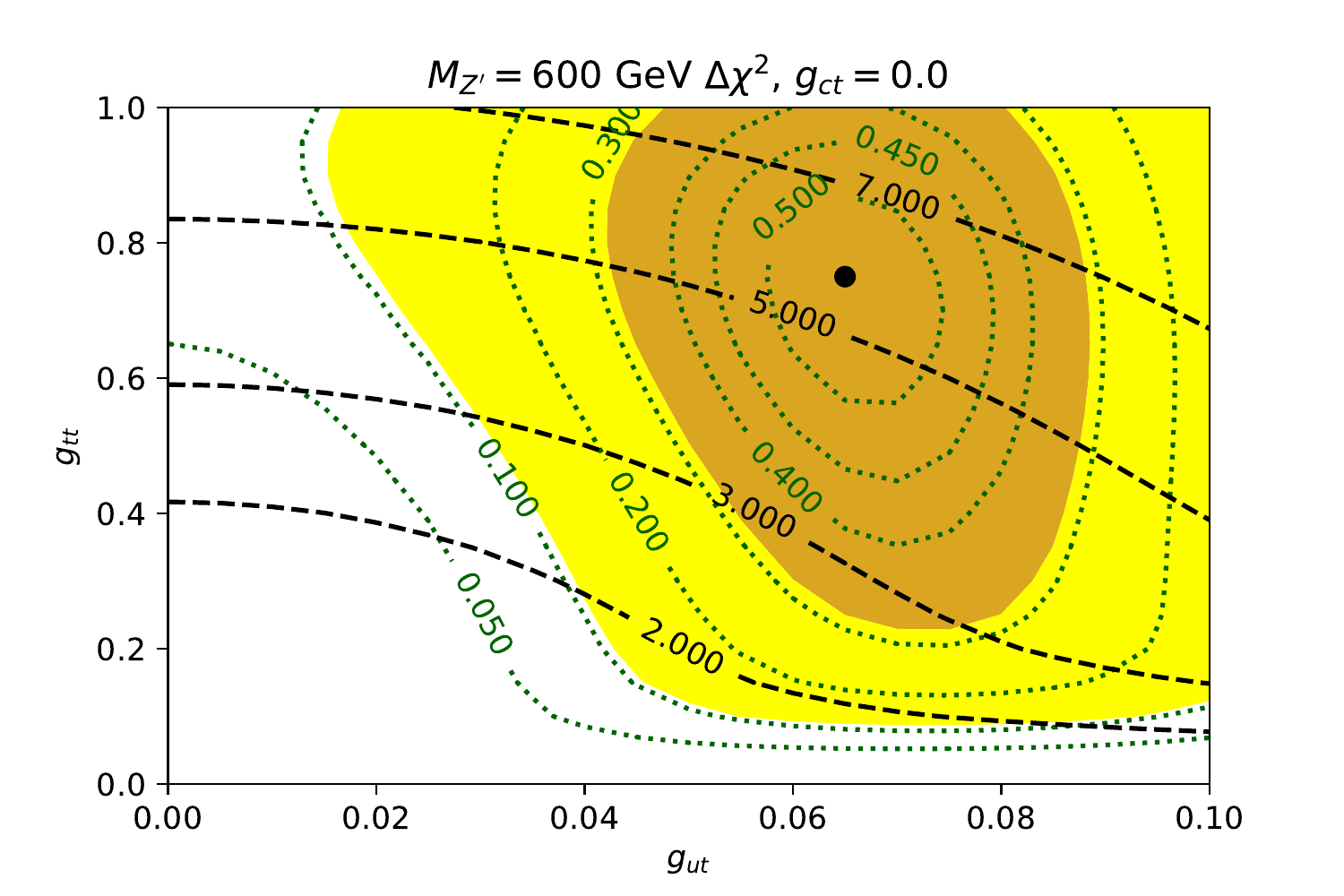}}\hspace{3mm}
	\subfloat[]{\includegraphics[width=0.45\textwidth]{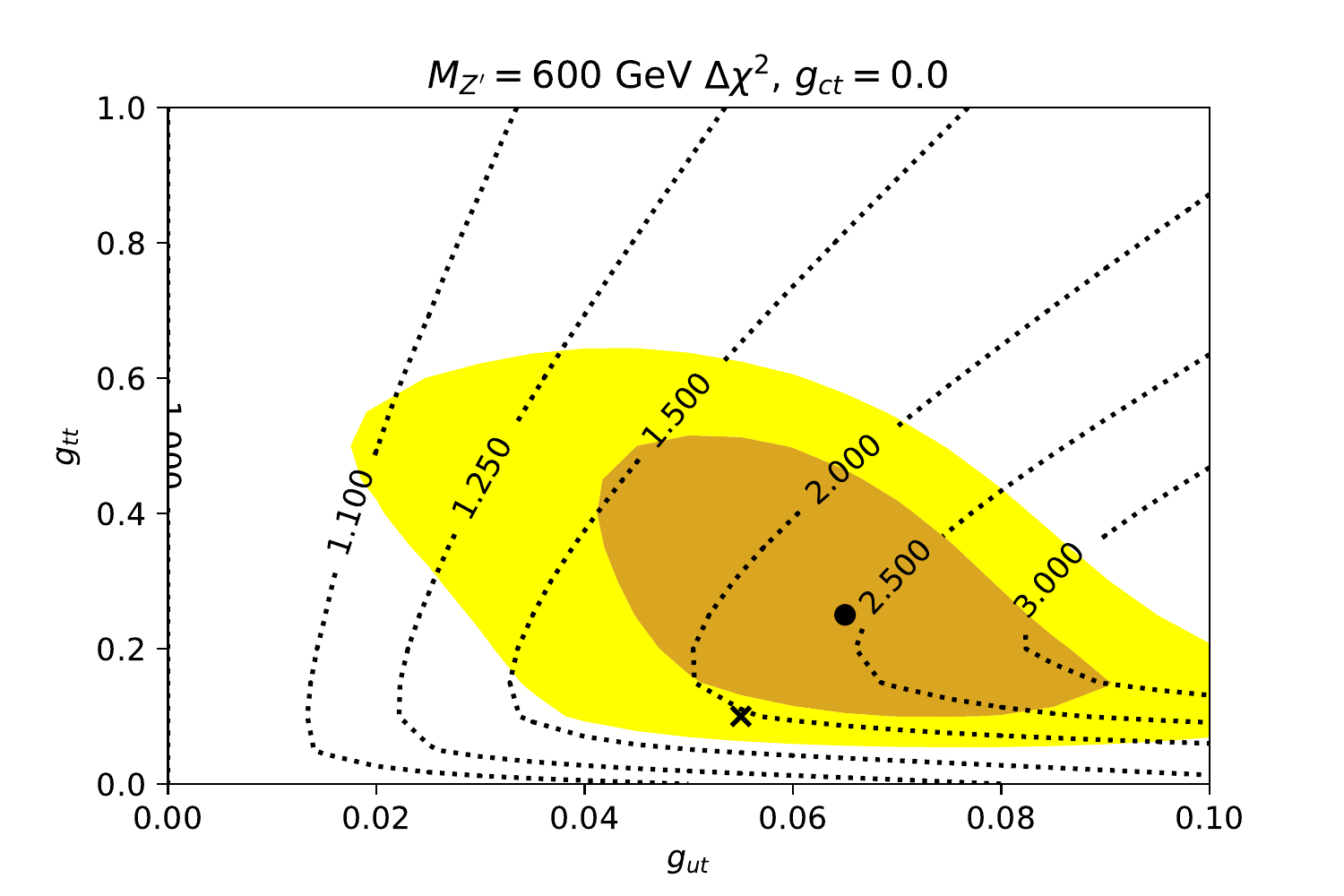}}\\
	\end{center}
	\caption{Idem as in Fig.~\ref{zp-fits-1}, but for $M_{Z'}>2m_t$. In this case, the opening of the decay $Z'\to t\bar t$ enhances the channel  $ug \to t Z' \to t \bar t t$ which induces a charge imbalance that favours  considerably the fit in $t\bar t H$.   This can be appreciated in the goodness-of-fit (green dotted), which is considerably better in this case than for those in Fig.~\ref{zp-fits-1}.}
\label{zp-fits-2}
\end{figure}

\begin{figure}[h!]
	\begin{center}
	\subfloat[]{\includegraphics[width=0.45\textwidth]{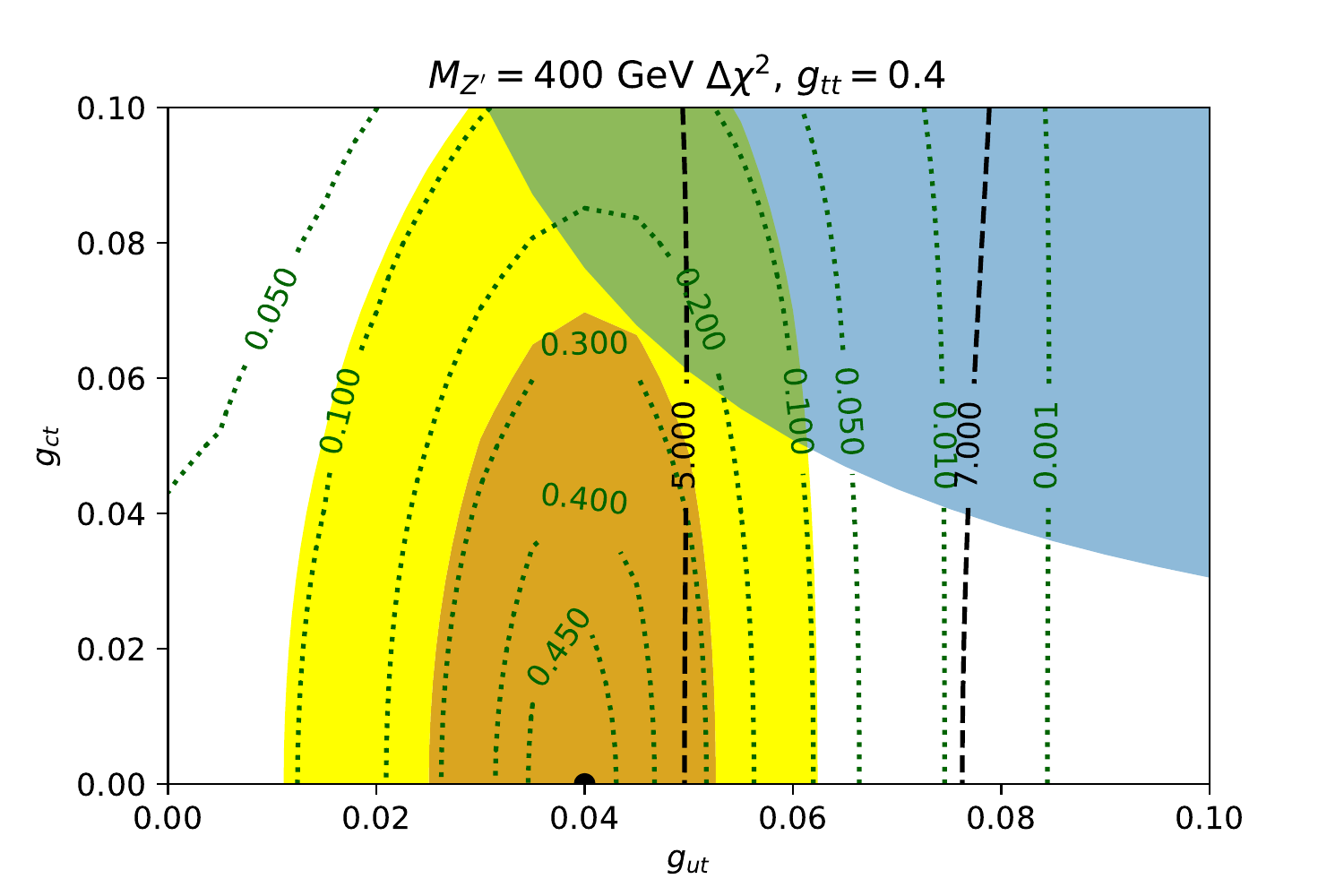}}\hspace{3mm}
	\subfloat[]{\includegraphics[width=0.45\textwidth]{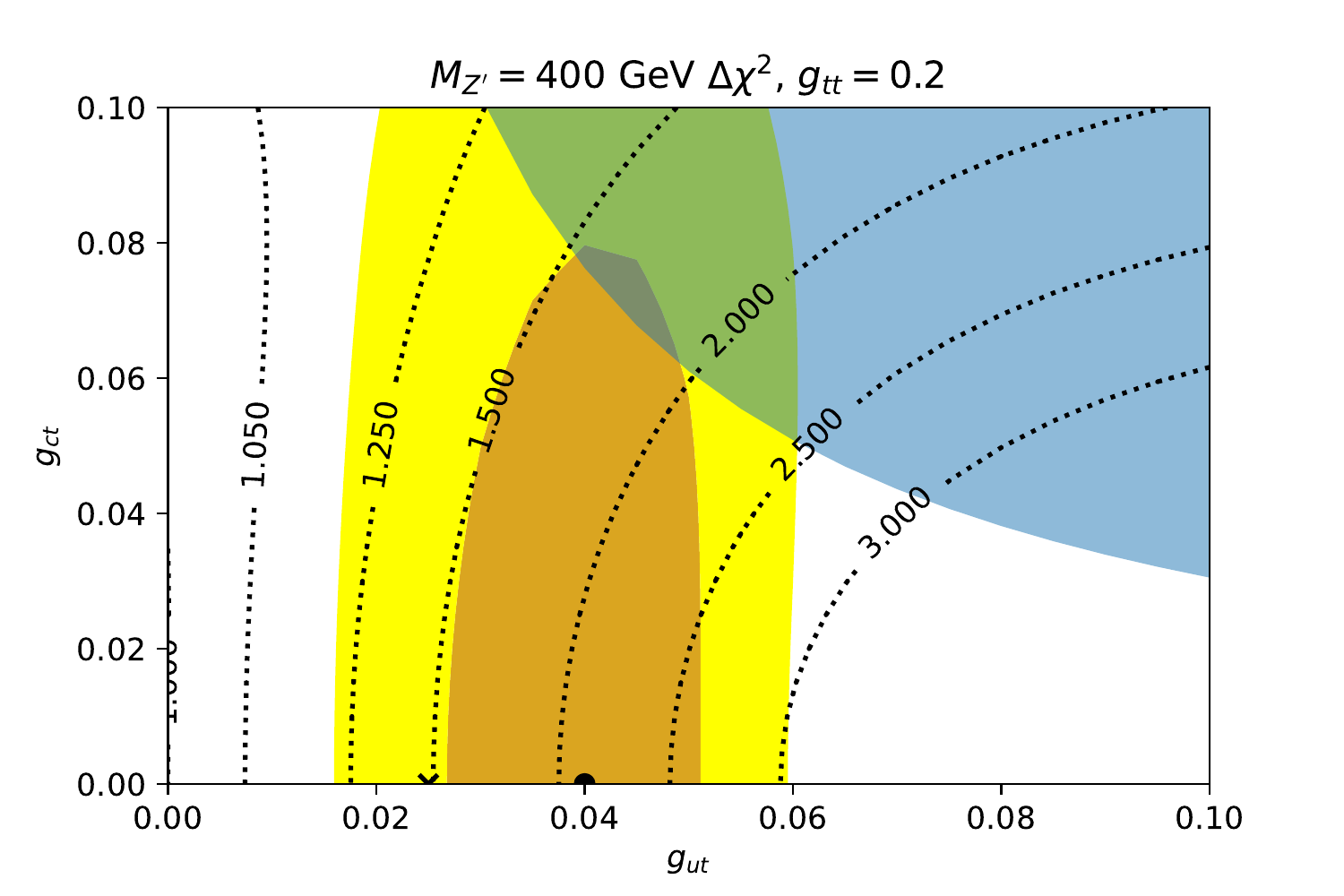}}
	\end{center}
	\caption{Idem as in Fig.~\ref{zp-fits-2}, but for different points in the $(g_{ut},g_{ct})$ plane. The blue regions correspond to regions in parameter space disfavoured by $D^0 \leftrightarrow \bar D^0$ mixing as detailed in Sect.~\ref{section:2}. 
		}
\label{zp-fits-gct}
\end{figure}

In Figs.~\ref{zp-fits-1}-\ref{zp-fits-gct} we compute the impact of different points in parameter space for different $M_{Z'}$ masses on the two experimental analysis. In the left column of these plots, we obtain the point that minimizes the Negative Log-Likelihood (NLL)~\cite{10.1093/ptep/ptaa104} for the reported observed events in Fig.~2 of the $t\overline{t}H$ search \cite{ATLAS-CONF-2019-045}, considering the pre-fit SM background contributions.  After finding this minima, we plot the 1 and 2 standard deviations (s.d.) regions and the goodness-of-fit contour lines \cite{Cousins:goodnessfit}. We include in this same left column the contour lines indicating the four-top-quark BSM and SM to SM only event ratio, $N^{\text{four-tops}}_{SM+BSM}/N^{\text{four-tops}}_{SM}$. These events need to pass the four-top-quarks-like selection cuts described in Table~\ref{table:tttt_sel} and include SM four-top-quarks and BSM $t\overline{t}t+t\overline{t}\overline{t}$, $t\overline{t}tj+t\overline{t}\overline{t}j$ and four-top-quarks processes. We can see the interplay between a good $t\overline{t}H$ fit, which requires asymmetry (non-negligible $g_{ut}$) but also a high $b$-jet multiplicity (large $g_{tt}$), and the four-top-quarks fit (not so large $g_{tt}$). We combine both measurements in the right column where we minimize the $t\overline{t}H+$ four-top-quarks data
\begin{eqnarray}
\text{NLL}(t\overline{t}H)+\frac{(N^{\text{four-tops}}_{SM+BSM}/N^{\text{four-tops}}_{SM}-N^{\text{four-tops}}_{\text{obs}}/N^{\text{four-tops}}_{SM})^{2}}{\sigma^{2}}\nonumber
\end{eqnarray}

For four-top-quarks we consider two $N^{\text{four-tops}}_{\text{obs}}/N^{\text{four-tops}}_{SM}$ possibilities: the reported ATLAS~\cite{Aad:2020klt} value alone $N^{\text{four-tops}}_{\text{obs}}/N^{\text{four-tops}}_{SM}=2.0^{+0.8}_{-0.6}$, choosing $\sigma=0.8$ when $N^{\text{four-tops}}_{SM+BSM}\geq 2.0 N^{\text{four-tops}}_{SM}$ and $\sigma=0.6$ otherwise, and in combination with the CMS~\cite{Sirunyan:2019wxt} reported value $N^{\text{four-tops}}_{\text{obs}}/N^{\text{four-tops}}_{SM}=1.1\pm 0.5$, which yields an average value of $N^{\text{four-tops}}_{\text{obs}}/N^{\text{four-tops}}_{SM}=1.4\pm 0.3$. In all cases we find that the best fit points are compatible and we chose to focus on the ATLAS result to tune our simulation and obtain further information. We plot the 1 s.d. and 2 s.d. regions for the ATLAS value and we showcase how the $Z'$ introduces an imbalance in the ratio of yields of $\geq$ 3 top-quarks events with positive and negative total leptonic charge, 
\begin{equation}
	r(4t)=\frac{N^{\text{four-tops},Q>0}_{SM+BSM}}{N^{\text{four-tops},Q<0}_{SM+BSM}}\frac{N^{\text{four-tops},Q<0}_{SM}}{N^{\text{four-tops},Q>0}_{SM}},
	\label{r4t}
\end{equation}
in the four-top-quarks search.  Observe that the ATLAS four-top-quarks analysis \cite{Aad:2020klt} does not distinguish three- and four-top-quarks and therefore this imbalance is also induced by diagrams as in Fig.~\ref{feyndiag}a because of an $up$-quark in the initial state and, as expected, grows with $g_{ut}$.  We plot the contour levels for $r(4t)$ in the figures since it provides a qualitative insight on the behaviour of the asymmetry coming from the $g_{ut}$-mediated BSM contributions in a four-top-quarks-like selection.  We also observe that this parameter is not currently reported by the experimental collaborations, and could provide clues of BSM contributions. 
 
To better identify interesting features, we plot in Fig~\ref{zp-fits-1} the region $M_{Z'}<2m_{t}$ and $g_{ct} = 0$; in Fig.~\ref{zp-fits-2} we study $M_{Z'}>2m_{t}$ and $g_{ct} = 0$; and in Fig.~\ref{zp-fits-gct} we explore the $g_{ct} \neq 0$ region while fixing $g_{tt}=0.2$ and $0.4$.  In the following paragraphs we discuss each one of these figures in detail.  Although we find many points in parameter space compatible with the experimental results, we observe that $M_{Z'}=400$ GeV has a slightly better accordance with the data. 

For $M_{Z'}<2m_{t}$ (Fig.~\ref{zp-fits-1}), the 1 s.d. regions in all plots include $g_{ut}\approx 0$. This corresponds to no charge asymmetry and is indicative of the fact that none of the points in the parameter space are a particularly good fit for the data in this region, as quantified by the poor goodness-of-fit, not being significantly better than that of the SM hypothesis ($g_{ut}=g_{ct}=g_{tt}=0$). This is because $Z' \to t\bar t$ is suppressed and thus the events providing charge asymmetry with large $b$-jet multiplicity are suppressed as well. This leaves $Z'\rightarrow t\overline{u}+\overline{t}u$ as the dominant decay mode, worsening the fit. When comparing the right column ($t\bar t H$ and $t\bar t t \bar t$) to the left column (only $t\bar t H$), one should keep in mind that yields in four-top-quarks searches are obtained mainly through the three-top-quarks production diagrams, which are proportional to $(g_{tt})^{2}BR(Z'\rightarrow t\overline{u}+\overline{t}u)$ and $(g_{ut})^{2}BR(Z'\rightarrow t\overline{t}^{*}+t^{*}\overline{t})$. For $M_{Z'}=200$ GeV, the second process is negligible and therefore the effect of incorporating four-top-quarks to the NLL is to constrain $g_{tt}$. Whereas for $M_{Z'}=300$ GeV, there is a slight opening of $Z'\rightarrow t\overline{t}^{*}+t^{*}\overline{t}$ and thus the 1 s.d. allowed region enlarges to the medium $g_{ut}$ region from Fig.~\ref{zp-fits-1}c to \ref{zp-fits-1}d.

The goodness-of-fit to $t\overline{t}H$ data increases when we consider $M_{Z'} > 2m_{t}$ (Fig.~\ref{zp-fits-2}).  In fact, for these $M_{Z'}$ values the $Z' \to t\bar t$ channel opens up and diagrams as in Fig.~\ref{feyndiag}a provide multilepton events with charge asymmetry and large $b$-jet multiplicity, which are crucial to improve the fit.  This important improvement can be explicitly seen by comparing the goodness-of-fit (green dotted contours) between the $M_{Z'} < 2m_{t}$ (Fig.~\ref{zp-fits-1}) and  $M_{Z'} > 2m_{t}$ (Fig.~\ref{zp-fits-2}) results in the left plots.  We can see in the left plots that as $M_{Z'}$ increases from 400 GeV to 600 GeV, the $tZ'$ and $t\bar t Z'$ production cross-sections decrease and therefore larger couplings are needed for the best-fit regions.    We see however in these plots that the best-fit regions are in potential tension with four-tops-quark production cross-section (black dashed contours), which indicate a preference for lower values for $g_{tt}$. In the right plots of Fig.~\ref{zp-fits-2} we include four-top-quarks data to the NLL and we observe that the best-fit point has a noticeable lower $g_{tt}$ and similar $g_{ut}$.  This is because four-top-quarks searches are more sensitive to $g_{tt}$ than to $g_{ut}$.  We do not consider masses above 600 GeV because we find that larger masses would yield $H_T$ distributions that enter into conflict with those reported in four-top-quarks analyses, as discussed below in Fig.~\ref{ht_global}.

We plot the $g_{ct}\neq 0$ cases in Fig.~\ref{zp-fits-gct} for $M_{Z'}=400$ GeV and two values of $g_{tt}$. In all cases we see that the best-fit point is in $g_{ct}=0$, which indicates that the main handles to accommodate the data are $g_{ut}$ and $g_{tt}$. Nevertheless, we observe that $g_{ct}\neq 0$ is allowed at the 1 s.d. level in a large part of parameter space.   We show in blue the region disfavoured by $D^0 \leftrightarrow \bar D^0$ mixing which limits the $g_{ut}$ and $g_{ct}$ couplings as detailed in Sect.~\ref{section:3_other}.

\begin{figure}[h!]
	\begin{center}
	\subfloat[]{\includegraphics[width=0.6\textwidth]{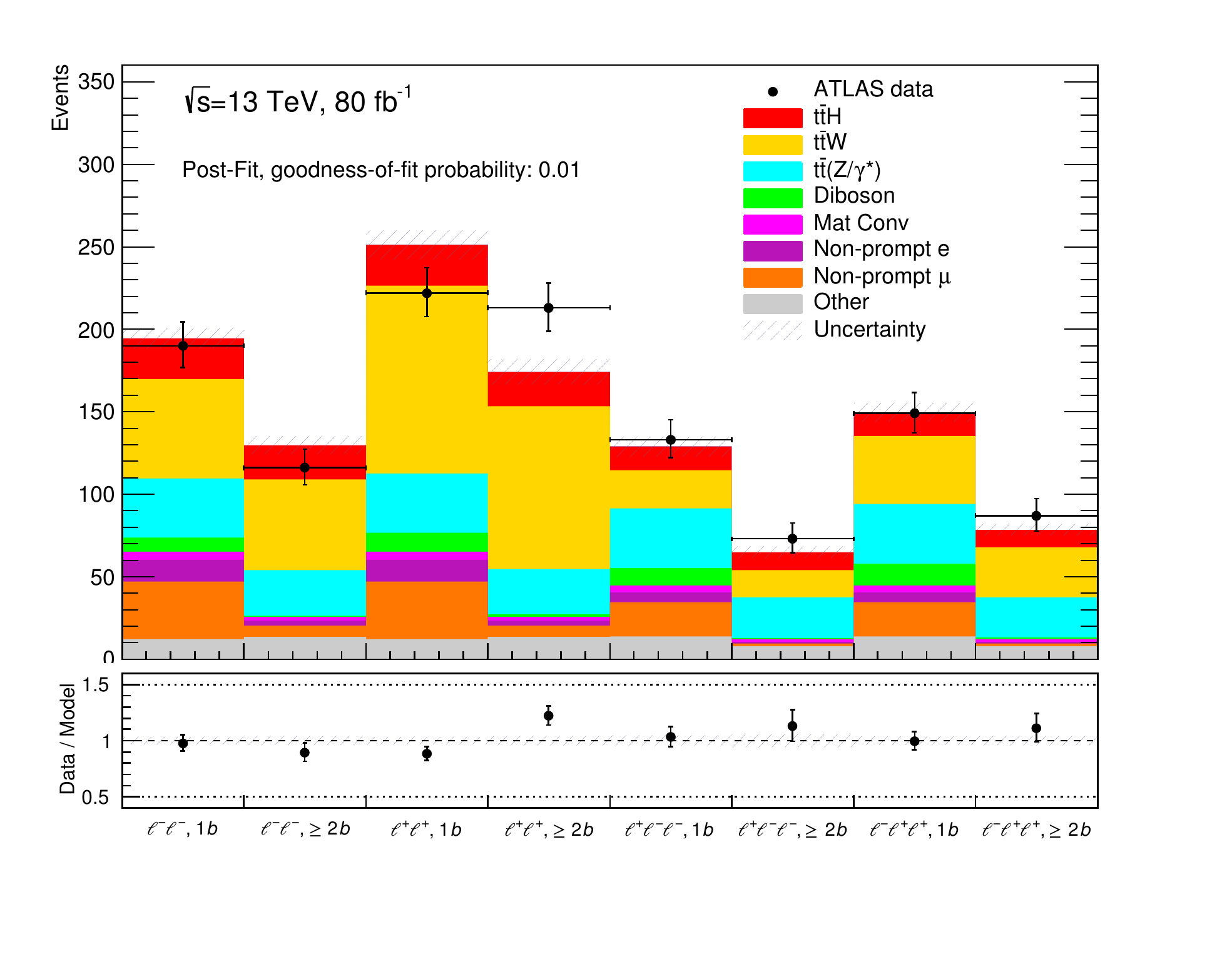}}\\
	\subfloat[]{\includegraphics[width=0.6\textwidth]{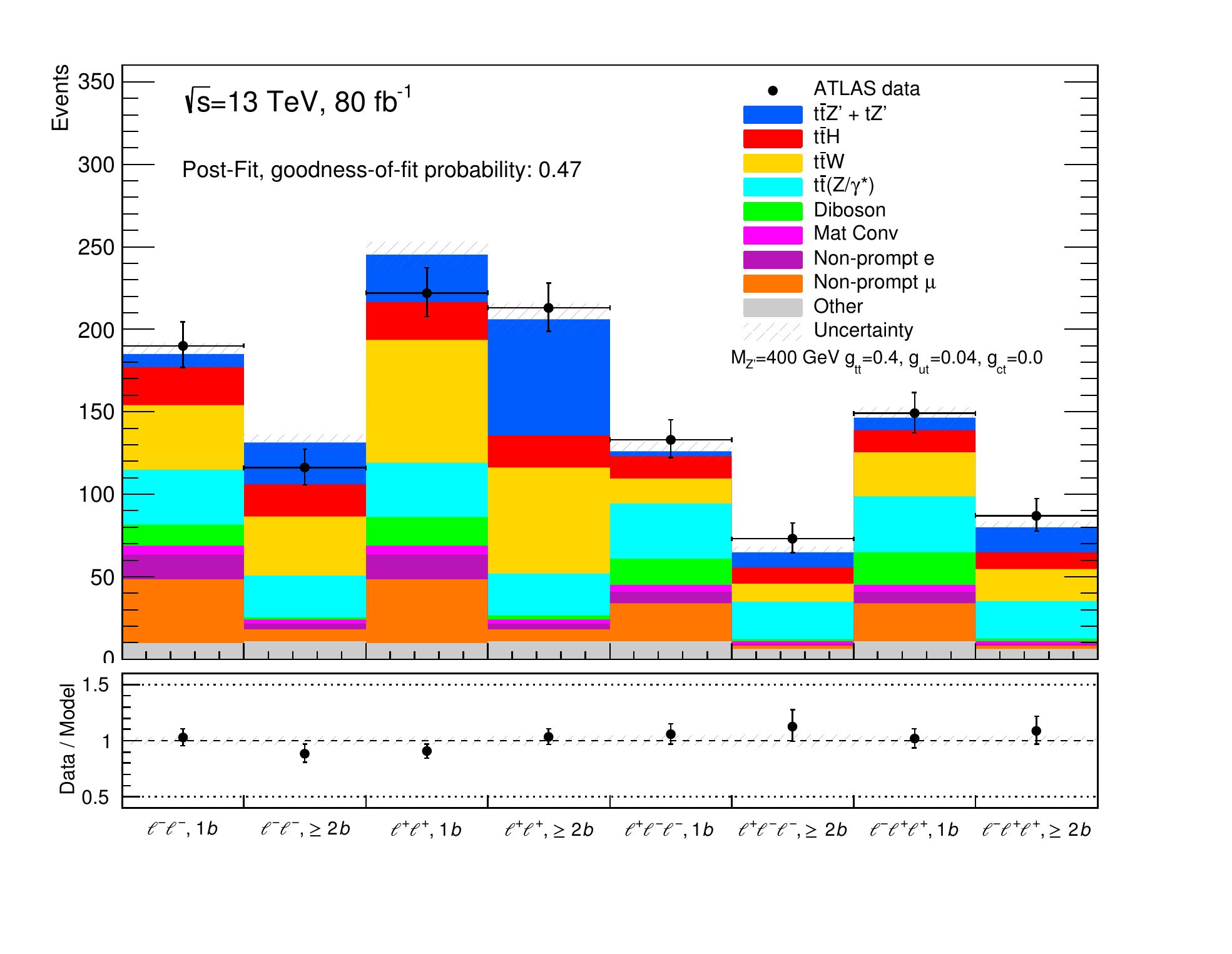}}\hspace{3mm}\\
	\end{center}
	\caption{Comparison between ATLAS data~\cite{ATLAS-CONF-2019-045} and a) SM background prediction b) BSM signal ($tZ'$ + $t\overline{t}Z'$) plus SM background prediction for the event yields in 2LSS and 3L channels in categories based on the total lepton charge and the $b$-jet multiplicity. The BSM signal corresponds to a $M_{Z'}=400$ GeV and couplings $g_{ut} = 0.04$, $g_{ct}=0.0$ and $g_{tt}=0.4$, which minimizes the NLL for the ATLAS data~\cite{ATLAS-CONF-2019-045} for $M_{Z'}=400$ GeV. In both cases, we show the SM background prediction with fitted NP to the data although the fitted values differ due to the presence of BSM events in b).
		}
\label{bins_tth}
\end{figure}

From these fits to the data we can conclude that the $Z'$ is not only compatible with the $t\overline{t}H$ and the four-top-quarks data, but also in many regions is more compatible that the SM. The $tZ'$ process can provide the necessary charge asymmetry and $b$-jet multiplicity while still being hidden in four-top-quarks production. To summarize how selecting a good benchmark point in parameter space reduces the tension in the reported results in $t\bar{t}H$ (Fig.~2 in Ref.~\cite{ATLAS-CONF-2019-045}), we show in Fig.~\ref{bins_tth} how these results are modified if the $Z'$ BSM is added to the SM yields. We consider the best-fit point for $t\overline{t}H$ data only for $M_{Z'}=400$ GeV:  $g_{ut} = 0.04$, $g_{ct}=0.0$ and $g_{tt}=0.4$. For comparison, we show the equivalent post-fit plot in the case where only SM processes are considered in the fit. In both fits, the same systematic model is used for the SM processes.

Although all masses above $2m_{t}$ provide good fits, we consider the best-fit point for $t\overline{t}H$ and four-top-quarks data corresponding to $M_{Z'}=400$ GeV and couplings $g_{ut} = 0.04$, $g_{ct}=0.0$ and $g_{tt}=0.2$ as our benchmark point for further studies.  As we show in Sect.~\ref{section:4}, this mass has the interesting feature of having an $H_{T}$ distribution similar to, although slightly softer than, that of SM four-top-quark production.

Finally, it is interesting to observe that the preferred region in parameter space by the fit indicates that $Z'$ associated production either with a single top-quark or with a top-quark pair can mimic $t\overline{t}W^{\pm}$ in the $t\overline{t}H$ search and four-top-quarks in the four-top-quarks search. In Sect.~\ref{section:4}, we study kinematic distributions and properties of the new signals compared to the main SM background processes in order to find ways to break the degeneracy.

\subsection{Constraints from other observables}\label{section:3_other}

In this section we analyse how other observables besides $t\bar t H$ and $t\bar t t \bar t$  constrain the previously studied parameter space, namely $0 \leq g_{ut} \leq 0.1$, $0 \leq g_{ct} \leq 0.1$ and $0 \leq g_{tt} \leq 1.0$; and $200 \leq M_{Z'} \leq 600 \text{ GeV}$.

As a low-energy physics phenomena, $D$-meson mixing is particularly sensitive to any BSM effects. In particular, since virtual $Z'$ and top quarks can contribute to $D^0 \leftrightarrow \bar D^0$ mixing at loop level, it is a sensitive observable to our model.  Since we approximate $g_{uc}\approx 0$, there is no tree level contribution to $D$-meson mixing.   We find that the Lorentz structure in Eq.~\ref{lagrangian1} provides a contribution to $D$-meson mixing that is translated to the $D$-meson mass difference as:
\begin{equation}
	\Delta M_D^{\mbox{BSM}} = - \frac{f_D^2 M_D B_D x}{64\, \pi^2 M_{Z'}^2} \left(  \frac{8}{3} f(x) \, (g_{ut})^2 \, (g_ {ct})^2 \right) .
	\label{deltaMD}
\end{equation}
where $x = (M_{Z'}/m_t)^2$. The deduction of this expression with the explicit expression for $f(x)$ and the numerical factors involved are detailed in Appendix~\ref{appendix_d0mixing}.

Although in principle one should compare the experimental value to the predicted value due to SM and BSM contributions, a detailed knowledge of $\Delta M_D^{\mbox{SM}}$ is currently lacking \cite{10.1093/ptep/ptaa104, Franco_2012, Isidori_2012}.  Therefore, as a naive estimation, we require that the BSM contribution to $\Delta M_D$ is smaller than the uncertainty in its measurement:
\begin{equation}
	\Delta M_D^{\mbox{BSM}} \lesssim \delta \left(  \Delta M_D^{\mbox{exp}} \right) .
\end{equation}
Using current available data~\cite{10.1093/ptep/ptaa104} we obtain that the above constraints are translated into the model parameters as
\begin{equation}
	g_{ut} \, g_{ct} < 2.0 \times 10^{-3} \mbox{ to } 4.5 \times 10^{-3} 
\end{equation}
for $M_{Z'}$ ranging from 200 GeV to 600 GeV, respectively.  These bounds are plotted as blue regions in Fig.~\ref{zp-fits-gct}.

High-energy collider physics is sensitive both to inclusive on-shell production of the $Z'+X$ and to non-resonant behaviour, such as the re-scaling of different top-physics cross-sections~\cite{Cox:2015afa,Fox:2018ldq,Alvarez:2016nrz, Cho_2020}. As mentioned before, taking $g_{uu},g_{cc} \approx 0$ makes our model insensitive to many of them. Other effects are greatly diminished.

Measuring top-quark rare decays opens a window to BSM effects. $Z'$-induced top-quark rare decays have been studied in Ref.~\cite{Aranda_2020} and can be confronted to the experimental bounds set in Refs.~\cite{Aad_2020, Aad:2012gd, Khachatryan:2015att, Khachatryan:2016sib}. Using the same matrix elements we can produce the relevant decay branching ratios for the model described in Eq.~\ref{lagrangian1}. After adding the smaller SM contributions, we find that rare decays branching ratios are below current limits, which are of $\mathcal{O}(10^{-5})$. The main difference between our model and those considered in Ref.~\cite{Aranda_2020} is the absence of $g_{uu}$ and $g_{cc}$ couplings and of left-handed couplings, which effectively reduces the $Z'$-induced rare decays and allows us to go to smaller masses.

As mentioned above, in high-energy collider searches our model yields no $pp\rightarrow Z'$ production. In $t\overline{t}$ production, it can only contribute through the t-channel and cannot be probed by $t\overline{t}$ resonance searches. The only constraints coming from $t\overline{t}$ are then those on the inclusive cross-section, where we find that $Z'$ effects are well below the current experimental bounds~\cite{Aad:2020tmz, Khachatryan_2016} for the parameter space we are exploring, being at most 45 $\text{fb}$  for $g_{ut}=0.1,g_{ct}=0.1,g_{tt}=1.0$ and $M_{Z'} = 200$ GeV.

Our $Z'$ model is also sensitive to same-sign top-quark pair searches, both to prompt same-sign top-quark pair production mediated by a $Z'$ in the t-channel  and to same-sign top-quark production with an associated jet. The former is sensitive to $(g_{ut})^{4}, (g_{ct})^{4},$ and $(g_{ut})^{2}(g_{ct})^{2}$, the dominant process being $uu\rightarrow tt$ due to the PDF imbalance between the up quark and the others. The latter corresponds to the same $tZ'$ production mechanism we discuss in Sect.~\ref{section:2_pheno}, but we now select events where the $Z'$ decays to $t\bar u + t\bar c$. Current experimental limits from prompt same-sign top-quark pair production are $\sigma_{tt}\leq 1.2\text{ pb}$~\cite{Sirunyan:2017uyt} and $\sigma_{tt}\leq 89\text{ fb}$~\cite{Aaboud:2018xpj}. However, the latter limit is not fully model-independent, as the signal regions are optimized for higher mass vector mediators (with $M_{V} \geq 1$ TeV). As for our $Z'$ model, the highest possible $\sigma_{tt}$ we can produce is $76\text{ fb}$, for $M_{Z'} = 200$ GeV and $g_{ut}=0.1$, which is just below the most stringent current experimental limits. $M_{Z'}$ is so low that we should expect the acceptance to be significantly affected by the experimental cuts.  In particular, the $H_{T} \geq 750$ GeV cut imposed in Ref.~\cite{Aaboud:2018xpj} could be too high for such a low $M_{Z'}$. The cross-section $\sigma_{tt}$ is proportional to $(g_{ut})^{4}$ and decreases as we increase the mass, e.g. $\sigma_{tt} = 15\text{ fb}$ for $M_{Z'} = 400$ GeV and $g_{ut}=0.1$. All things considered, we conclude that prompt same-sign top-quark pair production does not place relevant constraints on our parameter space.  

Ref.~\cite{Aaboud:2018xpj} also casts experimental limits on $ttj$, although they consider a heavier $Z'$. Taking into account the re-scaling due to the $\text{BR}(Z'\rightarrow tj)$ and the same caveats about the acceptances, we find that these limits are also avoided by our model. A dedicated search for a light $Z'$ in this channel would be interesting, as one could potentially reconstruct the $Z'$ mass. Relaxing the $H_{T}$ cut and selecting events with a hard light jet would potentially augment the signal acceptance, while the background would still be relatively small. Low-mass $tj$ resonances have been searched in the $t\bar t +$ jets channel~\cite{Aad:2012em, Aaltonen:2012qn}, albeit with very low sensitivity.

While $Z'$-mediated single top-quark production $tj+\overline{t}j$ is absent due to $g_{uu}, g_{cc} \approx 0$, there is the possibility of $Z'$ radiative production $tZ'j$, where $Z'$ is emitted from either the $t$/$\overline{t}$ or the jet (which can then be either $u$, $\overline{u}$, $c$ or $\overline{c}$). However, in our model this channel is also severely suppressed due to the chiral nature of our coupling choice. All radiative $tZ'j$ production requires a $W^{\pm}$, either in the s-channel or the t-channel, with the latter dominating\footnote{A similar scenario is explored for a scalar in Ref.~\cite{Craig_2017}.}. To interact with this $W^{\pm}$, we need left-handed quarks. However, in our model we require right-handed quarks to radiate a $Z'$, which yields a considerable suppression. In particular, due to PDF imbalance the most relevant diagram with $W^{\pm}$ for $tZ'j$ is $ud\rightarrow du \rightarrow dtZ'$. But this diagram requires a left-handed up quark to interact with the $W$ and a right-handed up quark to interact with the $Z'$. As the up quark is essentially massless, the most important contribution to $tZ'j$ essentially vanishes for right-handed couplings only. To illustrate this point, assuming $M_{Z'} = 400$ GeV, we find that the $tZ'j$ cross-section for $g^{L}_{ut}=0.0,g^{R}_{ut}=1.0$ is approximately 40 times smaller than for $g^{L}_{ut}=1.0,g^{R}_{ut}=0.0$.

\clearpage

\section{Global search and kinematic features}
\label{section:4}

In Sect.~\ref{section:3} we show how a specific $Z^{'}$ vector boson can affect $t\bar t H$ and four-top-quarks analyses. Although all BSM diagrams to some extent affect both observables, the charge asymmetric contribution mimics mainly $t\overline{t}W^{\pm}$, whereas the charge symmetric contribution mimics mostly four-top-quarks behaviour in different searches and channels.  In this section we show how a more global study of the $Z^{'}$ could help to break this degeneracy and disentangle signal from SM background processes.  Then, based on the studied discriminating observables, we define signal-enriched regions that have either $t\overline{t}W^{\pm}$ or four-top-quarks as the main SM background and thus are sensitive to different parameters of the model. With the help of a few kinematical variables in these regions we aim to improve signal vs background discrimination, showing the potential of a future optimised Multivariate Analysis (MVA) discriminant.

\subsection{Expected sensitivity of the global analysis}

The event selection used for the global analysis is similar to that of the four-top-quarks search~\cite{Aad:2020klt}, but with some modifications. We define the regions as:
\begin{eqnarray}
	\mbox{2LSS:}&& \mbox{Two same-sign leptons, at least 3 jets and at least 1 $b$-jet,} \nonumber \\
	\mbox{3L:}&& \mbox{(Exactly) Three leptons, at least 2 jets and at least 1 $b$-jet.} \nonumber
\end{eqnarray}
No $H_{T}$ cut is applied in either of the regions.

\begin{figure}[h]
	\begin{center}
\captionsetup[subfigure]{labelformat=empty}
		\subfloat[]{2LSS\hskip 1cm}\hspace{0.4\textwidth}\subfloat[]{3L}\\
\captionsetup[subfigure]{labelformat=parens}
		\addtocounter{subfigure}{-2}
	\subfloat[]{\includegraphics[width=0.45\textwidth]{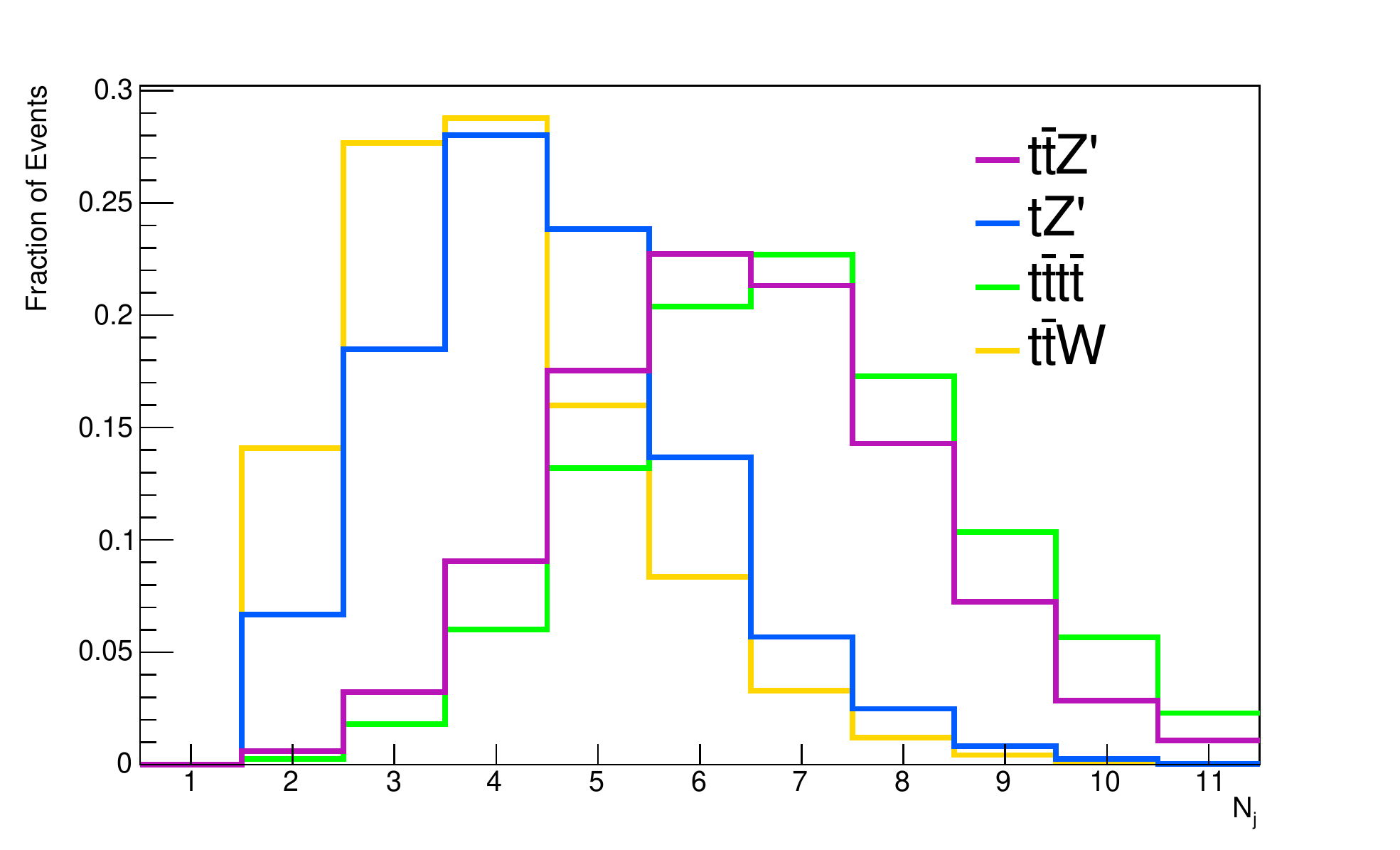}}\hspace{3mm}
	\subfloat[]{\includegraphics[width=0.45\textwidth]{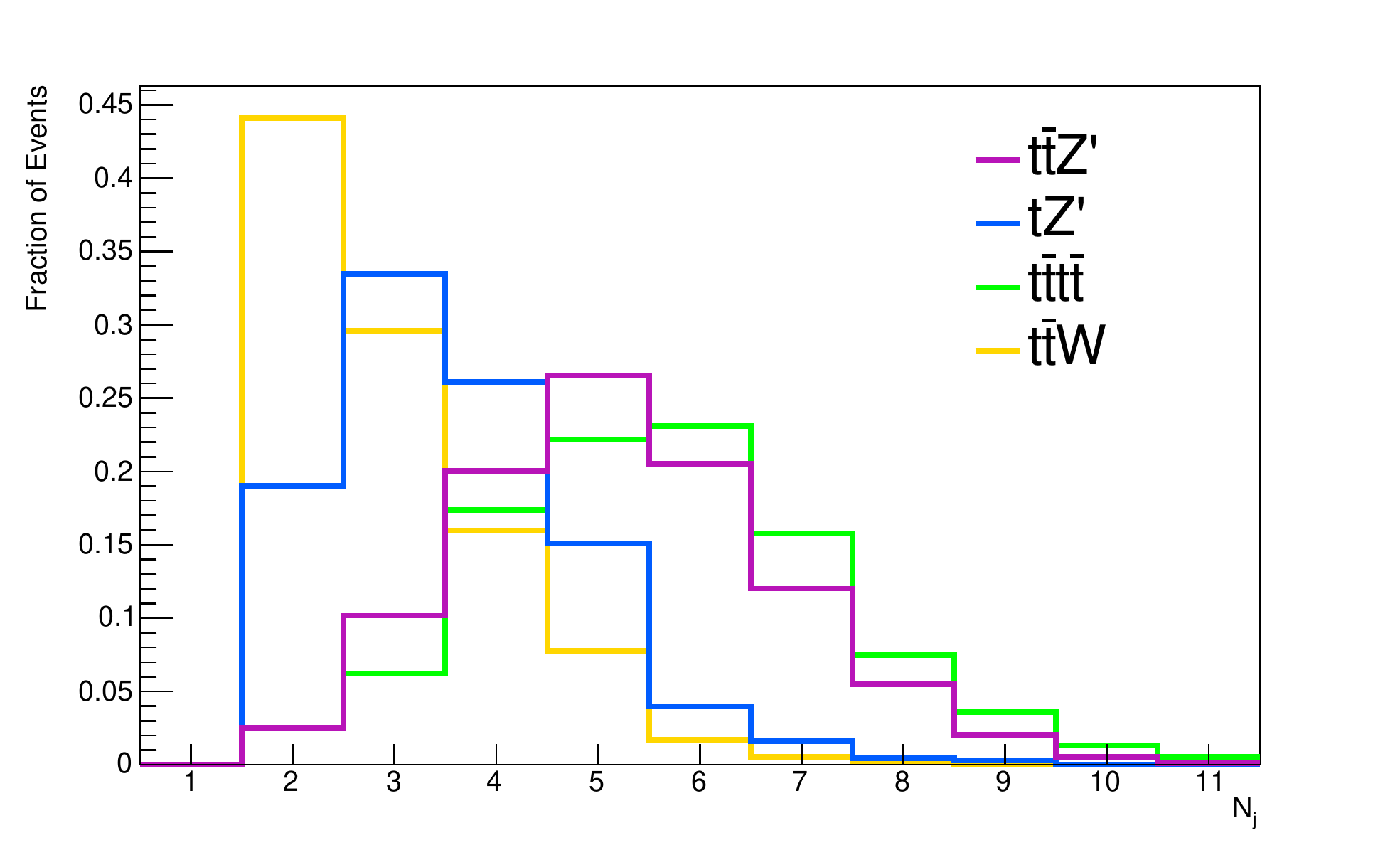}}\hspace{3mm}\\
	\subfloat[]{\includegraphics[width=0.45\textwidth]{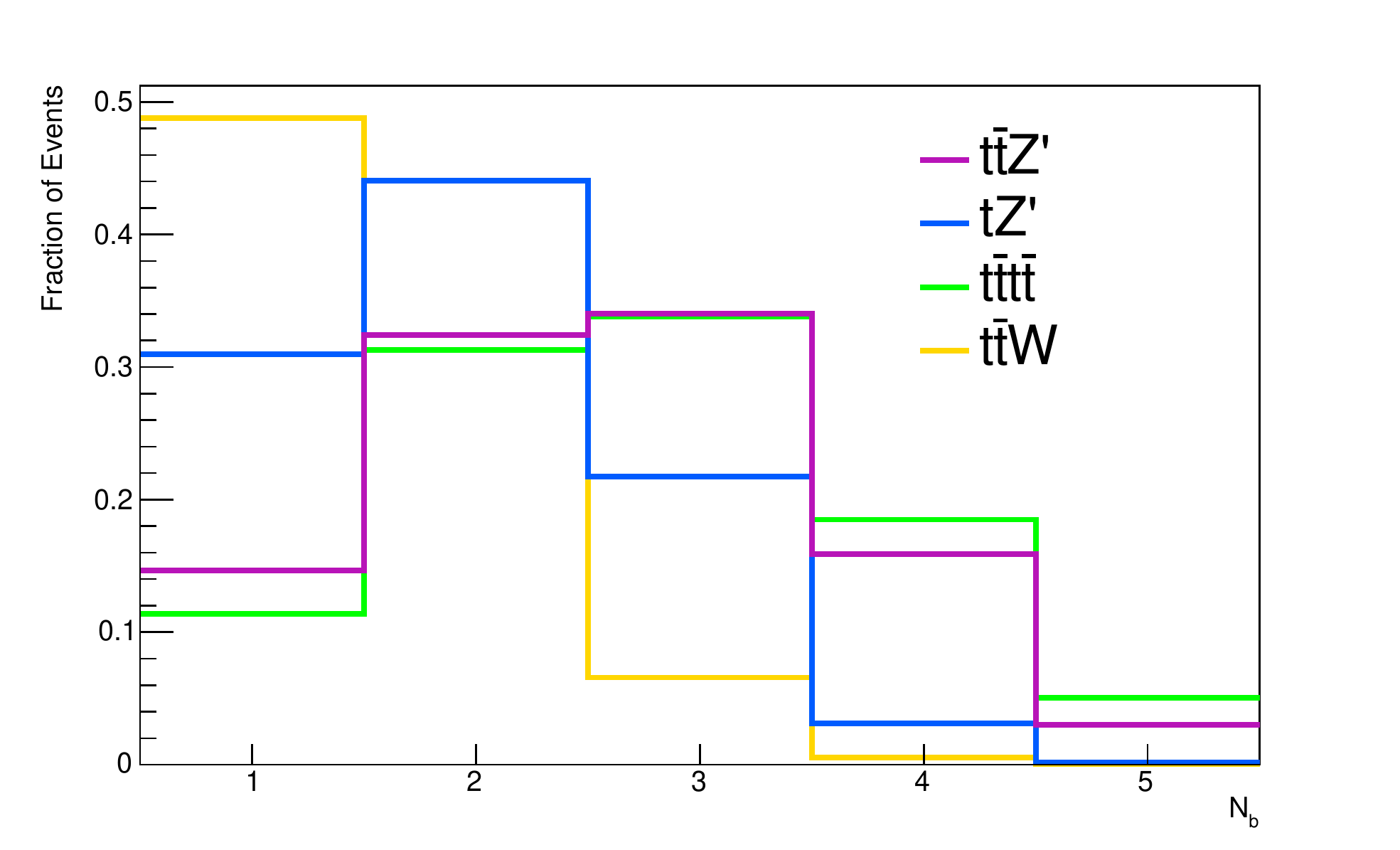}}\hspace{3mm}
	\subfloat[]{\includegraphics[width=0.45\textwidth]{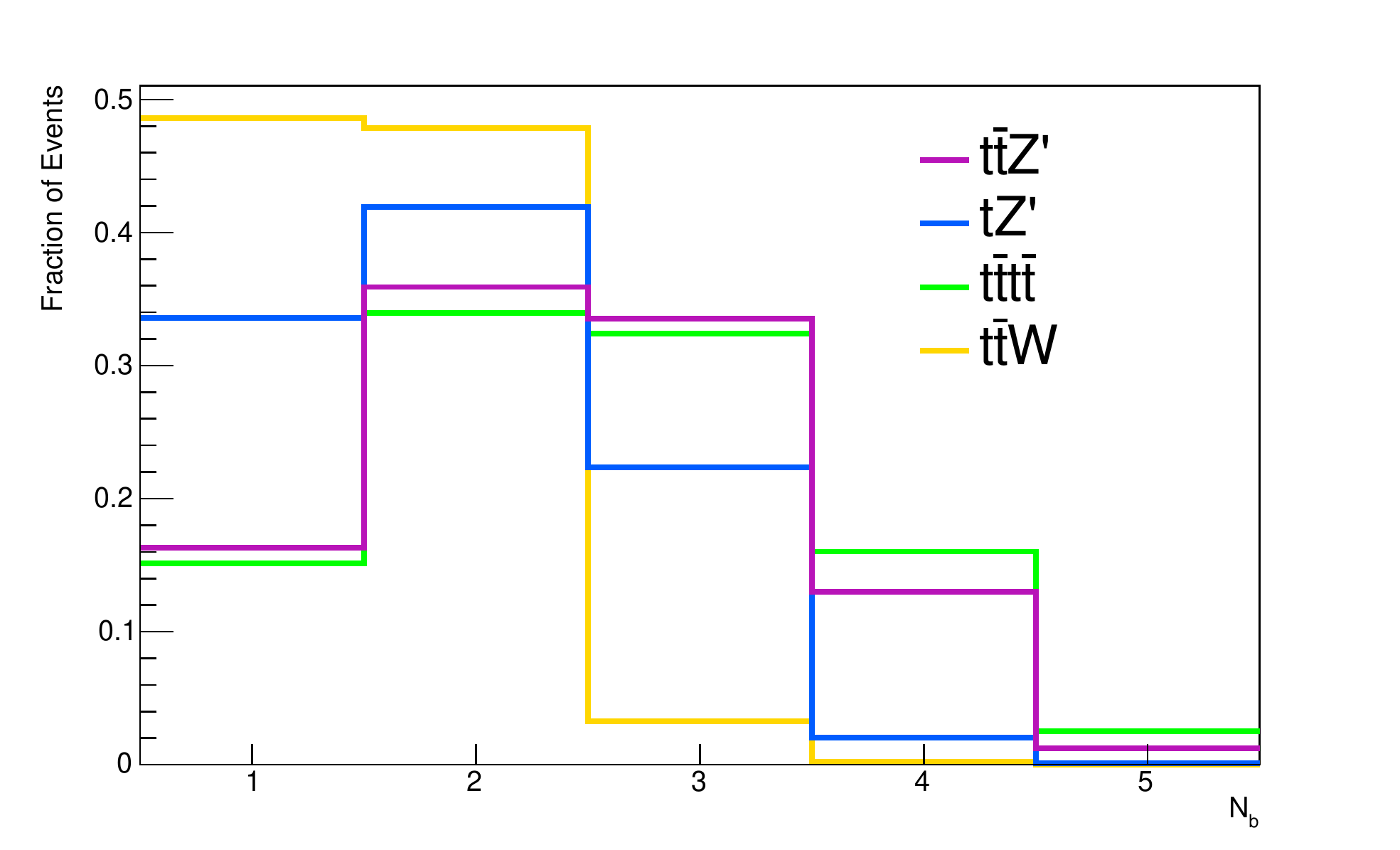}}\\
         \subfloat[]{\includegraphics[width=0.45\textwidth]{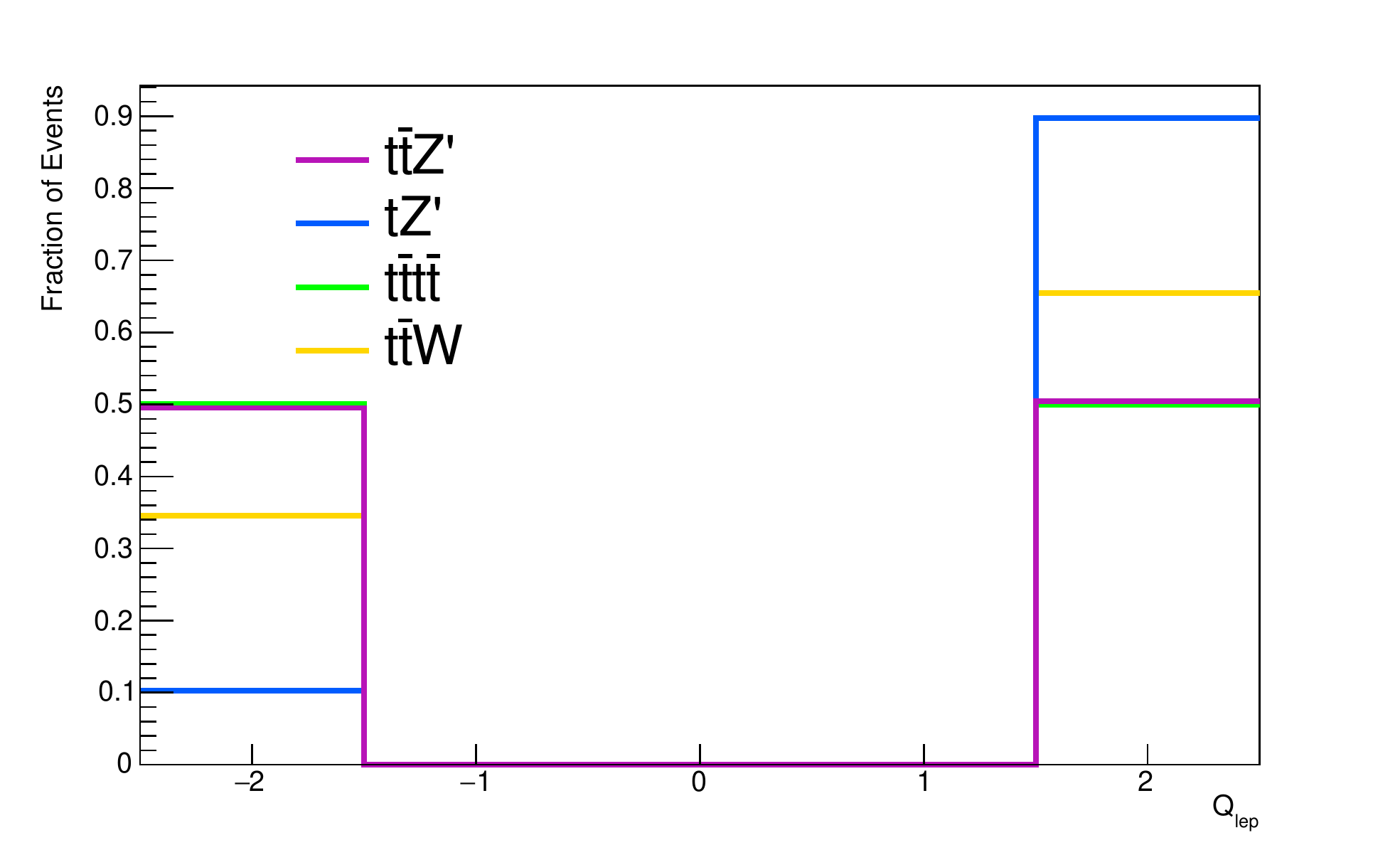}}\hspace{3mm}
	\subfloat[]{\includegraphics[width=0.45\textwidth]{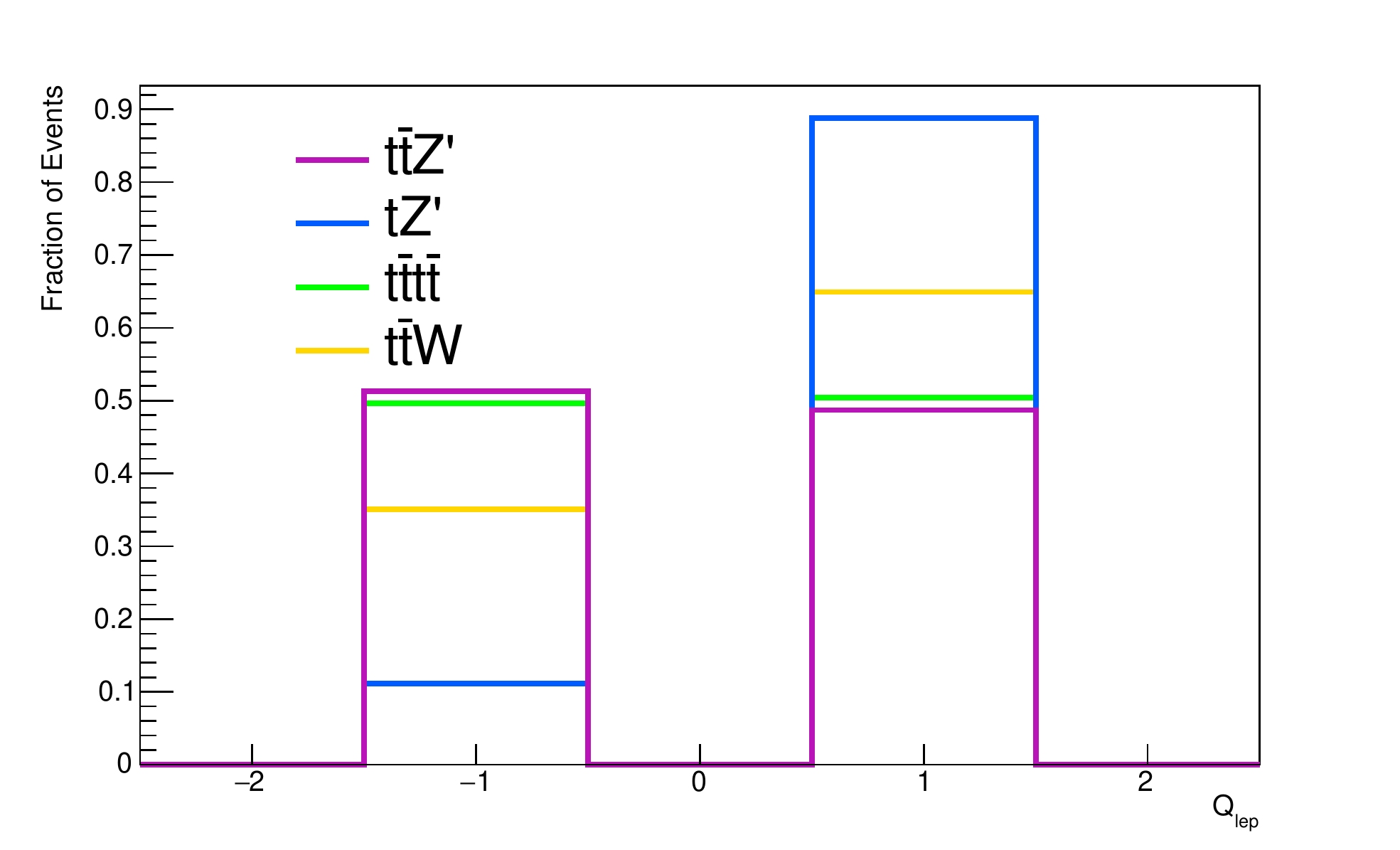}}\hspace{3mm}\\
	\end{center}
	\caption{$N_j$, $N_b$, and $Q_{\text{lep}}$ distributions for the 2LSS (left) and 3L (right) categories used for the global analysis.  Observe how $tZ'$ mimics $t\bar t W$ and $t\bar t Z'$ mimics $t\bar t t \bar t$, particularly in the $N_j$ distribution.}
\label{q_nj_nb}
\end{figure}

The event categories in the global analysis are defined for the 2LSS and 3L selection separately, based on the discriminating variables: number of jets ($N_j$), number of $b$-jets ($N_b$), and total leptonic charge ($Q_{\text{lep}}$). In Fig.~\ref{q_nj_nb} we show the distributions for these discriminating variables in each category. 

\begin{figure}[h!]
	\begin{center}
	\subfloat[2LSS]{\includegraphics[width=0.7\textwidth]{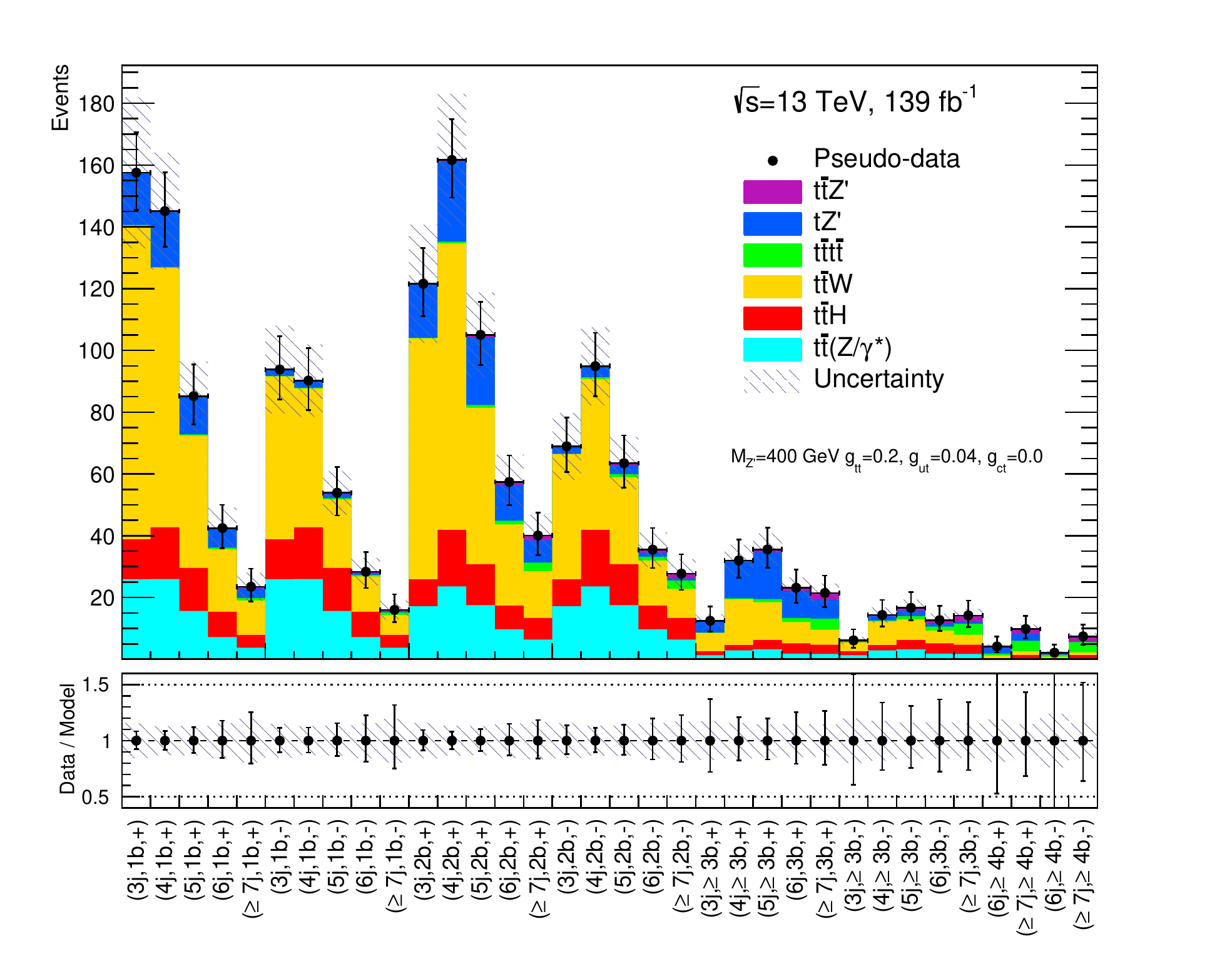}}\\
	\subfloat[3L]{\includegraphics[width=0.7\textwidth]{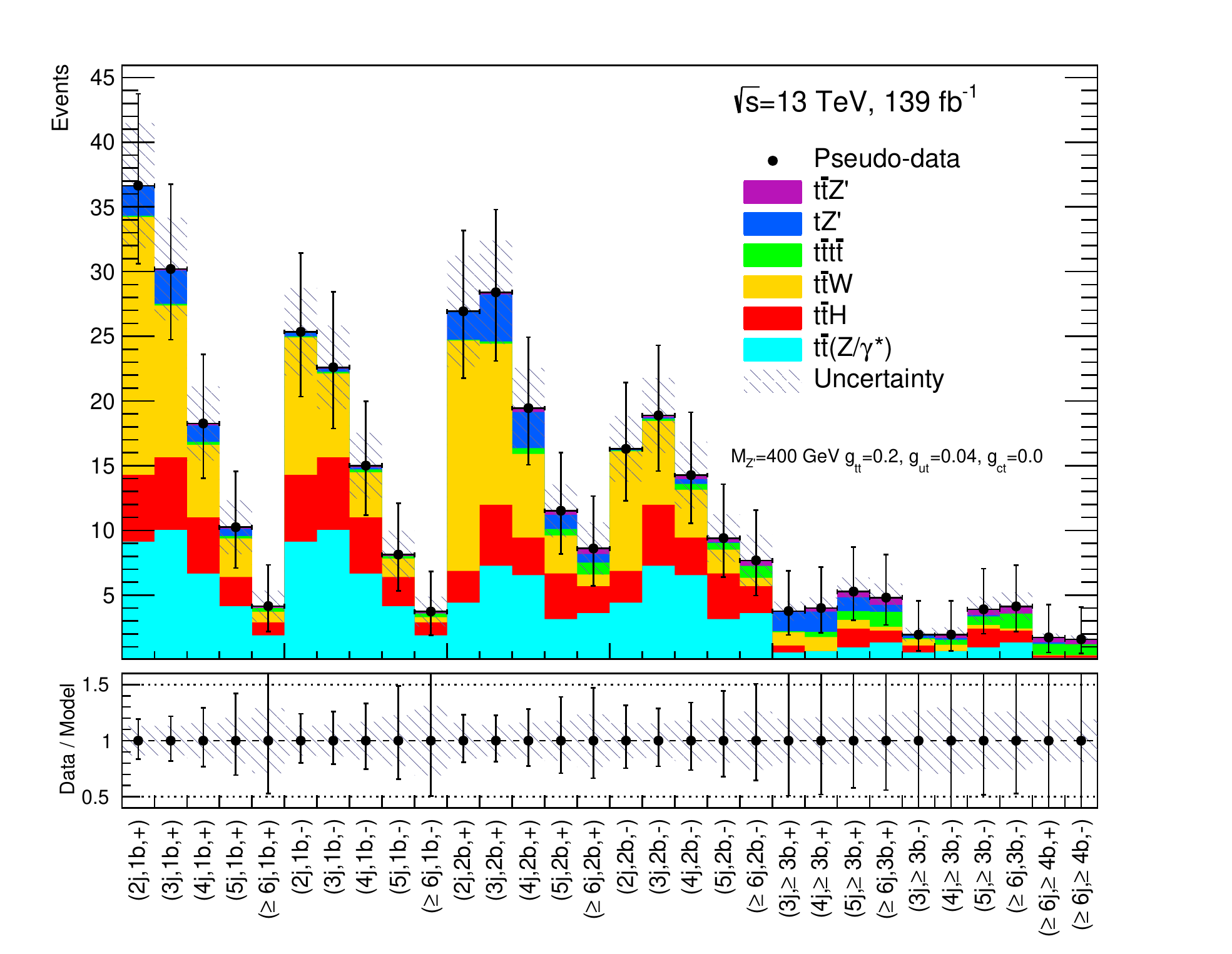}}\hspace{3mm}\\
	\end{center}
	\caption{Global event yields for $tZ'$, $t\overline{t}Z'$ and the main irreducible backgrounds for the benchmark point $M_{Z'}= 400$ GeV, $g_{ut}=0.04$, $g_{ct}=0.0$ and $g_{tt}=0.2$. Each bin consists of a global selection criteria on $N_{j}$, $N_{b}$, and $Q_{\text{lep}}$, as explained in the text.
		}
	\label{global}
\end{figure}

\begin{figure}[h!]
	\begin{center}
	\includegraphics[width=0.9\textwidth]{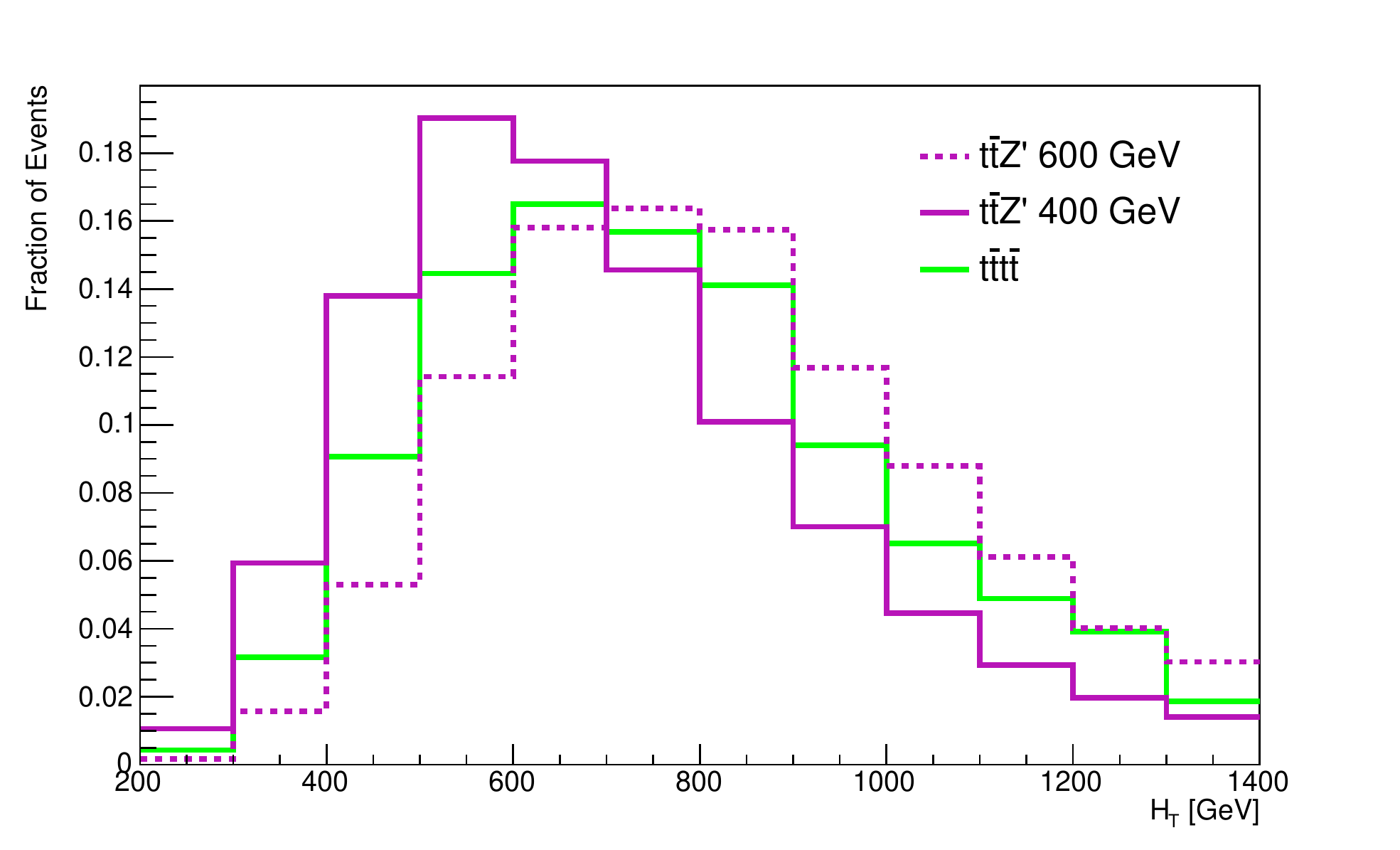}
	\end{center}
	\caption{$H_{T}$ kinematic distribution for events belonging to the global analysis selection.
		}
\label{ht_global}
\end{figure}

In order to perform a global analysis we define bins as $(N_{j},N_{b},Q_{\text{lep}})$ with $Q_{\text{lep}}=\pm$ as a shorthand for $\pm 2 (1)$ for 2LSS (3L) and $\geq N_{j}$ $(N_{b})$ meaning at least $N_{j}$ jets ($N_{b}$ $b$-jets). There are in total 34 (30) bins for the 2LSS (3L) channel. Low $N_{j}$ and $N_{b}$ bins will have a larger contamination of $t\overline{t}W^{\pm}$ and higher $N_{j}$ and $N_{b}$ bins will be more sensitive to four-top-quarks. The expected events for each one of these categories in the global analysis are shown in Fig.~\ref{global}, where we use as a signal benchmark the combined $t\overline{t}H$ and four-top-quarks best-fit point for $M_{Z'}=400$ GeV from Fig.~\ref{zp-fits-2}. We also plot the main irreducible backgrounds simulated at NLO accuracy with the same generator settings as for the signal, with cross-sections scaled to those reported in Ref.~\cite{Aad:2020klt}.  From the global picture we see how the signal mimics $t\overline{t}W^{\pm}$ in low $N_{j}$, low $N_{b}$ bins and four-top-quarks in high $N_{j}$, high $N_{b}$ bins. There are some bins that provide a clear distinction between signal and background. More insight into how $t\overline{t}Z'$ can resemble SM four-top-quarks is shown in Fig.~\ref{ht_global}, where both SM and BSM processes have a similar $H_{T}$ kinematic distribution in the events selected for the global histogram.

\captionsetup[subfigure]{labelformat=empty}
\captionsetup[subfigure]{labelformat=parens}

\begin{figure}[h!]
	\begin{center}
	\subfloat[]{\includegraphics[width=0.45\textwidth]{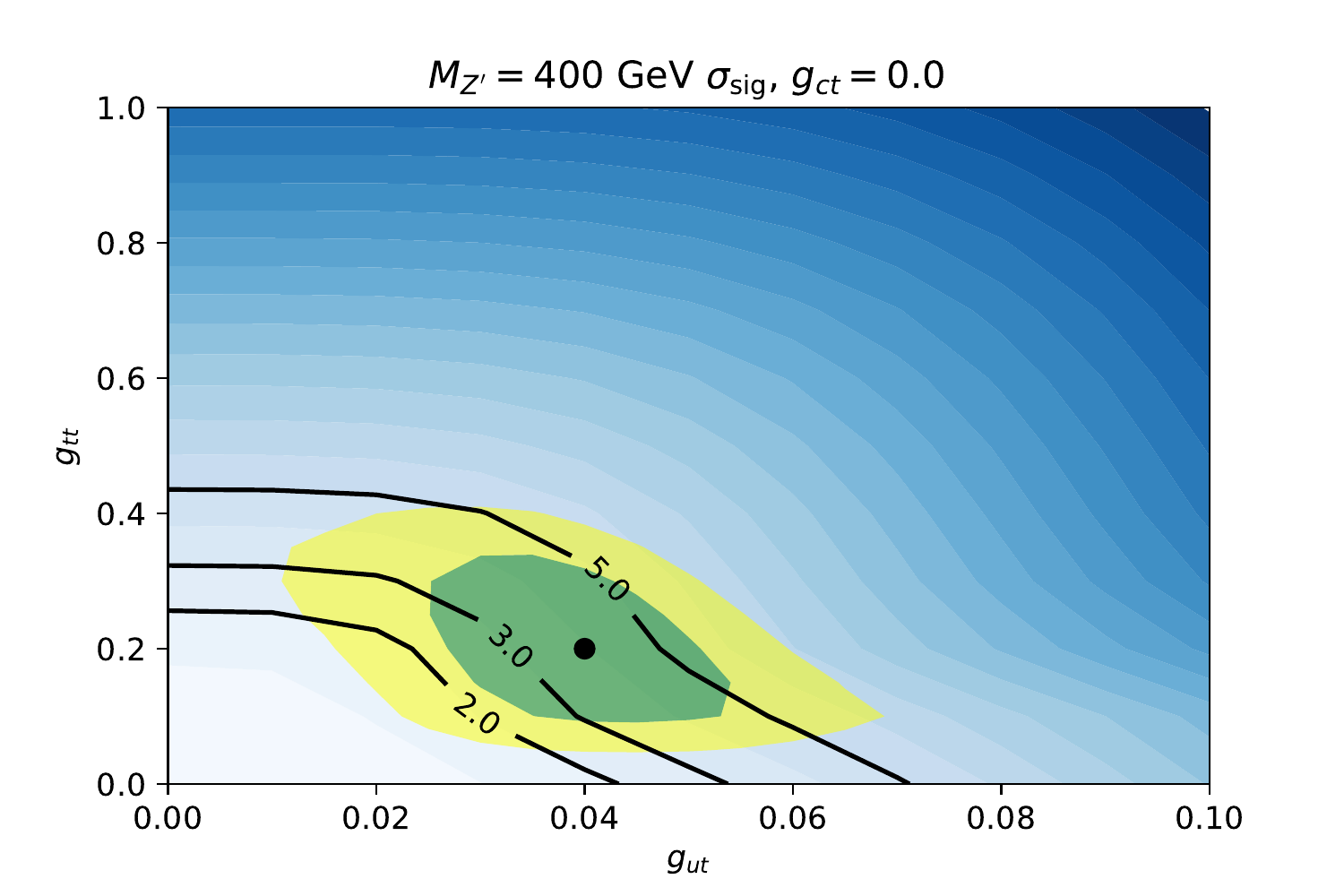}}\hspace{3mm}
	\subfloat[]{\includegraphics[width=0.45\textwidth]{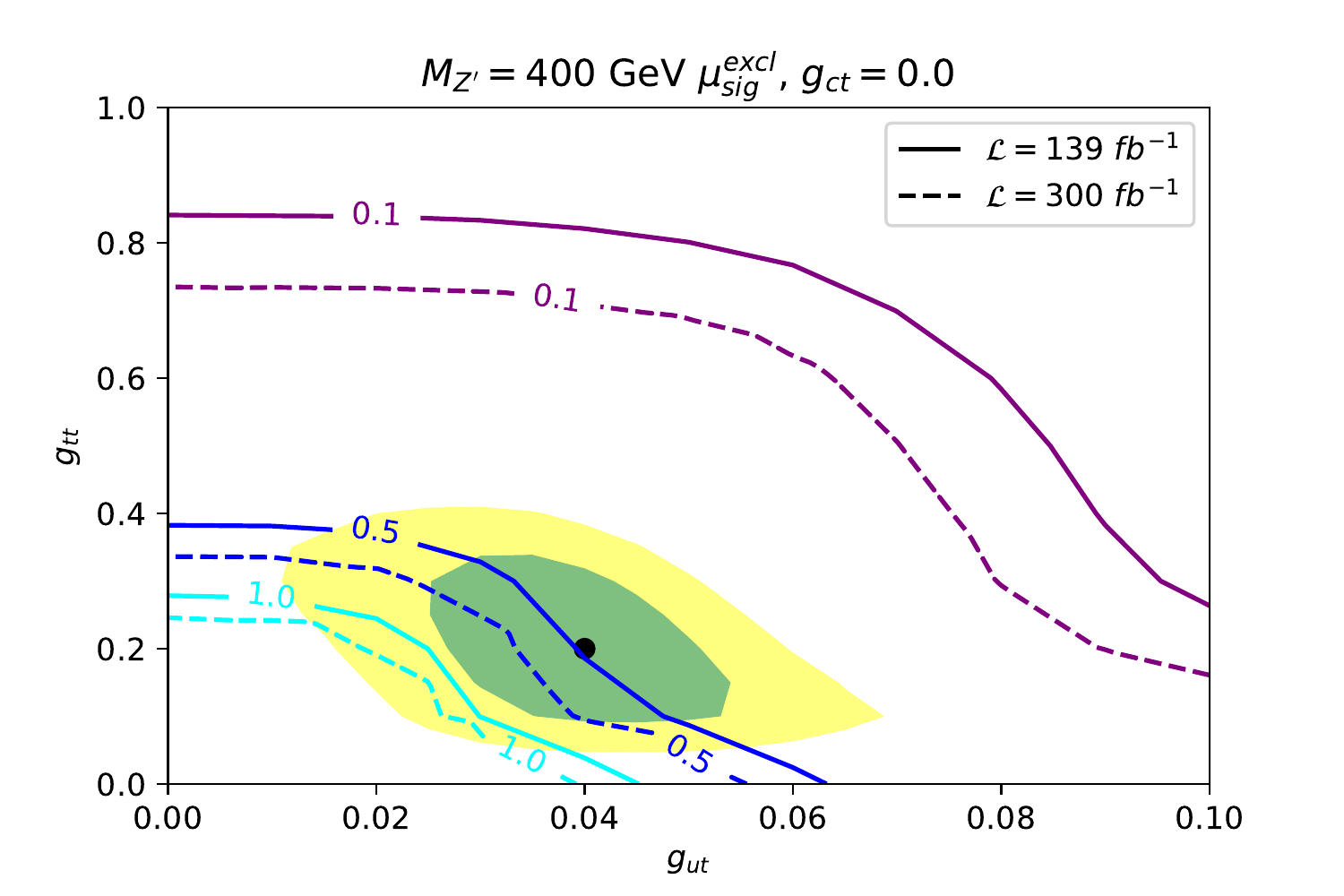}}\hspace{3mm}
	\end{center}
	\caption{Left: Expected significance using global bins for 139 $\text{fb}^{-1}$. Right: Expected exclusion limits on the ($tZ'$+$t\overline{t}Z'$) signal strength $\mu$. Solid (dashed) lines are limits for 139 $\text{fb}^{-1}$ (300 $\text{fb}^{-1}$). We also show the corresponding 1 and 2 s.d. regions of Fig.~\ref{zp-fits-2}b.}
\label{global-stat}
\end{figure}

The global analysis shown in Fig.~\ref{global} can be used to estimate a potential discovery or the exclusion limit for the BSM $Z'$ model.  This is done for $M_{Z'} =400$ GeV and each point in the $(g_{ut},g_{ct},g_{tt})$ parameter space. Using Eq.~\ref{eq:interpol} as detailed in Sec.~\ref{section:3}, we generate global histograms for an arbitrary point in the parameter space using a fixed set of simulated points. These processes are not independent of each other and thus need to be considered simultaneously.  An Asimov dataset~\cite{Cowan:asimov} is used to calculate the expected significance and exclusion limits. The systematic uncertainties included in the global fit consist of 20\% overall normalization uncertainty on the $t\overline{t}H$, $t\overline{t}W$, $t\overline{t}Z$, four-top-quarks processes, and signal, as well as $N_{j}$- and $N_{b}$-dependent uncertainties on $t\overline{t}H$, $t\overline{t}W$, and $t\overline{t}Z$ to account for larger modelling uncertainties in the production of additional light/heavy-flavour jets.  We show in Fig.~\ref{global-stat} the expected significance and the expected exclusion limits for the case where $M_{Z'}=400$ GeV and $g_{ct}=0$ for the corresponding integrated luminosity of the Run-2 dataset (139 $\text{fb}^{-1}$) and of the expected Run-2 plus Run-3 datasets (300 $\text{fb}^{-1}$).

\subsection{Disentangling signal and background in signal-enriched regions}

In order to define signal-enriched regions both for $tZ'$ and $t\overline{t}Z'$ we require $N_{b}\geq 3$. Moreover, given the top-quark content in each of these processes we can use $Q_{\text{lep}}$ and $N_{j}$ to further separate them. We therefore define two signal-enriched regions:
\begin{eqnarray}
	\mbox{Region I, 2LSS (3L):} && \mbox{$N_{j} \geq 4 (3)$, $N_{b} \geq 3$ and $Q_{\text{lep}}= +2 (+1)$,} \nonumber \\
	\mbox{Region II, 2LSS (3L):} && \mbox{$N_{j} \geq 7 (5)$, $N_{b} \geq 3$ and $Q_{\text{lep}}=\pm 2 (\pm 1)$.} \nonumber
\end{eqnarray}

\begin{figure}[h]
	\begin{center}
	\subfloat[Region I]{\includegraphics[width=0.48\textwidth]{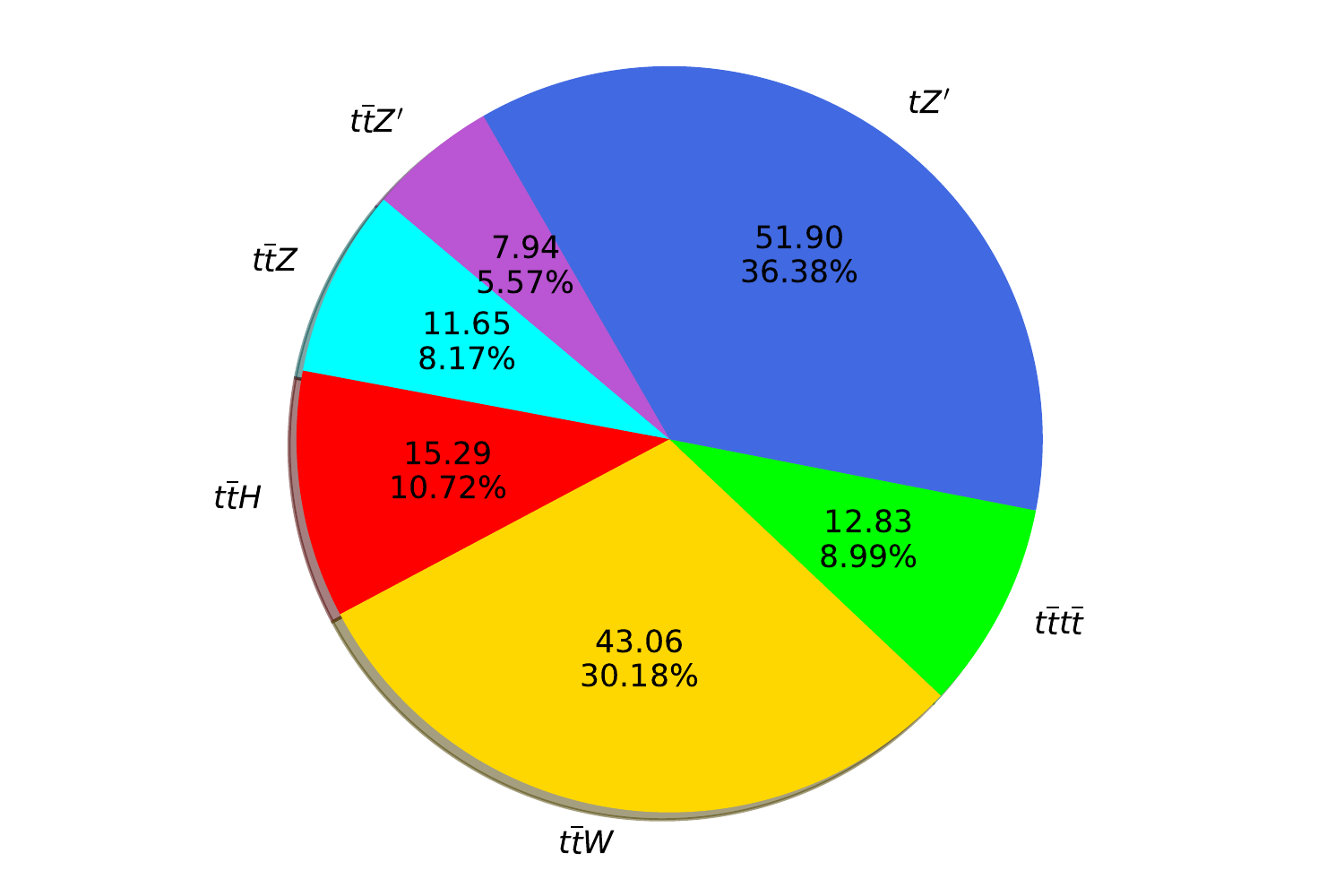}}\hspace{3mm}
	\subfloat[Region II]{\includegraphics[width=0.48\textwidth]{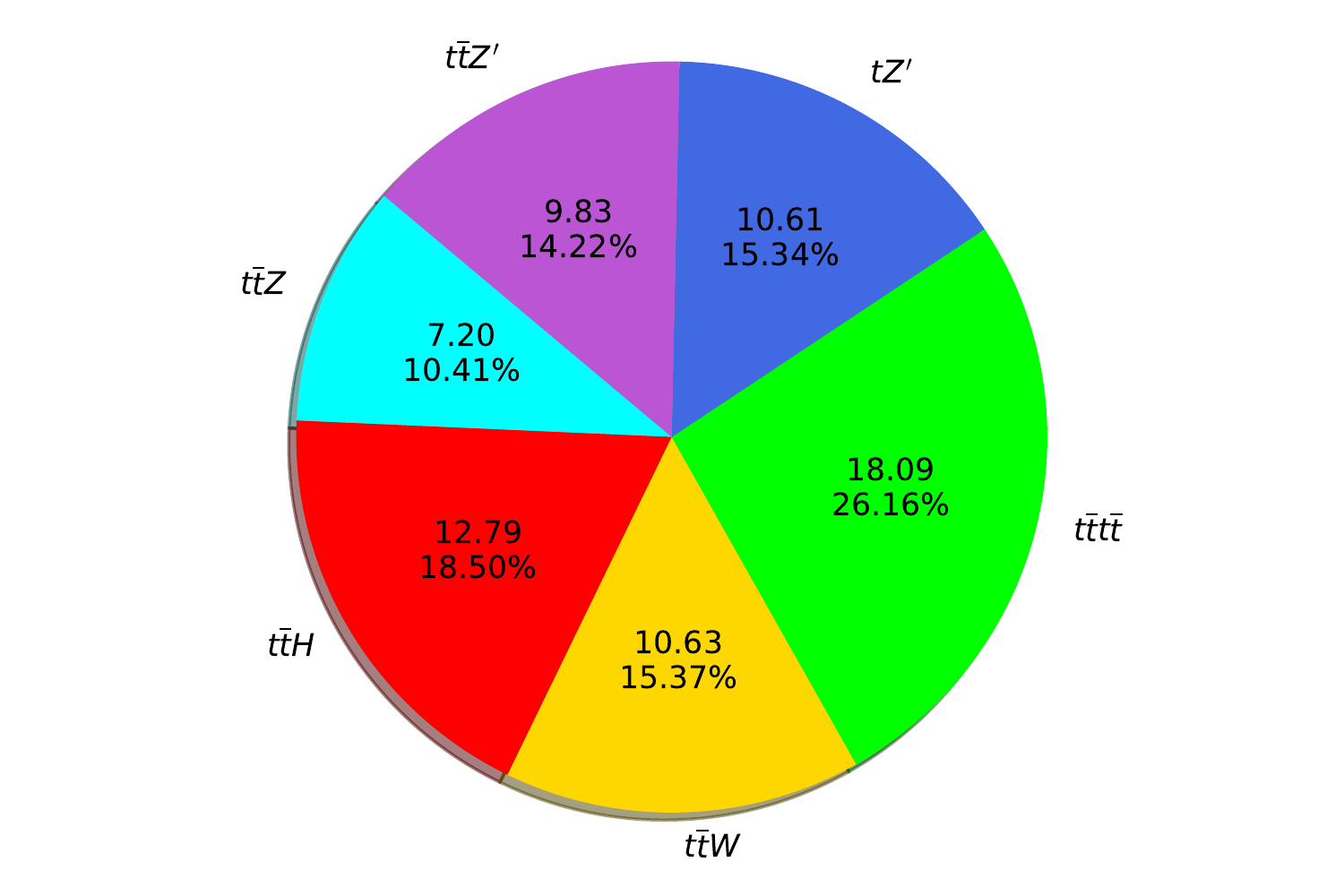}}
	\end{center}
	\caption{Event yields and composition of Regions I and II.}
\label{pie-charts}
\end{figure}

Region I is more sensitive to $tZ'$ while Region II is more sensitive to $t\overline{t}Z'$. We show the composition of the events that fulfil the selection of each Region in Fig.~\ref{pie-charts}.  We aim to examine observables that can disentangle signal from background in each one of the signal-enriched regions.

\captionsetup[subfigure]{labelformat=empty}
\captionsetup[subfigure]{labelformat=parens}

In the following, we plot only the fraction of events for the relevant processes in each region, i.e. $t\overline{t}W$ and $tZ'$ for Region I and four-top-quarks and $t\overline{t}Z'$ for Region II. We show  for $Z'$ the relevant masses 400 GeV and 600 GeV.  We also find compelling to include in the study a BSM model in which we replace the spin-1 $Z'$ with a spin-0 scalar field $\Phi$ in the Lagrangian in Eq.~\ref{lagrangian1}, while coupling to the full chirality fermions $u$ and $t$.   For this scalar BSM model we use a mass $M_\Phi=400$ GeV and the same couplings as the best-fit point of $Z'$ for this mass. The two relevant processes in this model are then  $t\Phi$ and $t\overline{t}\Phi$. 

To compare the relevant studied processes we calculate the separation between processes, defined as~\cite{TMVA:guide}:
\begin{equation}
\text{Separation}= \langle S^2 \rangle = \frac{1}{2}\sum_{i=1}^{N_{\text{bins}}}\frac{\left(f^{\text{sig}}_{i}-f^{\text{bkg}}_{i}\right)^{2}}{f^{\text{sig}}_{i}+f^{\text{bkg}}_{i}},
\end{equation}
where $f^{\text{sig}}_{i}$ ($f^{\text{bkg}}_{i}$) is the signal (background) fraction of events in bin $i$.

\begin{figure}[h]
	\begin{center}
	\subfloat[Region I]{\includegraphics[width=0.45\textwidth]{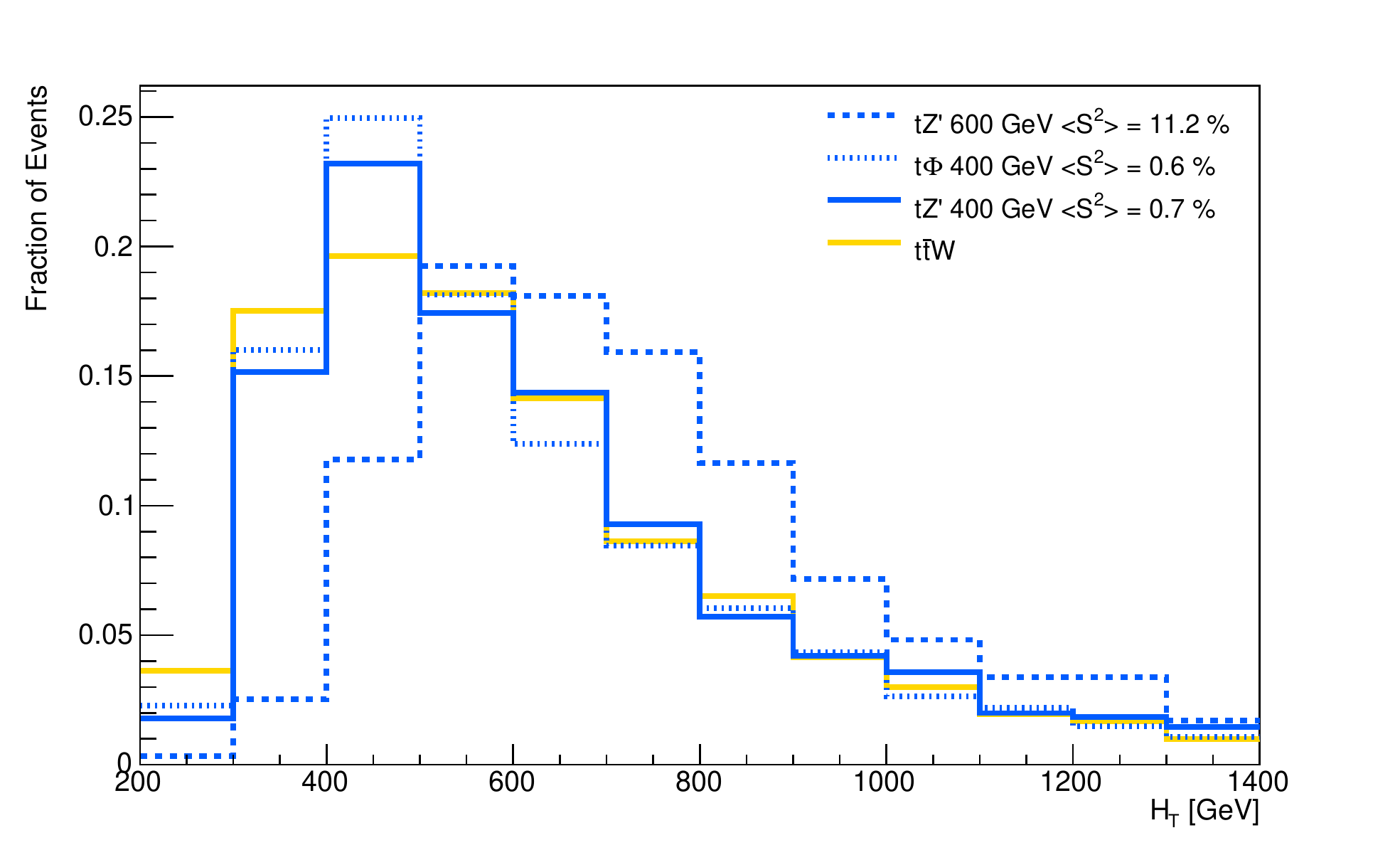}}\hspace{3mm}
	\subfloat[Region II]{\includegraphics[width=0.45\textwidth]{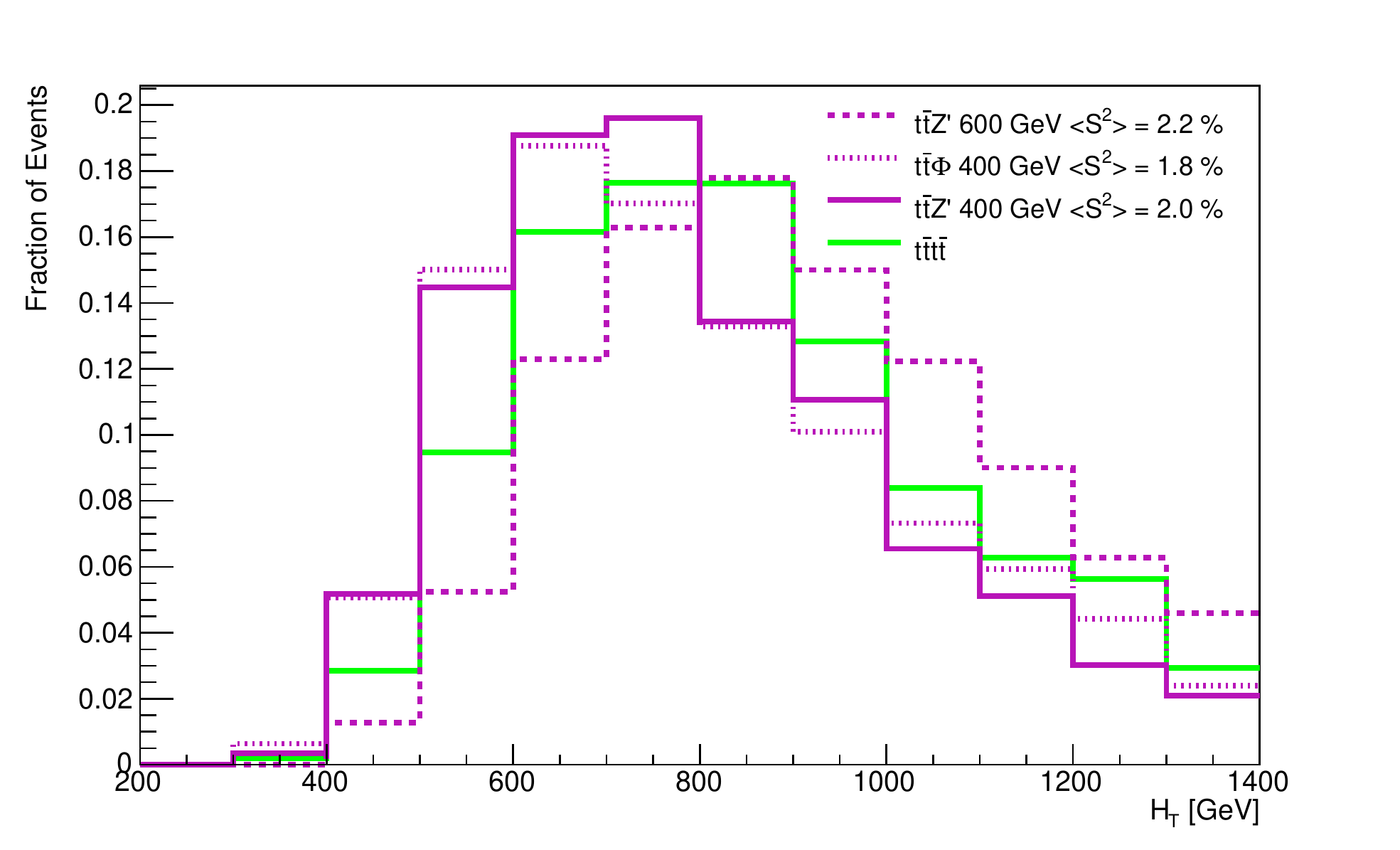}}\\
	\end{center}
	\caption{$H_{T}$ fraction of events for both signal-enriched regions.  As it can be seen, the 600 GeV BSM $H_T$ distribution has a separation an order of magnitude larger than the 400 GeV BSM scenarios for Region I, whereas for Region II all BSM scenarios are hardly distinguishable from SM.}
\label{hT}
\end{figure}

We plot in Fig.~\ref{hT} the $H_{T}$ distribution for both regions. We can see in both regions how the low mass (400 GeV) BSM scenarios mimic the SM distribution regardless of their spin ($Z'$ or $\Phi$), whereas the 600 GeV case begins to show some deviation from the SM expected distribution.

\captionsetup[subfigure]{labelformat=empty}
\captionsetup[subfigure]{labelformat=parens}

\begin{figure}[h]
	\begin{center}
	\subfloat[Region I]{\includegraphics[width=0.45\textwidth]{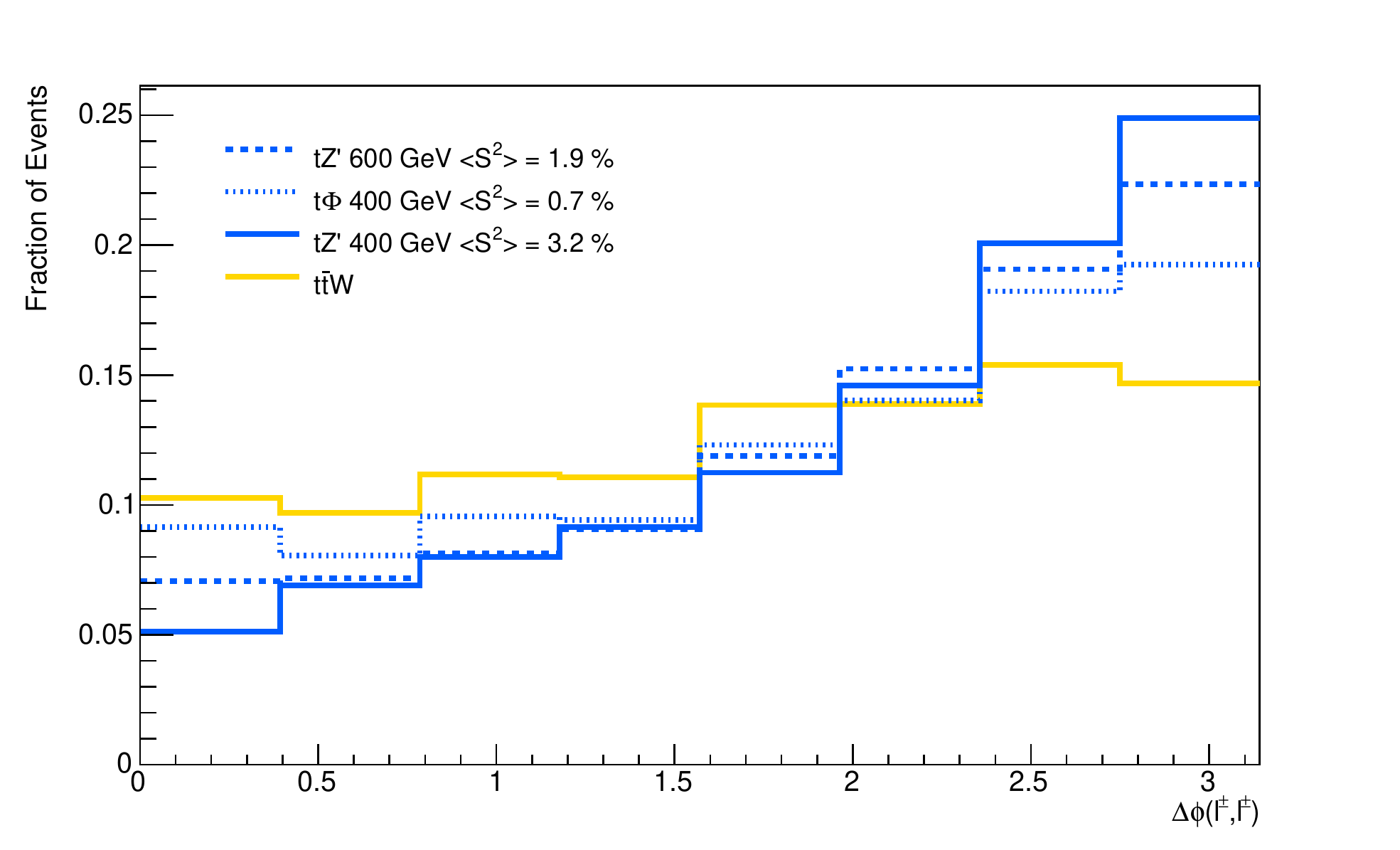}}\hspace{3mm}
	\subfloat[Region II]{\includegraphics[width=0.45\textwidth]{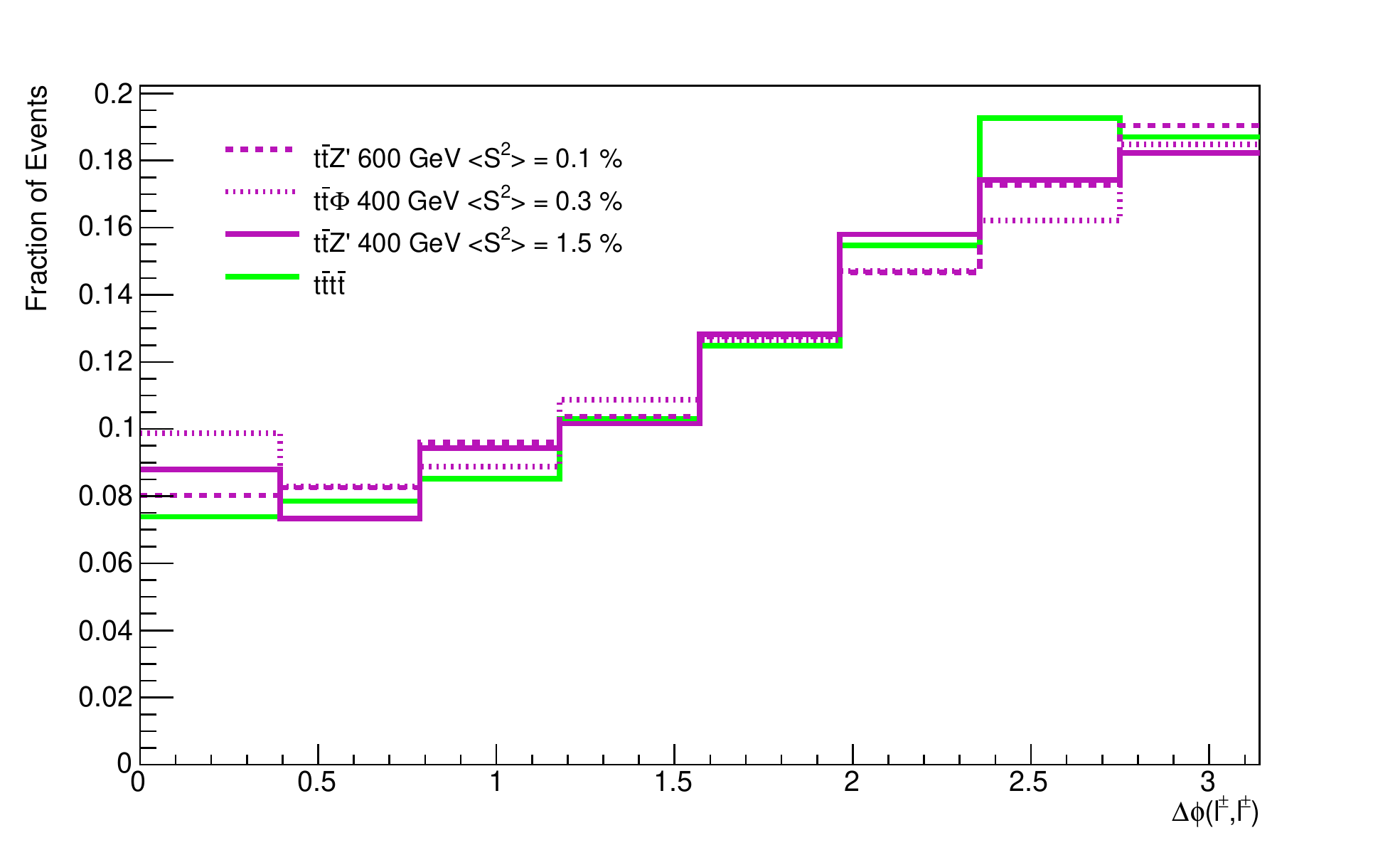}}\hspace{3mm}
	\end{center}
	\caption{Fraction of events of the azimuthal angular distance between the same-sign leptons for the main processes in Regions I and II.  We find it useful to disentangle $tZ'/\Phi$ from $t\bar t W$ in Region I.}
\label{deltaphill}
\end{figure}

The above results show that disentangling the 400 GeV BSM from the SM is challenging.  Considering $t Z'$ and $t\bar t W$ events, we see that in the diagrams the lines connecting both same-sign leptons have different Lorentz structure.  We therefore explore the azimuthal angular separation between the leptons $\Delta \phi (\ell^\pm,\ell^\pm)$, which is a spin-correlation-sensitive observable.   We observe in Fig.~\ref{deltaphill} that this observable can distinguish to some extent the different contributions in Region I, whereas no distinction occurs in Region II.  

\begin{figure}[h]
	\begin{center}
	\subfloat[Region I]{\includegraphics[width=0.45\textwidth]{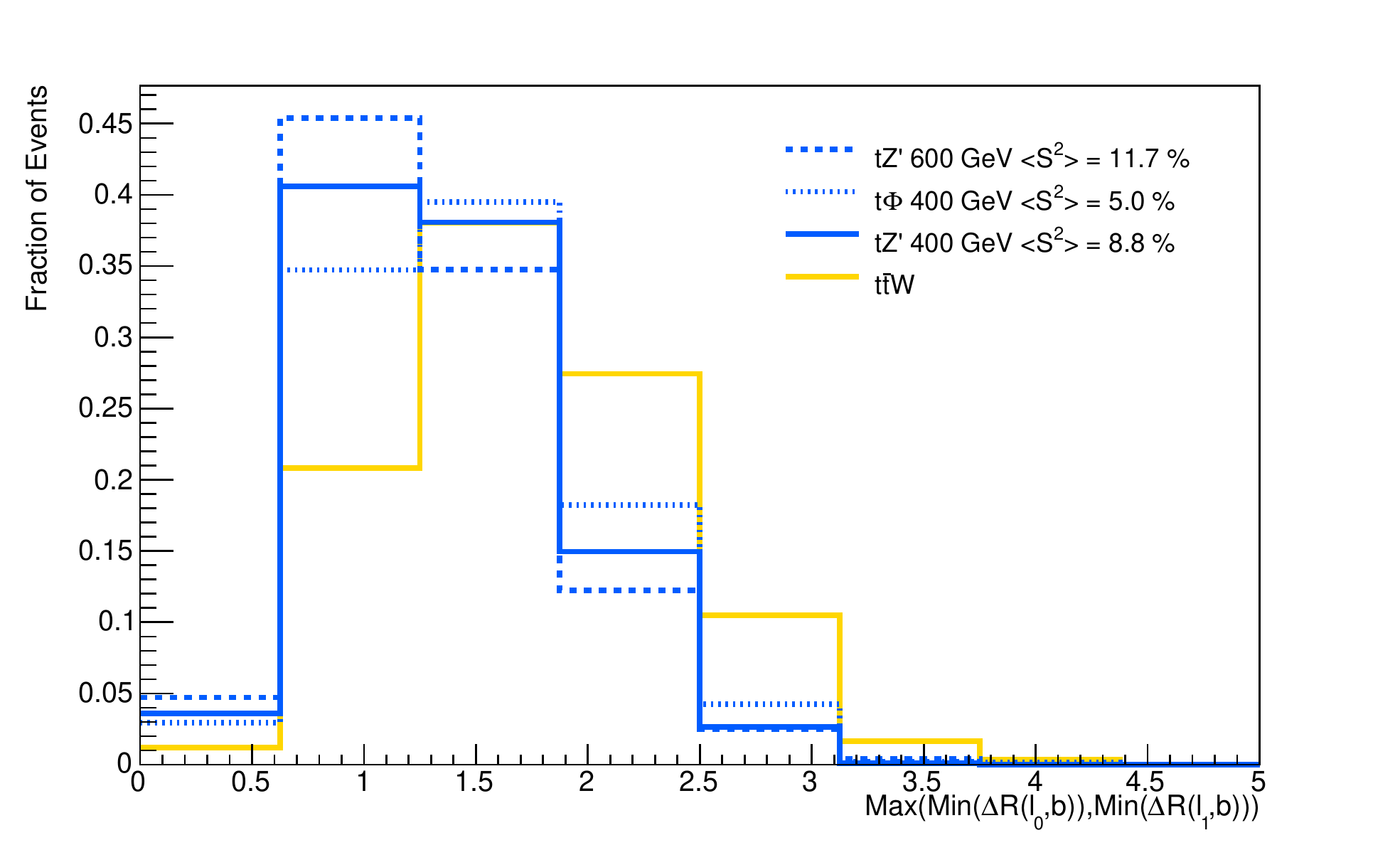}}\hspace{3mm}
	\subfloat[Region II]{\includegraphics[width=0.45\textwidth]{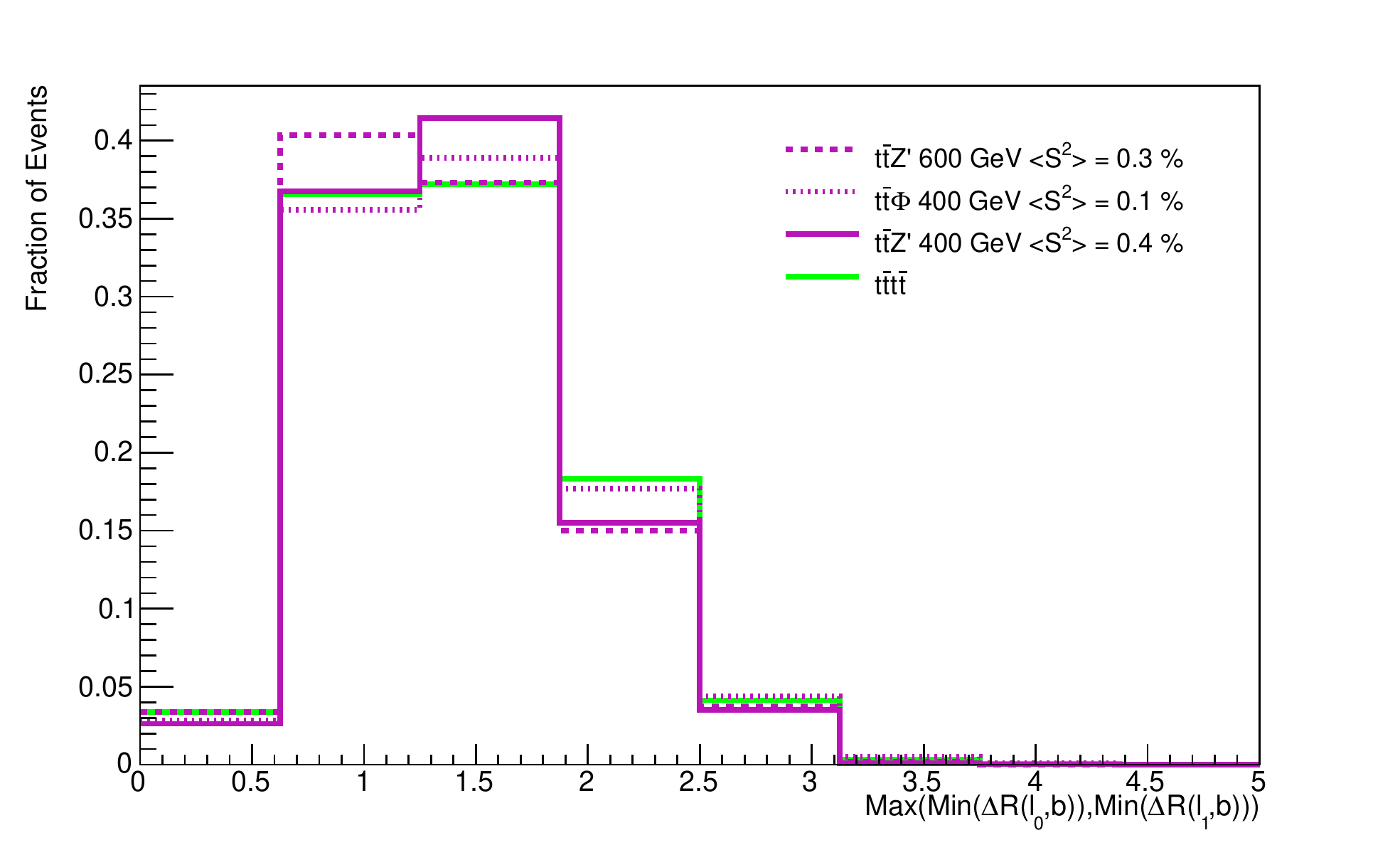}}\hspace{3mm}
	\end{center}
	\caption{Fraction of events of the MaxMin$(\ell,b)$ observable.  We find this observable to be useful for region I which is dominated by $tZ'/\Phi$ and $t\bar t W$, since the latter has a lepton coming from a prompt $W$ boson.}
\label{maxmindeltaR}
\end{figure}

We can further observe that all leptons in $tZ'$ come from top quarks, whereas in $t\bar t W$ there is one lepton coming from a prompt $W$ boson. In contrast to leptons coming from a top quark decaying to a $W$ boson and then to a lepton, the lepton coming from a prompt $W$ boson is not closely connected to a $b$-jet.  Motivated by this, we propose a new kinematic variable defined as:
\begin{eqnarray}
	\mbox{MaxMin($\ell,b$)} &=& \mbox{The maximum of the minimum $\Delta R$-distances}\nonumber \\
	&&\mbox{between the same-sign leptons and a $b$-jet.}
	\label{maxmin}
\end{eqnarray}
In the extreme case of boosted top quarks, we expect the $tZ'$ signal to have smaller MaxMin($\ell,b$) than the $t\bar t W$ background.  For less boosted top quarks we expect this qualitative behaviour to still hold, although at a lesser extent.  We show the  MaxMin($\ell,b$) distribution for both regions in Fig.~\ref{maxmindeltaR}, where the expected behaviour is verified.  Moreover, we find a slightly larger separation than for the $\Delta\phi(\ell^\pm, \ell^\pm)$ observable.  We have verified that the MaxMin($\ell,b$) observable has better separation for larger $H_T$, as expected from its construction.  

\captionsetup[subfigure]{labelformat=empty}
\captionsetup[subfigure]{labelformat=parens}

\captionsetup[subfigure]{labelformat=empty}
\captionsetup[subfigure]{labelformat=parens}

\begin{figure}[h]
	\begin{center}
	\subfloat[]{\includegraphics[width=0.45\textwidth]{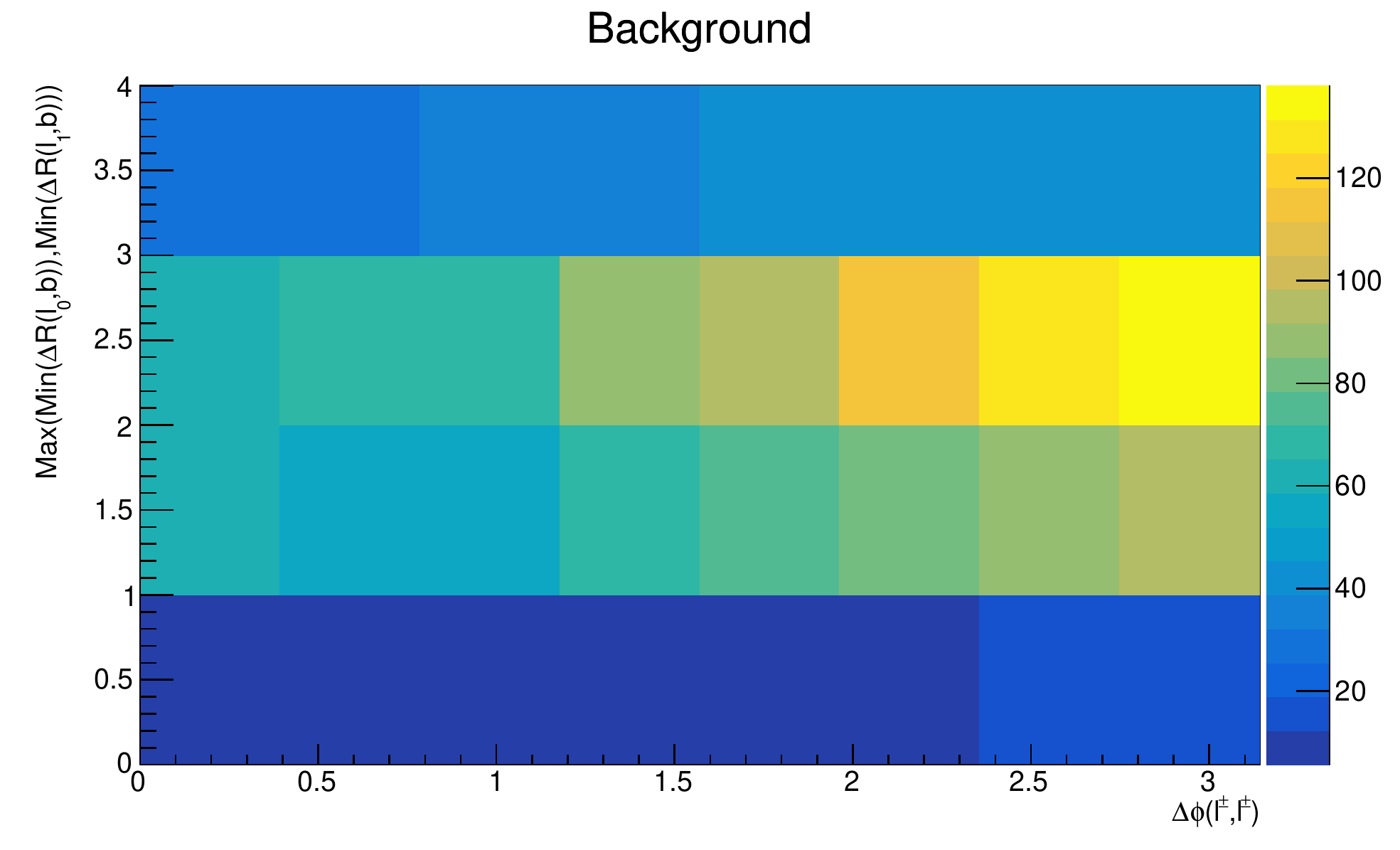}}\hspace{3mm}
	\subfloat[]{\includegraphics[width=0.45\textwidth]{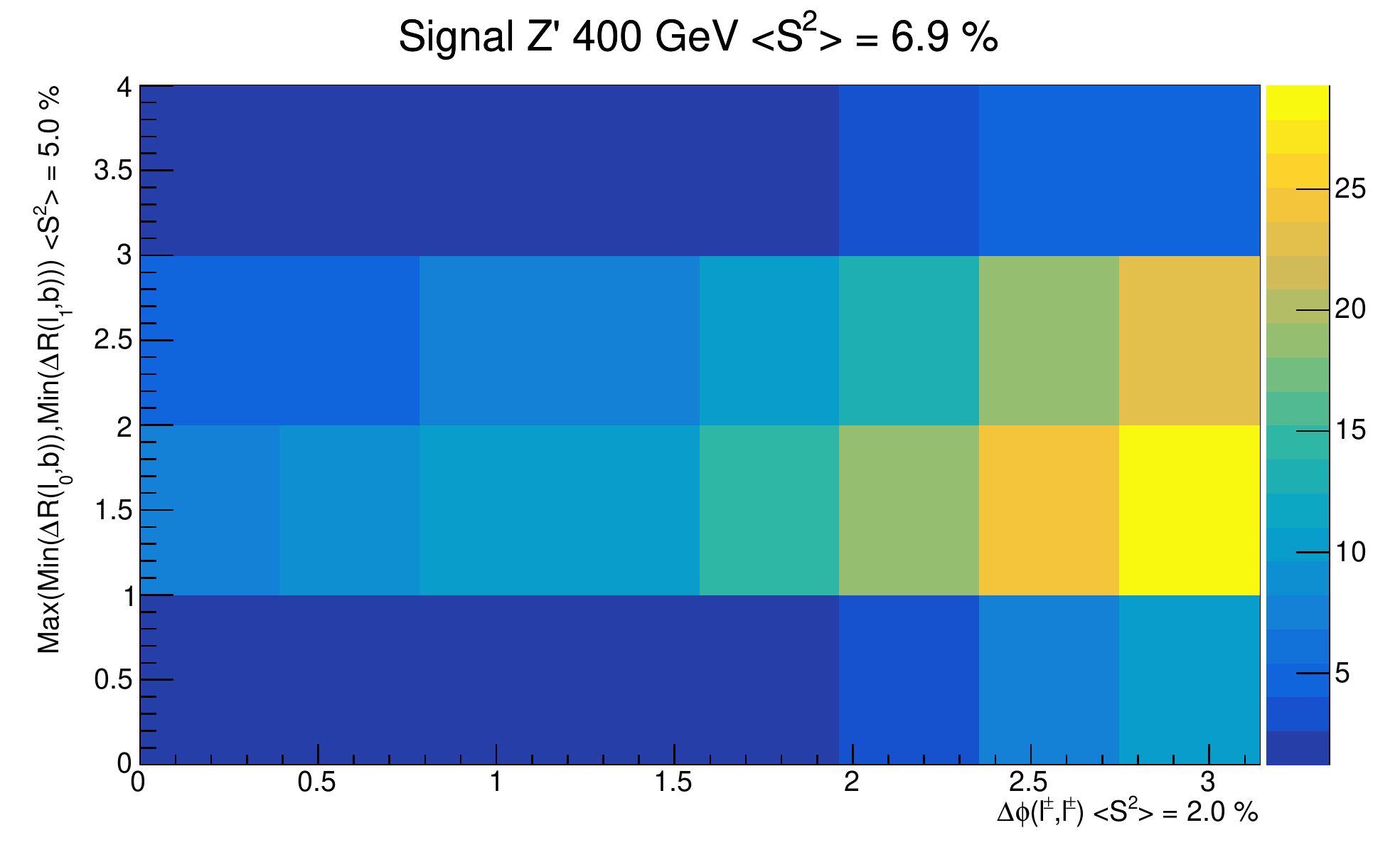}}\hspace{3mm}\\
	\subfloat[]{\includegraphics[width=0.45\textwidth]{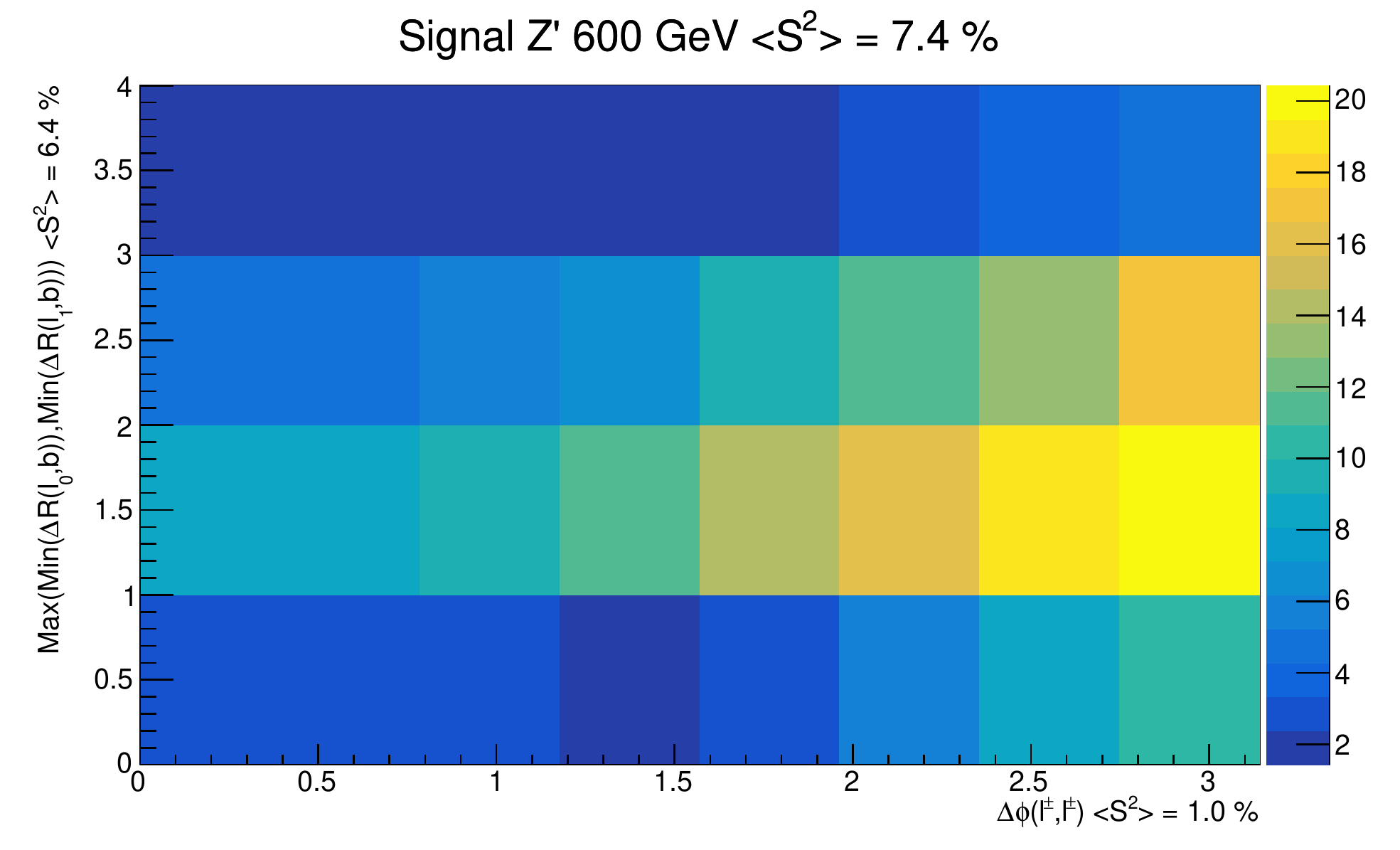}}\hspace{3mm}
	\subfloat[]{\includegraphics[width=0.45\textwidth]{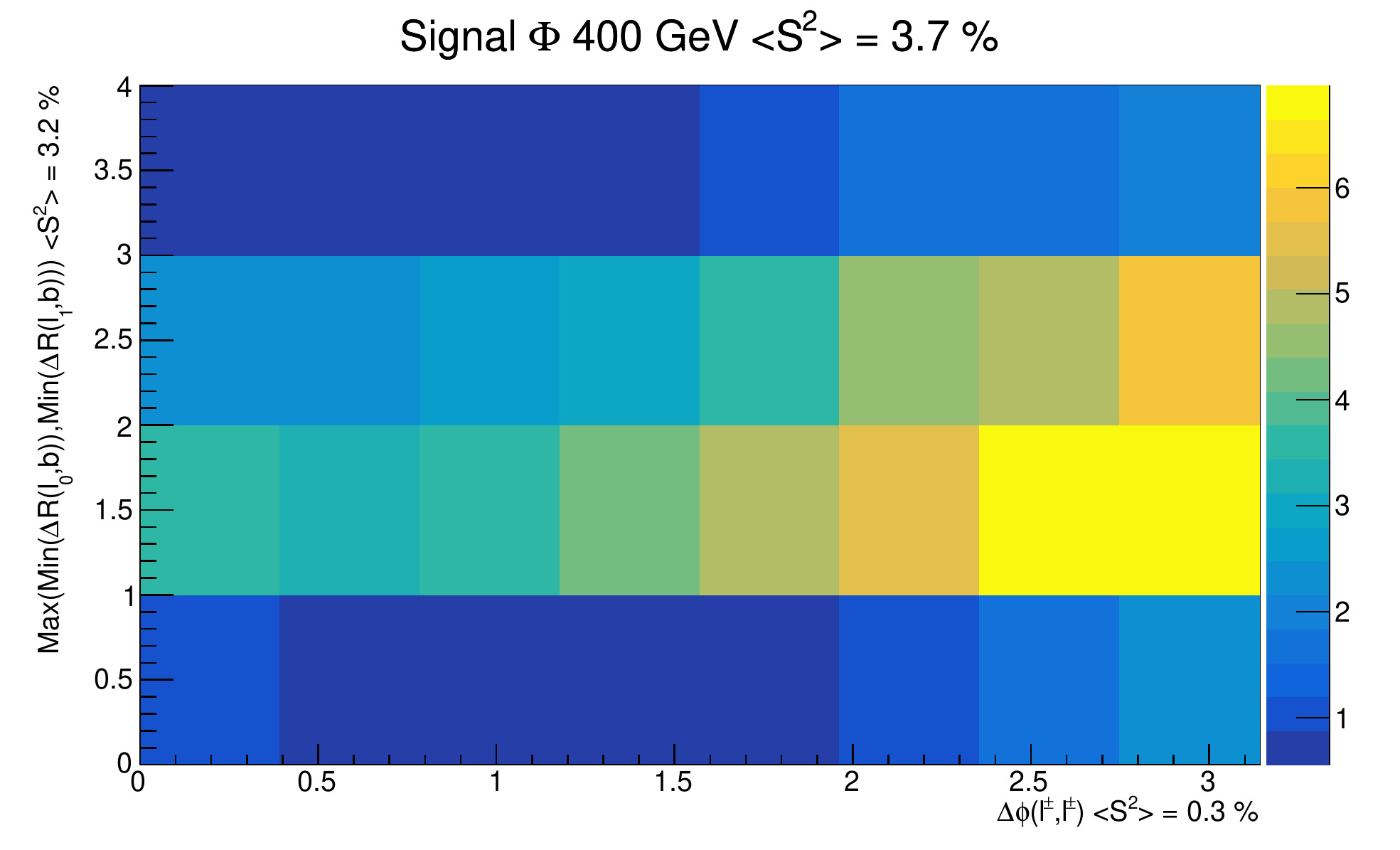}}\hspace{3mm}\\
	\end{center}
	\caption{Simulated event yields for different processes in the two dimensional space formed by the azimuthal angular distance between the two same-sign leptons (x-axis) and the MaxMin$(\ell,b)$ observable (y-axis). When comparing the reported separations to those in Figs.~\ref{deltaphill} and~\ref{maxmindeltaR}, one should take into account that this is for the global selection, not Region I.}
\label{hist2d}
\end{figure}

The previous observables, $\Delta \phi(\ell^\pm \ell^\pm)$ and MaxMin$(\ell,b)$, have shown a separation power between $tZ'/\Phi$ and $t\bar t W$.    We are interested in understanding  how independent is the separation power of these observables.  We show therefore in Fig.~\ref{hist2d} a two-dimensional event histogram where we plot the distribution over both observables to distinguish between signal and background. To avoid low Monte Carlo statistic, we use the global analysis selection, without further splitting in Region I and II. There, we count events for all background processes and for the three different signal processes. We see from the figure that the two studied observables are not strongly correlated and can therefore be exploited simultaneously in a future MVA analysis. 

\clearpage
\section{Summary and outlook}
\label{section:5}
We have addressed the pattern of mild but persistent anomalies in multilepton plus $b$-jets events at the LHC from a phenomenological point of view. More precisely, we have studied a BSM model with a $Z'$ in a mass range 200 GeV to 600 GeV that couples hierarchically to the right-handed up-quarks $u_R$, $c_R$ and $t_R$.  The hierarchy differentiates the third generation from the other two, allowing for $t\bar t$ diagonal couplings and FCNC $t\bar u$ and $t\bar c$ couplings.  Although we briefly discuss possible UV completions for this model, we consider that a more in-depth analysis in this direction, probing the additional expected new processes, would be very interesting and compelling.

The motivation for this BSM model is based on the multilepton plus $b$-jets anomalies pointing towards excesses in four-top-quarks analyses, as well as excesses in the positively-charged multilepton asymmetries resembling the $t\bar t W^\pm$ background. 

Throughout this article we focus primarily on the ATLAS $t\bar t H$ analysis in Ref.~\cite{ATLAS-CONF-2019-045} in which the post-fit yields of $t\bar tW^\pm$ show an excess over the expected cross-section by a factor of $\mu_{t\bar tW} = 1.39^{+0.17}_{-0.16}$.  We study the case in which the $t\bar t W^\pm$ cross-section is not free-floating in the fit, allowing the BSM model to fill this excess.  We explore this scenario by implementing the BSM model, simulating relevant processes up to detector level following the ATLAS analysis in two lepton same-sign (2LSS) and trilepton (3L) final states, and finding the BSM parameters that best fit the ATLAS results. We perform a new simultaneous fit to the observed ATLAS data in the eight bins corresponding to combinations of dilepton charge and $b$-jet multiplicity in 2LSS and 3L final states (Fig.~2 in Ref.~\cite{ATLAS-CONF-2019-045}). The region in parameter space that best fits the ATLAS data is shown in Figs.~\ref{zp-fits-1}-\ref{zp-fits-gct}.  We show in Fig.~\ref{bins_tth} how the BSM model improves the interpretation of the ATLAS results in Ref.~\cite{ATLAS-CONF-2019-045} without the need of re-scaling the $t\bar t W^\pm$ cross-section.  These results indicate that for this observable the data would prefer over the SM a $Z'$ with mass 400 GeV to 600 GeV and couplings of the order $g_{tt}\sim {\cal O}(10^{-1})$, $g_{tu} \sim g_{tc} \sim {\cal O}(10^{-2})$. When the BSM model is also tested to reproduce the slight excess in the ATLAS four-top-quarks analysis in Ref.~\cite{Aad:2020klt}, the preferred region for $g_{tt}$ is reduced to approximately half its value, since otherwise the model would populate this analysis with too many events.

We have verified that the best-fit regions are safe to other observables such as $D$-meson mixing, $Z'$-induced top quark rare decays, $t\bar t$ production, same-sign top-quark pair searches, $tj$ resonance searches in $t\bar t j$ events, and $Z'$-mediated single top-quark production.

In the second part of this paper we explore how the proposed BSM model could be distinguished from similar SM physics processes in the multilepton plus $b$-jets final state. In a first stage we define 34 (30) kinematic regions for the 2LSS (3L) channel that have different signal-to-background ratios and therefore could be exploited to disentangle them.  We perform a simultaneous fit of the regions in this global analysis search to the Asimov dataset assuming 139 fb$^{-1}$ and 300 fb$^{-1}$ integrated luminosity, and determine the prospects for discovering or excluding the proposed BSM model (Fig.~\ref{global-stat}). In a second stage, we define two kinematic regions tailored to increase the fraction of $tZ'$ and $t\bar t Z'$ events, respectively, and study observables that could help discriminating signal from the main SM background processes in each region. We find that $H_T$ could potentially discriminate $tZ'$ and $t\overline{t}Z'$ for $M_{Z'}$ above $\sim$ 600 GeV, whereas the $H_T$ distribution of $t\bar t Z'$ is very similar to that of the SM four-top-quarks process. We also find that the azimuthal separation between same-sign leptons $\Delta \varphi (\ell^\pm,\ell^\pm)$ is also a good variable to separate signal from its main background.  We also include the possibility of replacing the $Z'$ with a scalar $\Phi$ field and the same flavour structure, and find that the $\Delta \varphi (\ell^\pm,\ell^\pm)$ observable has discriminating power to differentiate the three models: SM, $Z'$ and $\Phi$.  Finally, we also propose a new kinematic variable MaxMin$(\ell,b)$ (c.f.~Eq.~\ref{maxmin}), which is likely to have a larger value if there is a lepton coming from a prompt $W$ boson in comparison to a $W$ boson coming from a top quark, which has a close $b$-jet as the top quark is boosted. This variable has a good separation power between signal and background.  We find that it is easier to distinguish signal from background in Region I, in which $tZ'$ is the main signal. As a last test, we investigate for all events in the previous global analysis whether $\Delta \varphi (\ell^\pm,\ell^\pm)$ and MaxMin$(\ell,b)$ are correlated by plotting their distribution in a $2D$ coloured histogram, and find them to have little correlation and thus considering them together improves the separation power in comparison to each observable on its own.

In summary, we have proposed a phenomenological FCNC $Z'$ model that couples hierarchically to the up-type right-handed quarks to explain LHC discrepancies in multilepton plus $b$-jet final states.  We have found regions in parameter space that fit the data better than the SM and proposed different ways to explore the data to test the BSM $Z'$ model. We find that a sophisticated experimental search, along the lines of our proposed global analysis could in the near future shed light on the existence of such a BSM scenario.
\section*{Acknowledgements}

M.S.~thanks IFAE, ICREA and UAB for its kind hospitality during part of the development of this work.   We thank D.~de Florian and L.~Da Rold for useful conversations.
A.J. is supported in part by the Spanish Ministerio de Econom\'ia y Competitividad under projects
RTI2018-096930-B-I00 and Centro de Excelencia Severo Ochoa SEV-2016-0588.

\clearpage
\begin{appendix}

\section{Pythia and Delphes parameters}\label{appendix_tunes}
 As detailed in Sect.~\ref{section:3}, we employ tuned parameters for {\tt Pythia\;8} and {\tt Delphes}. For the former, we use the Monash tune~\cite{Skands:2014pea} and we change the parameters shown in Table~\ref{table:pythia_params}. 
\begin{table}[h!]
\centering
 \begin{tabular}{|p{7cm}|p{3cm}|p{3cm}|} 
 \hline
Parameter & Default & Value\\ [0.5ex] 
 \hline\hline
TimeShower:alphaSvalue & 0.1365 & 0.127\\
SpaceShower:pT0Ref & 2.0 & 1.56 \\
SigmaProcess:alphaSvalue & 0.130 & 0.140 \\
SpaceShower:pTmaxFudge & 1.0 & 0.91 \\
SpaceShower:pTdampFudge & 1.0 & 1.05 \\
SpaceShower:alphaSvalue & 0.1365 & 0.127 \\
BeamRemnants:primordialKThard & 1.8 & 1.88 \\
MultipartonInteractions:pT0Ref & 2.28 & 2.09 \\
MultipartonInteractions:alphaSvalue & 0.130 & 0.126\\
ColourReconnection:range & 1.8 & 1.71\\ [1ex] 
 \hline
 \end{tabular}
 \caption{This table lists the {\tt Pythia\;8} parameters changed from their Monash Tune default values~\cite{Skands:2014pea}.}
 \label{table:pythia_params}
\end{table}

In the {\tt Delphes} case, we tune a {\tt Delphes} card for the $t\overline{t}H$ search~\cite{ATLAS-CONF-2019-045} in order to match the expected $t\overline{t}W^{\pm}$ event yields with the reference cross-section $\sigma = 727$ fb. We use almost the same {\tt Delphes} card for the four-top-quarks search~\cite{Aad:2020klt}, with the only difference being the b-tagging efficiency. The detailed features of the first {\tt Delphes} cards are:

\begin{itemize}
	\item We set the electron identification efficiency as
\begin{eqnarray}
	0.0&\text{ for }& p_{T} \leq 10.0\nonumber\\
	0.80\cdot(0.01\cdot(p_{T}-20.0)+0.65)&\text{ for }&10.0<p_{T}\leq 45.0\text{ and } |\eta|\leq 1.5\nonumber\\
	0.75\cdot(0.01\cdot(p_{T}-20.0)+0.65)&\text{ for }&10.0<p_{T}\leq 45.0\text{ and } 1.5<|\eta|\leq 2.5\nonumber\\
	0.80\cdot0.90&\text{ for }&45.0<p_{T}\text{ and } |\eta|\leq 1.5\nonumber\\
	0.75\cdot0.90&\text{ for }&45.0<p_{T}\text{ and } 1.5<|\eta|\leq 2.5\nonumber\\
	0.0&\text{ for }&|\eta|>2.5\nonumber
\end{eqnarray}
	\item We set the muon identification efficiency as
\begin{eqnarray}
	0.0&\text{ for }& p_{T} \leq 10.0\nonumber\\
	0.98\cdot(0.006\cdot(p_{T}-20.0)+0.80)&\text{ for }&10.0<p_{T}\leq 45.0\text{ and } |\eta|\leq 1.5\nonumber\\
	0.99\cdot(0.006\cdot(p_{T}-20.0)+0.80)&\text{ for }&10.0<p_{T}\leq 45.0\text{ and } 1.5<|\eta|\leq 2.5\nonumber\\
	0.98\cdot0.95&\text{ for }&45.0<p_{T}\text{ and } |\eta|\leq 1.5\nonumber\\
	0.99\cdot0.95&\text{ for }&45.0<p_{T}\text{ and } 1.5<|\eta|\leq 2.5\nonumber\\
	0.0&\text{ for }&|\eta|>2.5\nonumber
\end{eqnarray}
	\item We set the electron and muon isolation parameters as
\begin{eqnarray}
	R&=&0.3\nonumber\\
	p_{T}^{\text{min}}&=&1.0\nonumber\\
	I_{\text{min}}&=&0.2\nonumber
\end{eqnarray}
	\item We cluster jets with the anti-kt jet clustering algorithm~\cite{Cacciari:2008gp}, with $\Delta R = 0.4$ and $p_{T}^{\text{min}}=25.0$
	\item We set the jet energy scale as 
	\begin{equation}
	\sqrt{\frac{(3.0 - 0.2\cdot|\eta|)^2}{ p_{T}} + 1.0 }\nonumber
	\end{equation}
	\item We set the b-tagging efficiencies as
\begin{eqnarray}
\text{Default misidentification rate}&\text{: }&\frac{440.0}{313.0}\cdot(0.002+7.3\cdot 10^{-06}\cdot p_{T})\nonumber\\
\text{c-jet misidentification rate}&\text{: }&0.20\cdot \text{tanh}(0.02\cdot p_{T})\cdot\frac{1}{1+0.0034\cdot p_{T}}\nonumber\\
\text{b-jet efficiency}&\text{: }&0.80\cdot \text{tanh}(0.003\cdot p_{T})\cdot\frac{30}{1+0.086\cdot p_{T}}\nonumber
\end{eqnarray}
	\item We modify the $\tau$-tagging module to match the medium working point in Ref.~\cite{ATLAS-CONF-2019-045}
\begin{verbatim}
module TrackCountingTauTagging TauTagging {

  set ParticleInputArray Delphes/allParticles
  set PartonInputArray Delphes/partons
  set TrackInputArray TrackMerger/tracks
  set JetInputArray JetEnergyScale/jets

  set DeltaR 0.2
  set DeltaRTrack 0.2

  set TrackPTMin 1.0

  set TauPTMin 20.0
  set TauEtaMax 2.5

  # instructions: {n-prongs} {eff}

  # 1 - one prong efficiency
  # 2 - two or more efficiency
  # -1 - one prong mistag rate
  # -2 - two or more mistag rate

  set BitNumber 0

  add EfficiencyFormula {1} {0.55}
  add EfficiencyFormula {2} {0.40}
  add EfficiencyFormula {-1} {0.02}
  add EfficiencyFormula {-2} {0.002}

}
\end{verbatim}
\end{itemize}

For the four-top-quarks {\tt Delphes} card we modify the b-tagging efficiencies to

\begin{eqnarray}
\text{Default misidentification rate}&\text{: }&\frac{440.0}{313.0}\cdot(0.002+7.3\cdot 10^{-06}\cdot p_{T})\nonumber\\
\text{c-jet misidentification rate}&\text{: }&\frac{8.1}{4.0}\cdot 0.20\cdot \text{tanh}(0.02\cdot p_{T})\cdot\frac{1}{1+0.0034\cdot p_{T}}\nonumber\\
\text{b-jet efficiency}&\text{: }&\frac{77.0}{70.0}\cdot0.80\cdot \text{tanh}(0.003\cdot p_{T})\cdot\frac{30}{1+0.086\cdot p_{T}}\nonumber
\end{eqnarray}

This {\tt Delphes} card allows us to recover the expected yields for $t\overline{t}W^{\pm}$, four-top-quarks production, $t\overline{t}H$ and $t\overline{t}Z$ reported in Table 3 of Ref.~\cite{Aad:2020klt} to a very good degree of approximation.

\section{$Z'$ induced $D^0-\overline{ D^0}$ mixing}\label{appendix_d0mixing}
We compute the one-loop contribution to $D^0-\overline{ D^0}$ mixing due to a generic $Z'$ with arbitrary flavour-changing non-zero couplings $g^{L,R}_{ut,ct}$.  We observe that the general tree level computation has already been performed in \cite{Golowich:2007ka}, whereas the one-loop contribution with equal couplings for both chiralities $g^{L}_{ut,ct} = g^{R}_{ut,ct}$ can be found in \cite{Aranda:2010cy}.  Along this appendix we follow and expand the $|\Delta C| =2$ calculations in previous references to the one-loop case with different couplings for each chirality.

\begin{figure}[h]
        \begin{center}
		\subfloat[]{\includegraphics[width=0.4\textwidth, keepaspectratio]{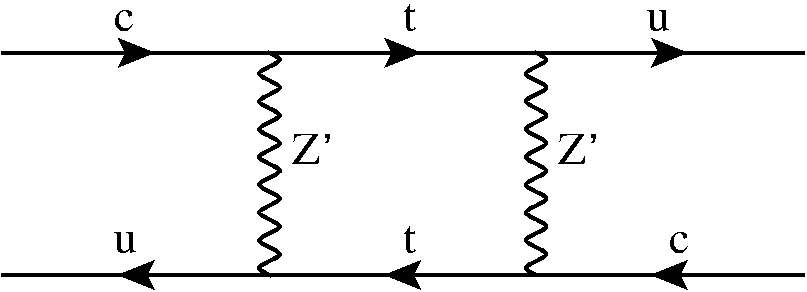}}\hspace{10mm}
        \subfloat[]{\includegraphics[width=0.4\textwidth, keepaspectratio]{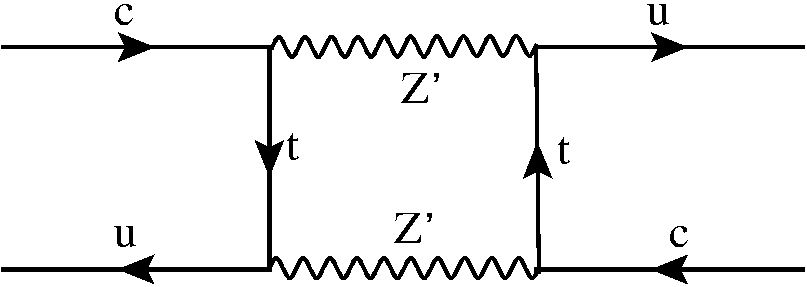}}
        \end{center}
	\caption{One-loop box Feynman diagrams contributing to $D^0-\overline{ D^0}$ mixing due to a generic $Z'$ with non-zero off-diagonal couplings to the top quark.}
\label{Dmix}
\end{figure}

The amplitude corresponding to the Feynman diagrams in Fig.~\ref{Dmix} reads
\begin{eqnarray}
	{\cal M} &=& 2 \int \frac{d^4k}{(2\pi)^4} \frac{\left[ \, \overline{ {\cal P}_{ut} u }\gamma^\lambda (\not k + m_t) \gamma^\nu {\cal P}_{ct} c \right] \left[ \, \overline{ {\cal P}_{ut} u }\gamma^\nu (\not k + m_t) \gamma^\lambda {\cal P}_{ct} c \right] }{(k^2-m_t^2)^2\, (k^2-M_{Z'}^2)^2} 
	\label{box}
\end{eqnarray}
where
\begin{eqnarray}
	{\cal P}_{ut} &=& g^L_{ut} P_L + g^R_{ut} P_R \\
	{\cal P}_{ct} &=& g^L_{ct} P_L + g^R_{ct} P_R 
\end{eqnarray}
and $P_{L,R} = (1 \mp \gamma^5) /2$.  We work in the low energy limit for $u$ and $c$.

In order to perform the $k$-integral in Eq.~\ref{box} we observe that by symmetry arguments any integral with odd number of $k$'s in the numerator vanishes.  We are then left with the two relevant terms proportional to  $\not k \not k$ and $m_t^2$.  We can then use the identity
\begin{equation}
	\frac{1}{A^m\, B^n} = \frac{\Gamma(m+n)}{\Gamma(m)\,\Gamma(n)} \int_0^\infty d\lambda \frac{\lambda^{m-1}}{(\lambda A + B)^{m+n}}
\end{equation}
to perform $k$-integrals through the usual $d$-dimensions integrals such as
\begin{equation}
	\int \frac{d^dk}{(2\pi)^d} \frac{k^\mu k^\nu}{(k^2 - \Delta)^n} = \frac{(-1)^{n-1} i}{(4\pi)^{d/2}} \frac{g^{\mu\nu}}{2} \frac{\Gamma(n-\frac{d}{2}-1)}{\Gamma(n)} \left( \frac{1}{\Delta}  \right)^{n-\frac{d}{2}-1},
\end{equation}
setting $d=4$ in all cases.  After some algebra and some Fierz transformations, we obtain an explicit expression for ${\cal M}$ without any integral.  By relating ${\cal M}$ terms to four-fermion effective operators $Q_{1...8}$ (see Ref.~\cite{Golowich:2007ka}) we obtain an expression for the effective Lagrangian ${\cal L}_{eff}$. Using $\langle Q_i \rangle = \langle \overline{ D^0} | Q_i | D^0 \rangle$ and the {\it modified vacuum saturation} hypothesis as defined in \cite{Golowich:2007ka}, we obtain an explicit expression for $\langle \overline{ D^0} | {\cal L}_{eff} | D^0 \rangle$.  Utilizing the definition for 
\begin{equation}
	\Delta M_D = - \frac{1}{M_D} \langle \overline{ D^0} | {\cal L}_{eff} | D^0 \rangle
\end{equation}
we obtain the one-loop $D$-meson mixing parameter for chiral off-diagonal couplings with the top quark
\begin{eqnarray}
	\Delta M_D &=& \frac{f_D^2\, M_D\, B_D\, x}{64 \pi^2 M_{Z'}^2} \left[ f(x) \left( -\frac{8}{3} {g_{ut}^{L}}^2 {g_{ct}^L}^2 -\frac{8}{3} {g_{ut}^R}^2 {g_{ct}^R}^2 + \frac{80}{3} g_{ut}^L g_{ct}^L g_{ut}^R g_{ct}^R \right) \right. \nonumber \\ &&\left.  + g(x) \left( \frac{2}{3} {g_{ut}^{L}}^2 {g_{ct}^R}^2 + \frac{2}{3} {g_{ut}^{R}}^2 {g_{ct}^L}^2 - \frac{14}{3} g_{ut}^L g_{ct}^L g_{ut}^R g_{ct}^R \right)  \right] .
	\end{eqnarray}
Here $x = (M_{Z'}/m_t)^2$, $f_D = 222.6$ MeV \cite{Artuso:2005ym} is the $D^0$-meson decay constant, $B_D \approx 1$ \cite{Aranda:2010cy,Gupta:1996yt} is the bag model parameter, and
\begin{eqnarray}
	f(x) &=& \frac{1}{2}  \frac{1}{(1-x)^3} [ 1-x^2 + 2 x \log x ] \\
	g(x) &=&  \frac{2}{(1-x)^3} [2(1-x)+(1+x) \log x ]
\end{eqnarray}
are the functions that appear when integrating in $\lambda$ in the above scheme.

As it can be easily seen, one retrieves the results in Ref.~\cite{Aranda:2010cy} if Left and Right couplings are set equal.
\end{appendix}
\clearpage
\bibliographystyle{JHEP}
\bibliography{biblio}
\end{document}